\documentclass[trackchanges]{aastex631}
\usepackage{amsmath}

\submitjournal{the Astronomical Journal (AJ)}

\shorttitle{Radio SETI Strategy}
\shortauthors{Houston et al.}

\begin{document}

\title{Strategies for Maximizing Detection Rate in Radio SETI}

\correspondingauthor{Kenneth Houston}
\email{khouston@berkeley.edu}

\newcommand{\UCB}{Department of Astronomy,  University of California Berkeley, Berkeley CA 94720}
\newcommand{\SETI}{SETI Institute, Mountain View, California}

\author{Kenneth Houston}
\affiliation{\UCB}

\author{Andrew Siemion}
\affiliation{\UCB}
\affiliation{\SETI}

\author{Steve Croft}
\affiliation{\UCB}
\affiliation{\SETI}

\begin{abstract}

The Search for Extraterrestrial intelligence (SETI) is a scientific and cultural effort seeking evidence of intelligent life beyond earth.  Radio SETI observes the radio spectrum for ``technosignatures" that could be produced by an advanced ET society.  This work models radio SETI as an end-to-end system, and focuses on narrow-band intentional transmissions.  We look at strategies to maximize the expected number of detections per year (DPY) of search.  Assuming that ET civilizations will be associated with star systems, we want to maximize the number of stars that may be observed at one time.  Assuming a representative star density, this requires maximizing the search volume in a cone defined by the detection range and field of view (FOV).  The parameter trades are modified from the case where one simply maximizes signal-to-noise ratio.  Instead, a joint optimization between FOV and sensitivity is needed.    Some implications: 1) Instead of focusing on the terrestrial microwave window of $1-10$\,GHz, frequencies below 1\,GHz may be optimal for detection rate due to the larger field of view;  2) Arrays of smaller dishes should be favored compared to a single dish of equivalent area;  3) Aperture arrays are desirable due to their large potential FOV.  Many radio telescopes under development will provide both high sensitivity and large FOV, and should offer much improved SETI detection rates.  Still higher DPY is needed, however, to achieve results in reasonable time horizons, which should be possible by greatly expanding computation capability to the next-generation wide-FOV antenna arrays. 

\end{abstract}


\section{Introduction} \label{sec:intro}

The universe is unimaginably vast.  According to recent estimates, there are potentially 2 trillion galaxies \citep{Conselice_2016ApJ} and $10^{22}$ to $10^{24}$ stars in the universe\footnote{
\url{https://www.esa.int/Science_Exploration/Space_Science/Herschel/How_many_stars_are_there_in_the_Universe}}.  With such numbers, it is inconceivable to many that our Earth and solar system would be unique and intelligent life would not have evolved elsewhere.  It is up to us to try to detect it.  

SETI may be viewed as a cost-effective form of space exploration when compared to manned or unmanned efforts \citep{Drake_1984}.  While there are many facets to SETI (observing at radio, optical, and infrared frequencies), this analysis focuses on strategies for ETI search of the radio frequency (RF) microwave spectrum near the “Terrestrial Microwave Window” over roughly $1-10$\,GHz \citep{Drake_1984}.  Radio SETI looks for “technosignatures”, which are signals with characteristics suggesting an intelligent extraterrestrial (ET) source that can be distinguished from natural phenomena and terrestrial radio-frequency interference (RFI).

Past SETI efforts, strategies and metrics are summarized in the 2018 NASA Technosignatures Workshop \citep{Wright_2018_NASA}, and appear in works by \citet{Drake_1961}, the SETI Science Working Group \citep{Drake_1984}, \citet{Tarter_SETI_2001},  \citet{Tarter_SETI_SPIE_2001}, \citet{Dreher_1997}, \citet{Shostak_2000}, \citet{Siemion_2015}, \citet{Tingay_2016}, \citet{Garrett_2017}, \citet{Enriquez_2017}, \citet{Wright_2018AJ}, \citet{Tingay_2018_Galactic_Anticenter}, \citet{Tingay_2018_Oumuamua}, \citet{Price_2020}, \citet{Wlodarczyk_Sroka_2020}, \citet{Sheikh_2020},    \citet{MWA_Tremblay_Vela_2020}, and many others.

Optimizing the SETI detection rate has many parallels to the problem of maximizing Fast Radio Burst (FRB) and pulsar detection rates \citep{Macquart_2011,Macquart_2014}.

This work looks at strategies to maximize the expected detection rate, which we express as the number of detections per year (DPY) of search.  Two strategies are examined: wide-field search and targeted search.  We analyze star counts and DPY for each.
In addition, 1) existing and future radio telescopes are compared for the number of visible stars; 2) search quality metrics are defined and compared to others; 3) the observation times required for a high probability of detection are estimated; and 4) a set of requirements for a radio SETI system is proposed.   

We find that wide-field search leveraging up-coming generations of radio telescopes should be able to dramatically improve DPY, and therefore the chances for a successful SETI program.  As detection rate is (to first order) independent of what part of the sky is scanned, radio SETI should be highly compatible with existing and future interferometric radio telescopes that perform wide sky surveys.  Therefore, commensal observations may be a cost-effective option for future RF SETI work.

\section{The SETI System} \label{sec:seti system}

\begin{figure}[ht!]
\plotone{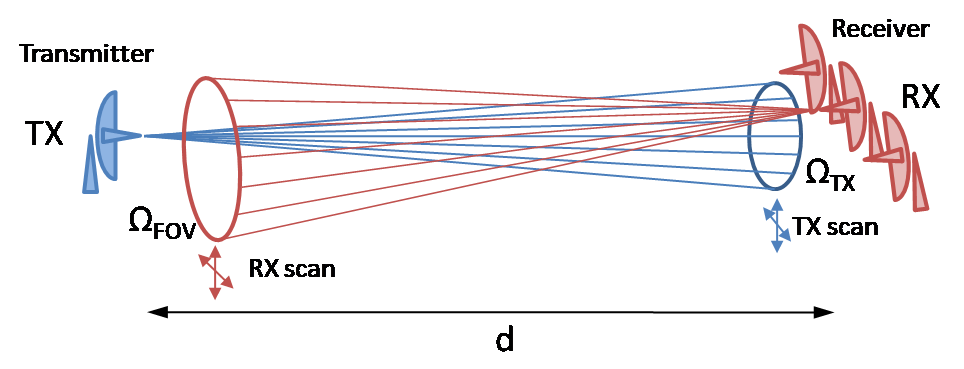}
\caption{SETI System Scenario}
\label{fig:SETISystem}
\end{figure}

An end-to-end SETI system may be visualized in Figure $\ref{fig:SETISystem}$.  Essential elements include:
\begin{itemize}
 \item{The transmitting (TX) party chooses transmitter parameters (power, beam width, waveform) and a scan strategy in spatial direction and frequency, and begins transmissions.}
 \item{The receiving (RX) party chooses receiver parameters (array configuration, frequency range, integration/detection parameters) and a scan strategy, and begins receiving.}
 \item{If TX and RX line up in space, time and frequency, with adequate receive power, a detection can occur.}
\end{itemize}
The SETI system design problem involves elements of strategy and game theory \citep{Kerins_2021} owing to the lack of communication between the TX and RX parties at the onset.  Our assumptions:
\begin{itemize}
\item{If the TX and RX parties understand the SETI problem and want to succeed in radio contact, both will do an end-to-end design of the SETI system and design their parts accordingly, based on their perception of the other’s likely strategy.}  
\item{The TX and RX parties will make similar, but not identical, system design decisions.  Though technology levels may be quite different between the societies, fundamental physics should force similar approaches.  }
\item{The TX design should allow for a minimal search space on receive so as to reduce receiver complexity and cost, while at the same time offering a waveform that can be distinguished from natural sources and local RFI.}
\item{A targeted search by both TX and RX parties might be prudent if astronomical observations suggest local star systems are promising for ET contact.  However, unless there is compelling \emph{a priori} information, the system design should assume a wide field survey with near $4\pi$ spherical coverage.  Regardless, we examine both sky surveys and targeted searches below.}
\item{The goal of both the TX and RX parties is to maximize the probability that an ET detection will occur, within a period of time commensurate with the attention span and patience of their host societies, and within constraints related to available RF equipment, power, computation, and cost.}
\end{itemize}
Other fundamental assumptions are listed below.
\begin{itemize}
\item{Transmit power will be severely limited, which favors intentional reception of systematically-scanned beacon signals over incidental reception of communications (TV, radio, data), radar and navigation links.  We therefore concentrate on intentional ``Active SETI" transmissions for our analysis.  Nothing precludes reception of incidental ET sources, but the following factors work against receiving them:
\begin{itemize}
\item{As will be described later, terrestrial signal sources have low power levels compared to what is required for reception at interstellar distances.  We assume that this will be the case for ET sources as well.}  
\item{Incidental sources on planetary bodies may not scan in a manner that covers the whole sky, reducing the chances of TX-RX alignment.} 
\item{If terrestrial technology evolution is a guide, transmissions will become more broad-band, noise-like, and power-efficient as technology progresses, making detection more difficult.}
\end{itemize}
}
\item{Beacons are highly speculative: propagation times may be measured in centuries, with an unknown probability of reception and response.  As a consequence, average power can’t be a burden to the TX society, and transmissions can’t cause excessive levels of local RFI.}
\item{Significant transmit antenna gain $G_{TX}$ is required to extend range with limited transmit power $P_{TX}$.  This increases Equivalent Isotropically Radiated Power (EIRP) = $P_{TX} G_{TX}$ at the expense of lower angular coverage and longer scan times.  To cover a full $4\pi$  sphere, we assume two independent TX sites ($N_{TXsite}=2$), each with hemispherical $2\pi$ coverage.  With a transmitted solid angle $\Omega_{TX}$, at least $N_{TXscan} = 4\pi/(N_{TXsite} \Omega_{TX})$ transmissions (dwell cycles) are needed before repeating.  Note that $G_{TX} = 4\pi/\Omega_{TX} = N_{TXsite} N_{TXscan}$, so TX antenna gain and scan time go hand-in-hand. }
\item{The TX system will be at one pointing direction for a time $T_{TXdwell}$.  The total scan time is $T_{TXscan}=N_{TXscan}\:T_{TXdwell}$.  $T_{TXdwell}$ might be on the order of tens of minutes, and $T_{TXscan}$ on the order of weeks or months, so $T_{TXscan} \gg T_{TXdwell}$.}
\item{Multiple receive sites may be required for $4\pi$ receive coverage, each working independently over a designated section of the sky.  The number of receiver scans before repeating will be $N_{RXscan} = 4\pi/(N_{RXsite} \Omega_{FOV})$, where $N_{RXsite}$ is the number of viewing sites and $\Omega_{FOV}$ is the receiver Field of View (FOV) for each site.}
\item{The TX and RX systems will be scanning asynchronously.  The TX system will be transmitting $T_{TX}$ seconds within the $T_{TXdwell}$ interval, with a net duty cycle of $\delta_{TX}=T_{TX}/T_{TXdwell}$. For simplicity, we assume that the RX system will be operating at nearly 100\% duty cycle.  Several scanning modes can be envisioned as will be described below.  The receive system can be based on antenna dishes or phased aperture arrays (AAs).}
\item{Regularity of reception will be required for the RX party to take notice, limiting the TX scan time and therefore antenna gain to values corresponding to “reasonable” time spans (weeks or months -?).  Note that ET civilizations may have different attention spans from us, depending on their patience but also their perception of time.}
\item{The need for reliable detection with long integration times severely limits waveform choices.  The transmitted signal will presumably have a beacon component for detectability and a limited message component.  There are many possible ways to integrate a message into the TX signal such as:
\begin{itemize}
\item{Modulating the beacon at a low rate, either by phase or amplitude}
\item{Use of frequency modulation}
\item{Use of modulated side bands adjacent to the beacon}
\item{Use of time division multiplexing (alternating between a beacon and a message)}
\end{itemize}
Spread-spectrum modulation for beacons and messaging is also possible \citep{Messerschmitt_Spread_Spectrum_2012,Messerschmitt_SETI_Comms_PwrEff_2013,Messerschmitt_SS_SETI_Comms_2015}, but is not considered here.  In general, discovery and decoding of these signals is much more difficult without knowledge of the encoding scheme. However, these waveforms may be attractive from the standpoint of TX energy efficiency. 

There will be trades between TX energy devoted to the beacon and message components that are beyond the scope of this paper.  We simply assume that there is a narrow-band beacon component that can be detected with a combination of coherent and non-coherent integration.
}
\item{Sinusoids and Linear FM chirps are obvious candidates for beacons over more complex waveforms due to their predictable phase which allows coherent integration.  Note that any constant-frequency sinusoid will be received as a chirp caused by relative TX/RX accelerations due to planetary rotation and orbital motion \citep{Siemion_2015,Sheikh_2019}.} 
\end{itemize}

\section{Expected ET Detection Rate } \label{sec:ET Det Rate}

A SETI program searches for the existence of technosignatures from ET civilizations, which we have assumed to be associated with star systems\footnote{While we have associated civilizations with star systems, partly as a means of developing a quantitative model, our approach of maximizing search cone volume should also optimize detection of interstellar civilizations, or the Milky Way's ``technological diaspora''.}.  A key metric to be maximized is the expected number of ET civilizations detected per year of search effort, or Detections per Year (DPY).  This is fundamental to the probability of success of a SETI effort.  
We can define two types of \text{SETI} strategies:

\begin{itemize}
\item{Wide Field Search (WFS) with near-full-sky scans.  This is best done on systems with a high scan rate, requiring a large FOV and high sensitivity at the same time.  There are two variants: }
\begin{itemize}
\item{Fast WFS (F-WFS): Scanning is rapid, so that the amount of time spent receiving in a specific scan direction is less than the TX dwell time $T_{TXdwell}$.   A new increment of the sky is observed for every averaging interval $\tau$, either by stepping or scanning at a constant angular rate. }
\item{Slow WFS (S-WFS): Scanning is incremental.  The amount of time spent receiving at a specific scan direction is much longer than the TX dwell time  $T_{TXdwell}$, but less than the transmitter scan time $T_{TXscan}$.  The receiver increments to a new scan direction, and ``stares'' over many integration periods.}
\end{itemize}
\item{Targeted Search (TS):  The receiver operates over a small solid angle.  Candidate stars are defined, and the radio telescope is successively steered to each star and observed.  Manual steering is employed for single dishes and ``digital pointing'' for arrays.  The key difference from S-WFS is that individual stars are targeted for observation.  To counter RFI, it may be necessary to implement a sequence of ``On-Off'' observations where ``On'' faces the star and ``Off'' points away from the star.  This is the conventional approach for single-dish systems. }
\end{itemize}

For each strategy we will evaluate the DPY, and place all approaches into a common framework.

\begin{figure}[ht!]
\plotone{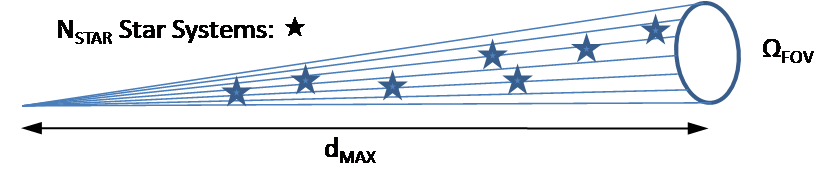}
\caption{Star Systems within Search Cone}
\label{fig:StarsInCone}
\end{figure}

\subsection{Wide-Field Search}

The most general survey method involves scanning across a wide solid angle and surveying a large number of candidate star systems.  The number of candidate solar systems within a search volume is the product of the volume and an average star density $\rho_{STAR}$, per Figure $\ref{fig:StarsInCone}$.  The search volume is a cone  (more precisely a spherical sector) of height $d_{MAX}$, the maximum detection range, with an angular extent based on the solid angle $\Omega_{FOV}$.  We assume a nominal star density of 0.1 star per cubic parsec in our neighborhood of the Milky Way, as did \citep{Siemion_2015}, though system trades are insensitive to the exact value.  This is a local density corresponding to the Milky Way's disk at the Sun's radius from the galaxy center \citep{Shostak_2000}.

We therefore get

\begin{equation} \label{eq:eqn1}
N_{STAR} = Expected\:Stars\:in\:Search\:Volume \approx \rho_{STAR}\:V_{CONE} = \onethird \rho_{STAR}\:\Omega_{FOV}\:{d_{MAX}}^{3}
\end{equation}

One may anticipate that there will be a trade between field of view and maximum detection distance.  Detection distance is a function of EIRP and sensitivity, which in turn depends on effective collecting area and system temperature $T_{sys}$.  There will be a complicated relationship of these quantities over frequency, with lower frequencies favoring FOV but not  $T_{sys}$.

\begin{figure}[ht!]
\epsscale{0.7}
\plotone{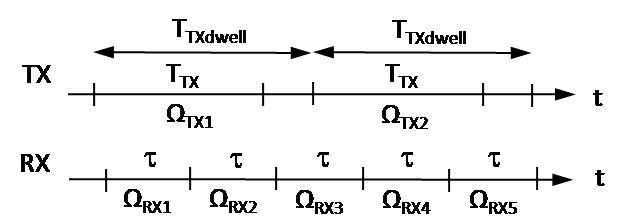}
\caption{Notional TX and RX Timeline}
\label{fig:SearchTimeline}
\end{figure}

A notional timeline is shown in Figure $\ref{fig:SearchTimeline}$.  The transmitter is scanning a section of sky  $\Omega_{TXn}$ over n=1 to $N_{TXscan}$, and transmitting for a time $T_{TX}$ within $T_{TXdwell}$.  For Fast WFS, the receiver is scanning  $\Omega_{RXn}$ over n=1 to $N_{RXscan}$, and integrating over an interval $\tau$ for each.  Depending on the antenna type, the receiver can be jumping FOV in increments (as shown) or scanning continuously (e.g. a drift scan or transit scan) so as to cover a new $\Omega_{RX}$ every $\tau$ seconds.  For Slow WFS,  $\Omega_{RXn}$ is constant over many TX dwell cycles and then will increment to a new look direction.

To estimate DPY, divide range into a large number of range bins $\{d_i\}$ of depth $\Delta d$ , and sum expected detections over all range bins:

\begin{eqnarray}
DPY_{WFS} &=& E(ET\:detections\:per\:year\:for\:Wide\:Field\:Search) \nonumber \\
&=& \sum\limits_{i} E(ET\:detections\:per\:TX\:dwell\:in\:d_i )\:N_{TXdwell/year} \nonumber \\
&=& \sum\limits_{i} P_{civTX} P_{TXtoRX}  N_{tpd}  P_D(d_i)  N_{STAR}(d_i ) \frac{T_{year}}{T_{TXdwell}} \nonumber
\end{eqnarray}

where summation is over range bins $d_i$, and

{\addtolength{\leftskip}{10 mm}
\setlength{\parindent}{0 mm}

$ $

$N_{STAR}(d_i) = \rho_{STAR}\:N_{RXsite} \: \Omega_{FOV} \: d_i^2 \:\Delta d$ = Average number of stars within the FOV of all receivers for range bin i

$ $

$P_{civTX}$ = Probability of a civilization existing in a star system and developing to a technology level where it conducts an Active SETI transmission program in the observation epoch.  This is closely related to the Drake equation and related work \citep{Drake_1961,Drake_1984,Siegel_2018}.  This might appropriately be called “the Drake fraction”.

$ $

$P_{TXtoRX} = (1/N_{TXscan})$ = Probability that TX and RX fields of view are aligned for any star within the receiver FOV, assuming TX and RX scans are independent. 

$ $

$N_{tpd}$ = Expected number of independent receiver trials per TX dwell interval (discussed below)

$ $

$P_D(d_i)$ = probability of detection in range bin $d_i$ for the TX and RX 
parameters

$ $

$T_{TXdwell}$ = TX Dwell time

$ $

$T_{year}$ = length of a year, in same time units as $T_{TXdwell}$

$ $

}
yielding

\begin{eqnarray}
DPY_{WFS}=\frac{P_{civTX}\:N_{tpd}\:N_{RXsite}\:\Omega_{FOV}\:\rho_{STAR}\:T_{year}}{N_{TXscan}\:T_{TXdwell}} \sum\limits_{i} P_D(d_i)\:d_i^2\:\Delta d  \nonumber
\end{eqnarray}

The probability of detection as a function of range $P_D(d_i)$ will be a sigmoid-like function (transitioning rapidly from 1 to 0 in an S shape) which for our purposes can be approximated by a binary 1-to-0 transition at $d=d_{MAX}$, the maximum detection range.  With this approximation

\begin{eqnarray}
\sum\limits_{i} P_D(d_i)\:d_i^2\:\Delta d \approx \frac{{d_{MAX}}^3}{3}   ,  \nonumber
\end{eqnarray}

so
\begin{eqnarray} 
DPY_{WFS}=\frac{P_{civTX}\:N_{tpd}\:N_{RXsite}\:T_{year}}{N_{TXscan}\:T_{TXdwell}} \;
	\frac{\rho_{STAR}\:\Omega_{FOV}\:{d_{MAX}}^3}{3}   \nonumber
\end{eqnarray}

Therefore, after substituting (\ref{eq:eqn1}), we get:

\begin{equation}  \label{eq:eqn2}
DPY_{WFS}=\frac{P_{civTX}\:N_{tpd}\:N_{RXsite}\:T_{year}} {T_{TXscan}}  N_{STAR}  \; .
\end{equation}

Let us also define

{\addtolength{\leftskip}{10 mm}
\setlength{\parindent}{0 mm}

$ $

$\tau$ = Total integration time in the receiver

$ $

$T_{TXscan}$ = $N_{TXscan}\:T_{TXdwell}$ = Total full sphere scan time for all transmitters working simultaneously

$ $

$T_{TX}/\tau$ = $\delta_{TX}\:T_{TXdwell}/\tau$ = number of receiver integration periods $\tau$ per transmission time $T_{TX}$ 

$ $

$T_{TX}$ = Transmit time in seconds

$ $

$\delta_{TX}$ = Transmitter duty cycle

$ $

}

We can obtain an alternative expression for (\ref{eq:eqn2}):	
\begin{equation}   \label{eq:eqn3}
DPY_{WFS}=\frac{N_{tpd}\:\delta_{TX}}{T_{TX}/\tau} \:\frac{P_{civTX}\:N_{RXsite}\:T_{year}}{N_{TXscan}} \:\frac{N_{STAR}}{\tau}
\end{equation}

\subsection{Targeted Surveys}

In a targeted survey, the receiver points to a specific candidate star and listens.   
At first glance, one might assume $N_{STAR}$ = 1 if  $d_{TgtStar} \le d_{MAX}$. However, there will be incidental detections (“bycatch”) so that $N_{STAR}$ = 1 + $\rho_{STAR} V_{cone}$. Recent papers have analyzed the “bycatch” factor and concluded it is significant \citep{Shostak_2000,Wlodarczyk_Sroka_2020}.  Mathematically, if $d_{TgtStar} \le d_{MAX}$, TS is identical to Slow WFS with a unity increment in the number of stars over the number expected from the search volume:

\begin{equation}  \label{eq:eqn4}
DPY_{TS}=\frac{P_{civTX}\:N_{tpd}\:N_{RXsite}\:T_{year}} {T_{TXscan}} (1+\onethird \rho_{STAR}\:\Omega_{FOV}\:{d_{MAX}}^{3})  \\
\approx DPY_{WFS} \: \text{for $N_{STAR} \gg 1$ }
\end{equation}

\noindent We will ordinarily expect $N_{STAR} \gg 1$, so the effect of an additional star should be negligible. TS can be treated as a special case of Slow WFS.  Targeted surveys typically observe a sequence of stars with the observation time on each star extending over many TX dwell intervals, as would be done in a Slow WFS.  We ignore the additional overhead of On-Off scanning: if $N_{STAR}$ were equal to one, On-Off scanning might typically reduce $DPY_{TS}$ by an additional factor of 2, but with $N_{STAR} \gg 1$, the ``Off'' scans are equally likely to detect as the ``On'' scans. \textbf{Henceforth we will not examine TS separately, and will reference ``$DPY_{WFS}$" as simply ``DPY".}  

\subsection{Trials per Dwell}

The value of $N_{tpd}$ or number of independent receiver trials per TX dwell cycle will depend on scanning and averaging methods in the receiver. $N_{tpd}$ accounts statistically for the asynchronous nature of TX and RX scanning and details of RX averaging windows.

\begin{itemize}
\item{Fast WFS: For fast wide field scanning, each $\tau$ interval is looking at a different section of sky, so each scan is independent.  The expected trials per TX dwell interval is (average $\#$ RX scans per dwell)*(probability of $\tau$ fitting completely within $T_{TX}$) = ($T_{TXdwell}/\tau)\:[(T_{TX}-\tau)/T_{TXdwell}] = (T_{TX}-\tau)/\tau$, so $N_{tpd} = T_{TX}/\tau-1$.  Typical values of $T_{TX}/\tau$ might be 2 or 3, and $T_{TX}>\tau$ is assumed.  A value of $\tau=T_{TX}/2$ will guarantee an average of 1 receiver trial per TX dwell cycle. }
\item{Slow WFS or TS: For slow scanning, successive $\tau$ intervals are not necessarily independent.  If a sliding “moving average” window is used for integration,  the number of full $\tau$ intervals coinciding with $T_{TX}$ will be floor($T_{TX}/\tau$), where floor(x) truncates x downward to the nearest integer.  For $\tau$=$T_{TX}$, coincidence is guaranteed, while if $T_{TX}=2\tau$ or greater, multiple detections per $T_{TXdwell}$ can occur.  However, to avoid distorting the DPY metric, we will want to set $N_{tpd}$=min(1,floor($T_{TX}/\tau$)), rather than $N_{tpd}$=floor($T_{TX}/\tau$).   If a burst of detections occurs for the same TX dwell interval, we want to count it as a single detection.  }
\end{itemize}

Since we require our averaging time to be less than the transmit time ($\tau \le T_{TX}$), $N_{tpd}$ becomes:	

\begin{equation}  \label{eq:eqn5}
N_{tpd}=
\begin{cases}
  T_{TX}/\tau-1 & \text{Fast WFS} \\
  1 & \text{Slow WFS and TS}
\end{cases}       
\end{equation}

In order to maximize sensitivity for Fast WFS, the value of $N_{tpd}$ will typically be on the order of unity.  

\subsection{Notes on Strategies to Maximize Detection Rate }

Other notes about equations  (\ref{eq:eqn2}), (\ref{eq:eqn3}) and  (\ref{eq:eqn4}):

\begin{itemize}
 \item{For a given site, the key parameters under control of the receiver are $N_{STAR}$, $\tau$ and coherent bandwidth $\Delta \nu$.  For given averaging parameters $\tau$ and $\Delta \nu$, maximizing $N_{STAR}$ is sufficient to maximize the expected number of ET detections per year.  Maximizing $N_{STAR}$ also avoids the problem of quantifying $P_{civTX}$.  The RX party will need to make loose assumptions on the overall transmit cycle time $T_{TXscan}$ and the TX dwell time so the RX scan and integration parameters are consistent.  }
 \item{If there is a transmitting star in the RX cone of one receiving station (e.g. $P_{civTX}$=1, $N_{STAR}$=1, $N_{RXsite}$=1) and  $\tau$=$T_{TX}/2$ ($T_{TX}=2\tau$, $N_{tpd}$=1), per  (\ref{eq:eqn2}) we should obtain one detection every $T_{TXscan}$, or $T_{year}/T_{TXscan}$ detections per year, as one would expect.}
 \item{We have assumed a uniform star density $\rho_{STAR}$ over all space, but the above may be modified with a range- or angle-dependent star density as required. For example, \citep{Shostak_2000} describes an angle- and range-dependent model of Milky Way star densities which predicts higher densities toward the galactic center. The GAIA star catalog\footnote{\url{https://www.cosmos.esa.int/web/gaia}} should offer up-to-date values.}
 \item{We could alternatively define a transmitter density  $\rho_{TX}=P_{civTX}\;\rho_{STAR}$ (TX per $pc^3$) and derive similar expressions for DPY.  The number of transmitters within the search cone is $N_{TX}=\rho_{TX}\,V_{cone}=P_{civTX}\,N_{STAR}$. We choose to keep $P_{civTX}$ and $\rho_{STAR}$ separate because star density can be readily estimated and optimized for a survey.  Nevertheless, it is the $P_{civTX}\;\rho_{STAR}$ product that is important. Ideally, we could estimate densities of certain types of stars according to the likelihood of having habitable zones, and estimate $P_{civTX}$ according to star type.  We would then calculate DPY for each star type and sum them, or equivalently use a weighted average for $P_{civTX}$ over all star types. }
 \item{$DPY$ in  (\ref{eq:eqn2}) for fast and slow scanning will be roughly comparable within a small factor $T_{TX}/\tau$-1.  Slow scanning has an advantage that it can use multiple sliding windows of varying duration, so matching $\tau$ to $T_{TX}$ is less critical.  Also, the scan method and rate of scan are less critical, so this may be better suited to commensal observations.  On the other hand, slow scanning does not attempt to cover the entire sky in a timely way.}
 \item{DPY has many factors working against detections:}
\begin{itemize}
 \item{$P_{civTX} \ll 1$ presumably. This is difficult to quantify, and even the order of magnitude is a guess.}
 \item{$N_{TXscan} \gg 1$ is needed to raise EIRP to sufficient levels for detection, given interstellar distances}
\end{itemize}
 \item{To counteract the above factors, we need:  }
\begin{itemize}
 \item{$T_{year}/\tau \gg 1$, i.e. many observations per year}
 \item{$N_{STAR} \gg 1$ , which requires wide FOV and high sensitivity simultaneously.}
\end{itemize}
\end{itemize}

Maximizing the number of visible stars $N_{STAR}$ is critically important to advance detection rates beyond what is currently possible. DPY and $N_{STAR}$ have a simple linear relationship.  Let us  define 

\begin{equation}  \label{eq:eqn6a}
N_{STAR1} = T_{TXscan}\:/\: (P_{civTX}\:N_{tpd}\:N_{RXsite}\:T_{year}) 
\end{equation}

Per equation (\ref{eq:eqn2}):

\begin{equation}  \label{eq:eqn6b}
DPY = \frac{N_{STAR}}{N_{STAR1}} \approx \frac{N_{STAR}}{41,100}\:\:\text{for the  parameters below}
\end{equation}
 
\noindent We see $N_{STAR1}$ is the number of stars needed to achieve one expected SETI detection per year.  If we set $P_{civTX}=10^{-6}$, $N_{tpd}=1.0$, $N_{RXsite}=2$, $T_{year}=365$ days and $T_{TXscan}=30$ days, $N_{STAR1} \approx 41,100$.  This gives a representative target value for a system design with 2 receiver sites.  The $P_{civTX}$ value of $10^{-6}$ is an “anchor point” for discussion, meaning that “one in a million” star systems will have an Active SETI transmitter, but clearly this number is unknown.  
\subsection{Expected Number of Detections} \label{ssec:NDet}

For later reference, we can easily determine the expected number of detections for a survey of duration $T_{obs}$.  With a constant detection rate, this scales as $T_{obs}/T_{year}$:

\begin{equation}  \label{eq:eqn641}
\overline{N_{Det}} = \frac{T_{obs}}{T_{year}}\,DPY
= \frac{T_{obs}}{T_{year}} \frac{T_{year}\,N_{RXsite}\,N_{tpd}\,P_{civTX}}{T_{TXscan}}\:N_{STAR}
= \frac{T_{obs}}{T_{TXdwell}} \frac{N_{RXsite}\,N_{tpd}\,P_{civTX}}{N_{TXscan}}\:N_{STAR}
\end{equation}

\noindent With observations over $N_{TXdwell}\!=\!T_{obs}/T_{TXdwell}$ dwell cycles, and the number of observations defined as  $N_{obs}\!=\!N_{TXdwell}\,N_{tpd}\,N_{RXsite}$, this reduces to:

\begin{equation}  \label{eq:eqn642}
\overline{N_{Det}} = \frac{P_{civTX}}{N_{TXscan}}\,N_{obs}\,N_{STAR} = 
\frac{T_{TXdwell}\,P_{civTX}}{T_{TXscan}}\,N_{obs}\,N_{STAR}
\end{equation}

\noindent The minimum number of observations required to obtain $\overline{N_{Det}}=1$ is:

\begin{equation}  \label{eq:eqn643}
N_{obs-min} = \frac{N_{TXscan}}{P_{civTX}\,N_{STAR}}\,=\,\frac{N_{TXscan}}{N_{TX}}
\end{equation}

\noindent That is, on average, if there is one transmitter within the search cone, we need to observe $N_{TXscan}$ times to achieve one detection.

\section{Number of Stars in the RX Cone for Wide Field Search} \label{sec:NstarsInRXCone}

We need to evaluate $N_{STAR}$ for WFS and characterize the resulting DPY. 
Let 

{\addtolength{\leftskip}{10 mm}
\setlength{\parindent}{0 mm}

$ $

$A_{eAP}$ = effective area of per aperture (dish antenna or aperture array station)

$N_{AP}$ = number of apertures (dishes or aperture arrays) 

$ $

}

\noindent With EIRP = $P_{TX}\:G_{TX}$ and $G_{RX} = 4\pi A_{eAP} / \lambda ^2$, Friis formula \citep{FRIIS_1946} gives the received power for an aperture (e.g. a dish or aperture array) as

\begin{equation}  \label{eq:eqn7}
P_{RX}= \frac{P_{TX}\:G_{TX}\:G_{RX}\:\lambda^2}{(4 \pi d)^2\:L} = \frac{EIRP\:A_{eAP}}{4 \pi d^2\:L}
\end{equation}

For simplicity, assume the loss L can be incorporated into the effective system temperature $T_{sys}$.  Also assume a dual-polarization receiver which combines   powers for arbitrary signal polarizations.  The effective SNR is the coherent SNR at the aperture increased by the array gain AG and non-coherent integration gain AvgG.  This must exceed a threshold DT for detection to occur:

\begin{equation}  \label{eq:eqn8}
SNR_{eff} = \frac{P_{RX}}{Noise_{RX}}\:AG \: AvgG = \frac{EIRP\:A_{eAP} \: AG \: {N_{avg}}^{1/2}} {4 \pi d^2\:(2\:k_B\:T_{sys}\:\Delta \nu)}  \ge DT
\end{equation}

The array gain AG will be a function of the signal processing, and will generally have the form $AG = \eta_{AG} {N_{AP}}^\gamma$.  For the ideal case of lossless, fully coherent beamforming (BF) across apertures, we will have AG=$N_{AP}$ ($\eta_{AG}$=1 and $\gamma$=1).  In general, there will be losses which cause $\eta_{AG} < 1$, such as ``scalloping loss'' and losses associated with imperfect calibration or parameter estimation.  Fully coherent beamforming may also be computationally impractical, especially for interferometric arrays with long baselines.  A method called incoherent beamforming (IBF) has AG=${N_{AP}}^{1/2}$ ($\gamma=\onehalf$), which achieves lower AG at considerably less computation.  Other array processing methods may be defined with AG values and computation in-between IBF and BF (\citet{Houston_SETI_Detect_URSI_2021}). 

In most imaging applications, fully coherent processing is implicit.  The total effective array area is $A_e = A_{eAP} N_{AP}$, and sensitivity is often described as the ratio $A_e/T_{sys}$.  For narrow-band detection in SETI, the effective area will still be relevant if we define it in terms of array gain: $A_e = A_{eAP} \: AG = A_{eAP} \: \eta_{AG} \: {N_{AP}}^\gamma$.

The non-coherent integration gain is given as $AvgG = {N_{avg}}^{1/2}$, where $N_{avg}$ is the number of averages.  We assume Fast Fourier Transform (FFT)-based polyphase filter banks \citep{Harris_Haines_PFB_2011,Price_2018} as the key coherent processing stage, so the noise bandwidth $\Delta \nu$ is the PFB bin width or the reciprocal of $T_{FFT}$, the time duration of the FFT at the input sample rate.  Non-coherent integration is done by summing the magnitude-squared FFT output bins from two polarizations (Stokes I) over $N_{avg}$ FFT cycles (non-overlapping), which implies the total averaging time is $\tau = N_{avg}\:T_{FFT}$, which must be less than the TX duration $T_{TX}$.  Note also that $N_{avg}=\Delta \nu\:\tau$, so that the quantity ${N_{avg}}^{1/2}/\Delta \nu$ in (\ref{eq:eqn8}) may be expressed as ${N_{avg}}^{1/2}/\Delta \nu$ = $T_{FFT}\:{N_{avg}}^{1/2}$ = $\tau / {N_{avg}}^{1/2}$ = ${(\tau / \Delta \nu)}^{1/2}$. 

The maximum detectable distance $d_{MAX}$ occurs for a given EIRP at equality:

\begin{equation}  \label{eq:eqn9}
{d_{MAX}}^2=\frac{EIRP\:A_{eAP}\:AG\:{N_{avg}}^{1/2}}{4 \pi (2\:k_B\:T_{sys})\:\Delta \nu \: DT}
\end{equation}

The number of stars in the RX cone will be

\begin{eqnarray}
N_{STAR}=\frac{1}{3}\:\rho_{STAR}\:\Omega_{FOV}\:{d_{MAX}}^3 = 
\frac{1}{3}\:\rho_{STAR}\:[\frac{\lambda^2\:N_{FOV}}{A_{eAP}}]\:
[\frac{EIRP\:A_{eAP}\: AG \:{N_{avg}}^{1/2}}{4 \pi (2\:k_B\:T_{sys})\: \Delta \nu \: DT}]^{3/2}   \nonumber
\end{eqnarray}

or 

\begin{equation}  \label{eq:eqn10}
N_{STAR}=\frac{1}{3}\:\rho_{STAR}\:
[\lambda^2\:{A_{eAP}}^{1/2}\:AG^{3/2}\:N_{FOV}]\:
[\frac{EIRP\:{N_{avg}}^{1/2}}{4 \pi (2\:k_B\:T_{sys})\: \Delta \nu \: DT}]^{3/2}  
\end{equation}

The total FOV is modeled as $\Omega _{FOV} = \Omega _{AP}\:N_{FOV} = (\eta _R \lambda^2/A_{eAP})\:N_{FOV} \approx  ( \lambda^2/A_{eAP})\:N_{FOV}$,  or the product of the nominal aperture FOV (assuming low antenna losses or radiation efficiency $\eta _R \approx 1$) and a new quantity $N_{FOV}$, which is the field-of-view amplification factor through use of the following:

\begin{itemize}
 \item{Focal plane arrays (FPAs), feed horn arrays, or phased array feeds (PAFs) \citep{Staveley_Smith_1996,VanArdenne_2009,Warnick_2016,SKA_PAF_Barker_2017}}
 \item{Fly’s-eye configurations (dividing antennas into groups, with each group pointing in a different direction) \citep{Siemion_2012ApJ}}
 \item{Multiple ``station beams" associated with aperture array stations \citep{Zarb-Adami_2010}, each with a separate pointing direction.  The set of station beams is notionally equivalent to the outputs of multiple dishes at each site, each of area $A_{eAP}$ and pointed in different directions.}
 \item{Incomplete coherent beamforming, where fewer tied-array beams are computed compared to what is needed to fully cover the primary FOV $\Omega _{AP}$.  In this case, $N_{FOV}<1$.}
\end{itemize}

$N_{FOV}$ is generally constant with frequency, but not always.  To account for frequency variations, define $N_{FOV}(\nu) = N_{FOV0} X_{FOV}(\nu)$, where $N_{FOV0}$ is a constant.  $X_{FOV}(\nu)$=1 in most cases.   $X_{FOV}(\nu)$ will vary with frequency if beams overlap, as in AAs and PAFs.  For example, in AAs, to control beam overlap and maintain a constant FOV below frequency $\nu_0$, the number of station beams may be adjusted downward at low frequencies so that $X_{FOV}(\nu)=min(1,(\nu/\nu_0)^2)$ and $N_{FOV0}$ = $N_{beam0}$ = the number of station beams at $\nu_0$.

Some notes:

\begin{itemize}
 \item{	All parameters in  (\ref{eq:eqn10}) are under control of the RX party except for $\rho_{STAR}$, $\lambda$ and EIRP.  A receiver will be designed to cover a wide frequency band.  A range of EIRP values can be assumed for further analysis.}
 \item{There are many different exponents for parameters, which are critical to system trades.}
 \item{In particular the exponents in  (\ref{eq:eqn10}) for $A_{eAP}$ and $N_{AP}$ are different for maximizing $N_{STAR}$ than when maximizing SNR and $d_{MAX}$ in  (\ref{eq:eqn8}) and  (\ref{eq:eqn9}) respectively, which is often the focus of SETI analyses. }
\end{itemize}

\section{Detection Rate Metrics} \label{sec:DetRateMetrics}

Per  (\ref{eq:eqn2}), $N_{STAR}$ is a key figure of merit (FOM) for the receive system.  Per  (\ref{eq:eqn10}), one can define additional FOMs which are factors of $N_{STAR}$:

\begin{equation}  \label{eq:eqn11}
Frequency\:FOM=FFOM= \lambda^2  X_{FOV}/ {T_{sys}}^{3/2}\;\;\;\;\; (m^2/K^{3/2})
\end{equation}

\begin{equation}  \label{eq:eqn12}
Array\:FOM = AFOM ={A_{eAP}}^{1/2}\:AG^{3/2}\:N_{FOV0} \;\;\;\;\;(m)
\end{equation}

\begin{equation}  \label{eq:eqn13}
Averaging\:FOM = AvgFOM = \frac{{N_{avg}}^{1/2}}{\Delta \nu \: DT} =
\frac{T_{FFT}\:{N_{avg}}^{1/2}}{DT} =
\frac{\tau}{DT\:{N_{avg}}^{1/2}}  = \frac{\tau^{1/2}}{DT\: \Delta \nu^{1/2}}  \;\;\;\;\; 	(sec)	
\end{equation}

so that 

\begin{equation}  \label{eq:eqn14}
N_{STAR}= \frac{\rho_{STAR}}{3\:(8 \pi k_B)^{3/2}}\:(FFOM)\:(AFOM) \: (AvgFOM)^{3/2}\:EIRP^{3/2}  
\end{equation}

The various FOMs allow us to examine frequency, array and averaging effects separately.
Equation  (\ref{eq:eqn2}) may be expanded to:

\begin{equation}  \label{eq:eqn15}
DPY = \frac{1}{N_{STAR1}}\: \frac{\rho_{STAR}}{3\: (8 \pi k_B)^{3/2}} 
\:(FFOM)\:(AFOM) \: (AvgFOM)^{3/2}\:EIRP^{3/2}
\end{equation}

\section{Examination of the DPY Metrics} \label{sec:ExamDPYMetrics}

\subsection{Frequency Figure of Merit}

\begin{figure}[ht!]
\epsscale{0.8}
\plotone{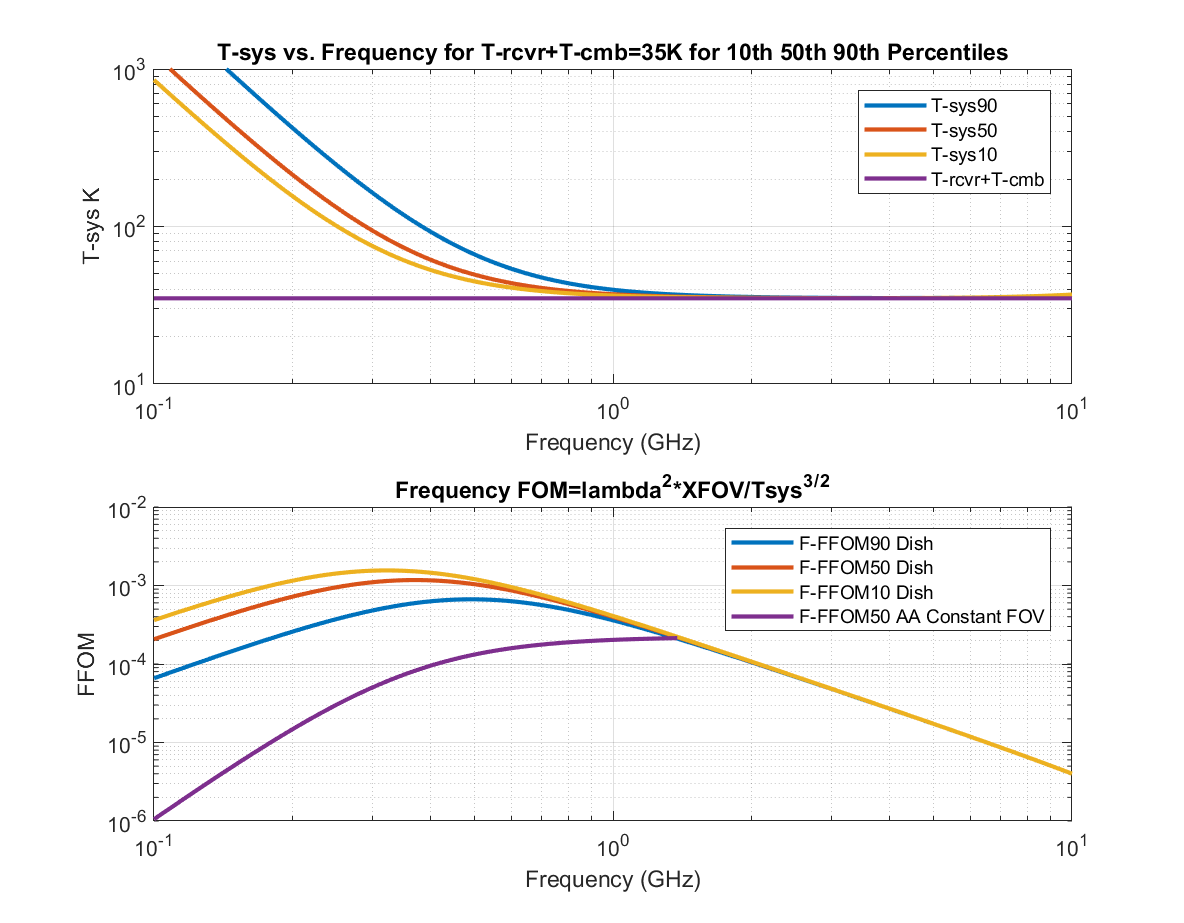}
\caption{Frequency FOM Example}
\label{fig:FFOMExample}
\end{figure}

Per the Frequency FOM  (\ref{eq:eqn11}), if $T_{sys}$ is approximately constant in a frequency band, then the number of potential stars in view will be proportional to $\lambda^2$, so on average at 10\,GHz there will be a factor of 100 fewer visible stars than at 1\,GHz. Furthermore, even though $T_{sys}$ increases rapidly due to synchrotron noise below 1\,GHz, the ratio  $\lambda^2/{T_{sys}}^{3/2}$ may still be favorable below 1\,GHz, as illustrated in Figure $\ref{fig:FFOMExample}$.

Figure $\ref{fig:FFOMExample}$ shows a hypothetical system noise model with constant noise levels at 35K in the microwave window, combined with galactic noise levels at $10^{th}$, $50^{th}$, $90^{th}$ percentiles \citep{Braun_2019}, $T_{gal}=T_{408} (.408/f_{GHz})^{2.75}$ K, $T_{408}$=[17.1, 25.2, 54.8] K, in the top plot.  The bottom plot shows FFOM for dishes at these percentiles.  For the  $50^{th}$ percentile, the maximum Frequency FOM is seen to occur around 370\,MHz, well below the $1 - 10$\,GHz microwave window!  Per Figure $\ref{fig:FFOMExample}$, $N_{STAR}$ may be a factor of $3-7$ times larger at 370\,MHz compared to 1.4\,GHz.  \emph{The ET TX party could choose 370\,MHz as a TX frequency based on their perception of where the RX party should be listening!}  At the least, this would suggest that SETI surveys should cover down to 300\,MHz in addition to $1 - 2$\,GHz.  Note also that even $100 -300$\,MHz has a better FFOM than 2\,GHz and above, where the FFOM drops rapidly.  Although dispersion and frequency spreading due to the interstellar medium is more significant at lower frequencies, the effects should be small for narrow-band signals, particularly at ranges below 1 kpc \citep{Cordes_ISM_NB_Signals_1991}.

In addition, in the bottom plot a curve for aperture arrays is shown where a constant FOV is maintained up to a maximum frequency of 1.4\,GHz, so $X_{FOV}(\nu)=(\nu/1.4\:GHz)^2$.  In this case, FFOM decreases monotonically below 1.4 GHz as $T_{sys}$ increases.  This $X_{FOV}(\nu)$ profile is entirely dependent on the beamforming allocation strategy.  A better profile for SETI might be to expand FOV to match $T_{sys}(\nu)^{-3/2}$, so as to achieve constant $FFOM(\nu)$ and $N_{STAR}(\nu)$.

Improving FFOM performance is not easily done, as $T_{sys}$ at frequencies below 1\,GHz is dominated by sky noise from galactic sources, as well as ``spillover'' from terrestrial sources into antenna sidelobes.  For dishes above 1\,GHz, the receiver noise can be lowered, but this is expensive for a large number of antenna feeds which are already cryogenic.  For aperture arrays, receiver noise is generally highly optimized.

\clearpage
\subsection{Array Figure of Merit}

Per the expressions for AFOM  (\ref{eq:eqn12}) and $N_{STAR}$  (\ref{eq:eqn10}), dish effective area $A_{eAP}$ has an exponent of 1/2, the number of apertures $N_{AP}$ has an exponent of $3\gamma/2$, and $N_{FOV}$ has a unity exponent.  This implies that increasing individual antenna area improves performance slowly, while increasing the number of antennas helps performance much more rapidly.  Because of this:

\begin{itemize}
 \item{	A single large dish will be less favorable than many small ones with the same total effective area.  Assume $N_{FOV}=1$ and $\eta_{AG}=1$.}
\begin{itemize}
 \item{BF (AG=$N_{AP}$): AFOM = $(A_e/N_{AP})^{1/2} {N_{AP}}^{3/2} = {A_e}^{1/2} N_{AP}$, which grows linearly with $N_{AP}$ and is lowest  when $N_{AP} =1$}
 \item{IBF (AG=${N_{AP}}^{1/2}$): AFOM = $(A_e/N_{AP})^{1/2} {N_{AP}}^{3/4} = {A_e}^{1/2} {N_{AP}}^{1/4}$, which is still minimized  when $N_{AP}=1$}
\end{itemize}
\item{	With a fixed number of antennas, Fly’s Eye clusters ($N_{FOV}$=$N_{cluster}$) will do worse with fully coherent beamforming, but better for incoherent beamforming: }
\begin{itemize}
 \item{BF: AFOM = ${A_{eAP}}^{1/2} (N_{APtotal}/N_{cluster})^{3/2} N_{cluster} = {A_{eAP}}^{1/2} {N_{APtotal}}^{3/2} {N_{cluster}}^{-1/2}$ which is maximized for $N_{cluster}$=1, with $AFOM={A_{eAP}}^{1/2} {N_{APtotal}}^{3/2}$ }
 \item{IBF: AFOM = ${A_{eAP}}^{1/2} (N_{APtotal}/N_{cluster})^{3/4} N_{cluster} = {A_{eAP}}^{1/2} {N_{APtotal}}^{3/4} {N_{cluster}}^{1/4}$ which is maximized for $N_{cluster}=N_{APtotal}$, with $AFOM={A_{eAP}}^{1/2} N_{APtotal}$. }
\end{itemize}
\end{itemize}

\noindent We can see that fully coherent beamforming clearly favors ``large N, small D" strategies (large numbers of small-diameter apertures). Strategies with IBF are more ambivalent.  A Fly's Eye approach (which for dishes involves no beamforming at all) is slightly superior to pointing all apertures in one direction and combining with IBF, so that the larger FOV of Fly's Eye makes up for lower sensitivity. 

If we have an antenna array with fixed $A_{eAP}$, we can define a variant of AFOM to be optimized:

\begin{equation}  \label{eq:eqn12a}
Array\:FOM2 = AFOM2 =AG^{3/2}\:N_{FOV0}
\end{equation}

\noindent A limited number of fully-coherent ``tied-array" beams are often formed that might be available for commensal SETI observations.  In this case, the number of beams required to fill the primary FOV is on the order of $N_{beam-full} = (D_{array}/D_{AP})^2$, where $D_{array}$ is the overall extent of the array and $D_{AP}$ is the diameter of the individual aperture.  With $N_{beam}$ beams computed, $N_{FOV0} = N_{beam}/N_{beam-full}$.  The value of $N_{beam-full}$ can be quite large. For example, for the full MeerKAT array with $N_{AP}$=64, $D_{array}$=8 km, $D_{AP}$=13.5 m, $N_{beam-full}$ = 350,000 for the full array.  If the ``array core'' is used, $N_{beam-full}$ reduces to 9300 and 3000 for 44 and 33 dishes, respectively (\citet{Houston_SETI_Detect_URSI_2021}), at the expense of AG.  Even with 100 beams computed and 33 dishes, $N_{FOV0} = 100/3000 = .033$, so $AFOM2=33^{1.5}*.033=6.3$, which compares to $(64^{1/2})^{1.5}=23$ for IBF (64 elements, 1 beam) and $64^{1.5}=512$ for BF (64 elements, 350,000 beams).  Therefore, IBF can still be effective when compute resources are limited. A dense non-imaging array can also be used with a much lower $D_{array}$, as is the case with systems optimized for detecting pulsars and FRBs.  The large numbers of  coherent beams required to fully cover the primary FOV is a significant problem requiring alternative approaches to achieving array gain.

AFOM values can vary widely for systems which might potentially be used for SETI.  This discussion is deferred to Section \ref{sec:SystemComparisons}.  

\subsection{Averaging  Figure of Merit}

Per  (\ref{eq:eqn13}) the AvgFOM is ${N_{avg}}^{1/2}/(\Delta \nu\:DT) = T_{FFT}{N_{avg}}^{1/2}/DT = \tau/({N_{avg}}^{1/2}DT)$ which is to be maximized.  For a fixed averaging time $\tau$, we want to increase coherent averaging time $T_{FFT}$ and reduce the number of non-coherent averages $N_{avg}$, though there are limitations.  First, for a constant false alarm rate, DT is a function of $N_{avg}$, and increases significantly for $N_{avg}\le 10$.  Second, as $T_{FFT}$ increases, the number of frequency bins of width $\Delta \nu$=1/$T_{FFT}$ increases proportionally, increasing storage and computation.  Third, and perhaps most important, we encounter rapidly increasing integration loss due to frequency rate (Doppler) or phase instability.  Balancing these effects is an important part of a system design.  Our assumed values of .33 second $T_{FFT}$ ($\Delta \nu$=3 Hz), 300 second $\tau$ ($N_{avg}$=900), and DT=10 are representative for SETI studies, and yield AvgFOM=1 second.

\subsection{Nominal EIRP}  \label{ssec:NominalEIRP}

We note from (\ref{eq:eqn15}) that DPY scales as $EIRP^{3/2}$, as opposed to $EIRP^{1/2}$ for $d_{MAX}$  (\ref{eq:eqn9}).  This is a consequence of the ${d_{MAX}}^3$ factor in the RX cone volume  (\ref{eq:eqn1}).  (\ref{eq:eqn15}) would be adequate if all emitters had identical EIRP values, but of course one would expect a range of EIRPs to be encountered.  A truncated power-law distribution for EIRP values is often assumed (\citealt{Drake_1973}, \citealt{Shostak_2000}, and others for SETI; \citealt{Macquart_2014} for fast transients) of the form

\begin{equation}  \label{eq:eqn16}
f_W(W)= 
\begin{cases}
  P_0\:W^{-\alpha} & EIRP_{MIN}<W<EIRP_{MAX} \\
  0 & \text{Otherwise}
\end{cases}       
\end{equation}

\noindent where $P_0$ is a normalizing constant.  We wish to know the expected value of $DPY$ and therefore $E(EIRP^{3/2})$.  We can define a nominal EIRP quantity $EIRP_0$ such that:

\begin{equation}  \label{eq:eqn17}
{EIRP_0}^{3/2} = E(EIRP^{3/2}) = \int_{EIRP_{MIN}}^{EIRP_{MAX}} W^{3/2} f_W(W) dW \\ = P_0 \:\int_{EIRP_{MIN}}^{EIRP_{MAX}} W^{3/2 - \alpha} dW
\end{equation}

\noindent where $P_0$ is a normalizing constant.  $EIRP_0$ values will be biased toward $EIRP_{MIN}$ for high values of $\alpha$, and $EIRP_{MAX}$ otherwise. While this approach is straightforward and can be justified for astrophysical phenomena, it is difficult to put limits on $EIRP_{MIN}$, $EIRP_{MAX}$ and $\alpha$ with any certainty for SETI transmissions.  \citet{Dreher_2002} argues that ET  sources may not follow a power law at all.  We follow his reasoning and ``hedge our bets" by simply choosing a representative value for $EIRP_0$.  A benchmark value that is often chosen for SETI analyses is $EIRP_0=10^{13}W$ (130 dBW), which corresponds to the former Arecibo planetary radar. Such a power level is difficult but achievable with our technology, though at a limited $\Omega _{TX}$. If ET transmitters prove to be higher power, so much the better.  

\section{System Comparisons} \label{sec:SystemComparisons}

We can examine AFOM, $d_{MAX}$, $N_{STAR}$ and DPY for various platforms (former, existing, under development, or proposed) potentially used for SETI.  In Table \ref{tbl:Table1}, 29 radio astronomy platforms are selected, and the system antenna parameters are listed and ranked according to their $N_{STAR}$ values.  Configurations are based on user guides or system publications, as listed in the Appendix \ref{sec:References}.  The systems include:

\begin{itemize}
 \item{Large single-pixel dishes (GBT, Parkes, Arecibo)}
 \item{Single-pixel, multi-dish interferometric arrays (ATA, JVLA, MeerKAT, MeerKAT Plus, SKA1-Mid, ngVLA, DSA-2000)}
 \item{Large dishes fortified with FPAs or PAFs (Parkes Multibeam, FAST, GBT FLAG)}
 \item{Interferometric dish arrays fortified with FPAs or PAFs (APERTIF, ASKAP)}
 \item{Compact array systems optimized for pulsar and FRB detections (CHIME\footnote{For comparison to other systems, CHIME is approximated as a single 80m diameter aperture array with a 4x256=1024 FOV multiplier.}, CHORD, HIRAX) }
  \item{Compact array systems optimized for planetary radar and other uses (Next Generation Arecibo Telescope)}
  \item{The L-Band Array of Small Arrays (LASA) concept presented by \citet{Lynch_2018}, which offers large FOV though modest sensitivity.  We consider 512 square tiles of antennas 1.7\,m on a side. }
  \item{Low-Frequency Aperture Array Systems (MWA, SKA1-Low, SKA2-Low) }
 \item{Hypothetical future Square Kilometre Array SKA2-Mid variations.  Note that the scale of SKA2 has been projected but is not definite.  }
\begin{itemize}
 \item{SKA2-Mid dishes expand the number of SKA1-Mid dishes, and include PAFs.  The full SKA2-Mid-2000 configuration includes 2000 15-m dishes, which offers 300,000\,$m^2 A_e$ and 20,000 $m^2/K$ sensitivities at $T_{sys} =15$\,K.  A second configuration has 1000 dishes.  We assume that the PAFs offer a nominal $15 \times$ improvement in FOV \citep{SKA1_Survey_Perf_2016}.}
 \item{Three Mid Frequency Aperture Array (MFAA) variants include 250, 150 and 100 stations of 46, 42 and 32 meter diameter, respectively.  Assuming $T_{sys} = 35$\,K, the largest meets the specified 10,000\,$m^2/K$ sensitivity, equivalent to SKA2-Mid-1000 above.  The 150 station has a factor of 2 lower sensitivity, while the smallest has 2,000\,$m^2/K$ sensitivity, equivalent to the 197 Dish SKA1-Mid configuration \citep{Torchinsky_2017_MFAA}.}
\end{itemize}
\end{itemize}
\noindent The aperture efficiency $\eta_{AP}$ values in Table \ref{tbl:Table1} are based on system literature if possible and estimated otherwise.  Note that high precision in $\eta_{AP}$ is not required to estimate $N_{STAR}$ and DPY. The MeerKAT, SKA1-Mid and SKA2-Mid dish values are based on formulas presented by \citet{Braun_2019}.  Dense aperture arrays (MFAA variants) are assigned $\eta_{AP}$=0.85, corresponding to an average $cos(\theta)$ loss due to steering away from array normal. Sparse aperture arrays (SKA1-Low and SKA2-Low) have an additional factor taking into account sensitivity loss at high frequencies due to sparseness $\eta_{AP}=0.85\; (\nu_1/\nu)^2$, where $\nu_1$ is a dense-sparse transition frequency \citep{SKA1_Baseline_Dewdney_2015,Braun_2006_SKA_memo_87_Dense_Sparse_AA}.

\begin{deluxetable*}{|l|c|c|c|c|c|c|c|c|}
\tablecaption{Parameter Summary for System Comparisons}
\tablewidth{0pt}
\label{tbl:Table1}
\tablehead{
\colhead{ } & \colhead{ }  & \colhead{ }  & \colhead{\textbf{Number}}  & \colhead{ }   & \colhead{ }   & \colhead{ }  & \colhead{\textbf{Area}}   & \colhead{ }  \\[-2mm]
\colhead{ } & \colhead{ }  & \colhead{\textbf{Site}}  & \colhead{\textbf{Antenna}}  & \colhead{\textbf{Nom.}}  & \colhead{\textbf{Beam}}   & \colhead{\textbf{FOV}}  & \colhead{\textbf{Effic.}}   & \colhead{ }  \\[-2mm]
\colhead{\textbf{Site Name}} & \colhead{\textbf{Site ID}}  &
\colhead{\textbf{Status}}  &
\colhead{\textbf{Stations}}  & \colhead{\textbf{Diam.}}  & \colhead{\textbf{Width}}   &  \colhead{\textbf{Mult.}}  & \colhead{\textbf{Factor}}   & \colhead{\textbf{$T_{sys}$}}
}
\startdata
 & & & $N_{AP}$ & $D_{AP}$ & $\Theta_{AP}$ & $N_{FOV0}$ & $\eta _{AP}$ &  \\  
 & & &  & m & Deg &  &  & K  \\ \hline 
\textbf{Sites at 1400 MHz} \\ \hline 
Parkes Observatory & Parkes & 1 & 1 & 64 & 0.23 & 1 & 0.60 & 23 \\ \hline 
Green Bank Telescope & GBT & 1 & 1 & 105 & 0.15 & 1 & 0.60 & 20 \\ \hline 
Parkes Multibeam & Parkes-MB & 1 & 1 & 64 & 0.23 & 13 & 0.60 & 50 \\ \hline 
GBT FLAG PAF & GBT-FLAG & 2 & 1 & 105 & 0.15 & 4 & 0.57 & 17 \\ \hline 
Allen Telescope Array & ATA-42A & 1 & 42 & 6 & 2.50 & 1 & 0.62 & 35 \\ \hline 
Giant Metrewave RT & uGMRT & 2 & 30 & 45 & 0.40 & 1 & 0.43 & 75 \\ \hline 
Arecibo Telescope & Arecibo & 1 & 1 & 225 & 0.05 & 7 & 0.65 & 30 \\ \hline 
Jansky Very Large Array & JVLA & 1 & 27 & 25 & 0.60 & 1 & 0.43 & 28 \\ \hline 
FAST 500m Aperture & FAST & 1 & 1 & 300 & 0.05 & 19 & 0.57 & 20 \\ \hline 
Westerbork APERTIF & APERTIF & 2 & 12 & 25 & 0.60 & 25 & 0.75 & 50 \\ \hline 
MeerKAT 64 & MeerKAT & 1 & 64 & 14 & 1.11 & 1 & 0.76 & 16 \\ \hline 
ASKAP & ASKAP & 2 & 36 & 12 & 1.25 & 26 & 0.76 & 70 \\ \hline 
MeerKAT Extension 84 & MK+ & 3 & 84 & 14 & 1.11 & 1 & 0.76 & 16 \\ \hline 
CHORD & CHORD & 5 & 512 & 6 & 2.50 & 1 & 0.70 & 30 \\ \hline 
Next Generation VLA & ngVLA & 4 & 214 & 18 & 0.83 & 1 & 0.80 & 26 \\ \hline 
SKA1 Mid (197 Dish) & SKA1-Mid-197 & 4 & 197 & 15 & 1.00 & 1 & 0.84 & 15 \\ \hline 
L-Band Array of Small Arrays & LASA-512 & 5 & 512 & 2 & 7.50 & 16 & 0.85 & 40 \\ \hline 
Next Gen Arecibo Telescope & NGAT & 5 & 1112 & 9 & 1.67 & 1 & 0.70 & 25 \\ \hline 
DSA-2000 & DSA-2000 & 5 & 2000 & 5 & 3.00 & 1 & 0.70 & 25 \\ \hline 
SKA2 MFAA 100 & MFAA-100 & 5 & 100 & 32 & 0.47 & 600 & 0.85 & 35 \\ \hline 
SKA2 Mid (1000 Dish) & SKA2-Mid-1000 & 5 & 1000 & 15 & 1.00 & 15 & 0.88 & 15 \\ \hline 
SKA2 Mid (2000 Dish) & SKA2-Mid-2000 & 5 & 2000 & 15 & 1.00 & 15 & 0.88 & 15 \\ \hline 
SKA2 MFAA 150 & MFAA-150 & 5 & 150 & 42 & 0.36 & 1000 & 0.85 & 35 \\ \hline 
SKA2 MFAA 250 & MFAA-250 & 5 & 250 & 46 & 0.33 & 1200 & 0.85 & 35 \\ \hline 

\textbf{Additional Sites at 700 MHz} \\ \hline 
CHIME & CHIME & 2 & 1 & 80 & 0.38 & 1024 & 0.85 & 50 \\ \hline 
HIRAX & HIRAX & 4 & 1024 & 6 & 5.00 & 1 & 0.70 & 50 \\ \hline 

\textbf{Additional Sites at 300 MHz} \\ \hline
Murchison Widefield Array II & MWA & 1 & 256 & 5 & 14.00 & 1 & 0.85 & 83 \\ \hline 
SKA1-Low & SKA1-Low & 4 & 512 & 38 & 1.84 & 0 & 0.09 & 98 \\ \hline 
SKA2-Low & SKA2-Low & 5 & 4880 & 38 & 1.84 & 0 & 0.09 & 98 \\ \hline 

\enddata
\tablecomments{Site status code: 1 - Facility supporting SETI (past, current, near-future); 2 - Existing facility; 3 - Facility under construction; 4 - Facility under development (detailed requirements/plans/prototype arrays); 5 - Proposed or concept}
\end{deluxetable*}

\begin{deluxetable*}{|c|c|c|c|c|c|c|c|c|c|}
\tablecaption{Key Metrics for Various Radio Telescope Systems}
\tablewidth{0pt}
\label{tbl:Table2}
\tablehead{
\colhead{ }  & \colhead{\textbf{Field}}  &  \colhead{\textbf{Eff.}}  & \colhead{\textbf{Total}}  & \colhead{ }  & \colhead{\textbf{Max.}}  & \colhead{ }  & \colhead{ }  & \colhead{ }  & \colhead{\textbf{Expected}}  \\[-2mm]
\colhead{ }  & \colhead{of}  &  \colhead{\textbf{Area}}  & \colhead{\textbf{Eff.}}  & \colhead{\textbf{Array}}  & \colhead{\textbf{Detect}}  & \colhead{\textbf{ }}  & \colhead{\textbf{Sensi-}}  & \colhead{\textbf{Visible}}  & \colhead{\textbf{Detections}}    \\[-2mm]
\colhead{\textbf{Site ID}}  & \colhead{\textbf{View}}  &  \colhead{\textbf{per AP}}  & \colhead{\textbf{Area}}  & \colhead{\textbf{FOM}}  & \colhead{\textbf{Range}}  & \colhead{\textbf{SEFD}}  & \colhead{\textbf{tivity}}  & \colhead{\textbf{Stars}}  & \colhead{\textbf{per Year}}
}
\startdata
 & $\Omega_{FOV}$ & $A_{eAP}$ & $A_{e}$ & AFOM & $d_{MAX}$ & Jy & $A_e/T_{sys}$ &  $N_{STAR}$ & DPY \\
 & $Deg^2$ & $m^2$ & $m^2$ & m & Parsec & & $m^2/K$ & & \\ \hline 
\textbf{Sites at 1400 MHz} \\ \hline 
Parkes & 0.0431 & 1930 & 1930 & 43.9 & 50 & 32.9 & 83.9 & 0.101 & 2.5e-06 \\ \hline 
GBT & 0.0177 & 5600 & 5600 & 74.8 & 92 & 9.86 & 280 & 0.213 & 5.2e-06 \\ \hline 
Parkes-MB & 0.561 & 1930 & 1930 & 571 & 34 & 71.5 & 38.6 & 0.412 & 1.0e-05 \\ \hline 
GBT-FLAG & 0.0707 & 4900 & 4900 & 280 & 93 & 9.58 & 288 & 1.02 & 2.5e-05 \\ \hline 
ATA-42A & 4.91 & 17.5 & 736 & 1140 & 25 & 131 & 21.0 & 1.40 & 3.4e-05 \\ \hline 
uGMRT & 0.127 & 684 & 20500 & 4300 & 91 & 10.1 & 274 & 1.69 & 4.1e-05 \\ \hline 
Arecibo & 0.0158 & 25800 & 25800 & 1120 & 161 & 3.21 & 859 & 1.74 & 4.2e-05 \\ \hline 
JVLA & 0.283 & 211 & 5700 & 2040 & 78 & 13.6 & 204 & 3.51 & 8.5e-05 \\ \hline 
FAST & 0.0373 & 40300 & 40300 & 3810 & 247 & 1.37 & 2010 & 10.9 & 2.6e-04 \\ \hline 
APERTIF & 7.07 & 368 & 4420 & 19900 & 52 & 31.3 & 88.4 & 14.4 & 3.5e-04 \\ \hline 
MeerKAT & 0.970 & 109 & 6980 & 5350 & 115 & 6.33 & 437 & 21.3 & 5.2e-04 \\ \hline 
ASKAP & 31.9 & 85.5 & 3080 & 51900 & 36 & 62.8 & 44.0 & 22.6 & 5.5e-04 \\ \hline 
MK+ & 0.970 & 109 & 9170 & 8040 & 132 & 4.82 & 573 & 32.0 & 7.8e-04 \\ \hline 
CHORD & 4.91 & 19.8 & 10100 & 51500 & 101 & 8.17 & 338 & 79.9 & 0.0019 \\ \hline 
ngVLA & 0.545 & 204 & 43600 & 44700 & 225 & 1.65 & 1680 & 85.8 & 0.0021 \\ \hline 
SKA1-Mid-197 & 0.785 & 149 & 29400 & 33800 & 244 & 1.41 & 1960 & 148 & 0.0036 \\ \hline 
LASA-512 & 707 & 2.67 & 1370 & 303000 & 32 & 80.8 & 34.2 & 305 & 0.0074 \\ \hline 
NGAT & 2.18 & 44.5 & 49500 & 247000 & 245 & 1.39 & 1980 & 504 & 0.012 \\ \hline 
DSA-2000 & 7.07 & 13.7 & 27500 & 332000 & 182 & 2.51 & 1100 & 676 & 0.016 \\ \hline 
MFAA-100 & 104 & 684 & 68400 & 1.57e+07 & 243 & 1.41 & 1950 & 19300 & 0.47 \\ \hline 
SKA2-Mid-1000 & 11.8 & 156 & 156000 & 5.93e+06 & 561 & 0.265 & 10400 & 26000 & 0.63 \\ \hline 
SKA2-Mid-2000 & 11.8 & 156 & 312000 & 1.68e+07 & 793 & 0.133 & 20800 & 73500 & 1.8 \\ \hline 
MFAA-150 & 100 & 1180 & 177000 & 6.30e+07 & 391 & 0.547 & 5050 & 77600 & 1.9 \\ \hline 
MFAA-250 & 100 & 1410 & 353000 & 1.78e+08 & 553 & 0.274 & 10100 & 219000 & 5.3 \\ \hline 

\textbf{Addl. Sites 700 MHz} \\ \hline 
CHIME & 113 & 4270 & 4270 & 66900 & 51 & 32.2 & 85.7 & 194 & 0.0047 \\ \hline 
HIRAX & 19.6 & 19.8 & 20300 & 146000 & 111 & 6.79 & 406 & 422 & 0.010 \\ \hline 

\textbf{Addl. Sites 300 MHz} \\ \hline
MWA & 154 & 16.7 & 4270 & 16700 & 40 & 53.5 & 51.6 & 123 & 0.0030 \\ \hline 
SKA1-Low & 2.67 & 107 & 54800 & 120000 & 130 & 4.93 & 560 & 64.9 & 0.0016 \\ \hline 
SKA2-Low & 2.67 & 107 & 523000 & 3.53e+06 & 402 & 0.517 & 5340 & 1910 & 0.046 \\ \hline 

\enddata
\tablecomments{Assumes fully coherent beamforming over the primary FOV, and parameters noted in the text.  Systems are ordered by $N_{STAR}$ count.}
\end{deluxetable*}


\begin{figure}[ht!]
\epsscale{0.8}
\plotone{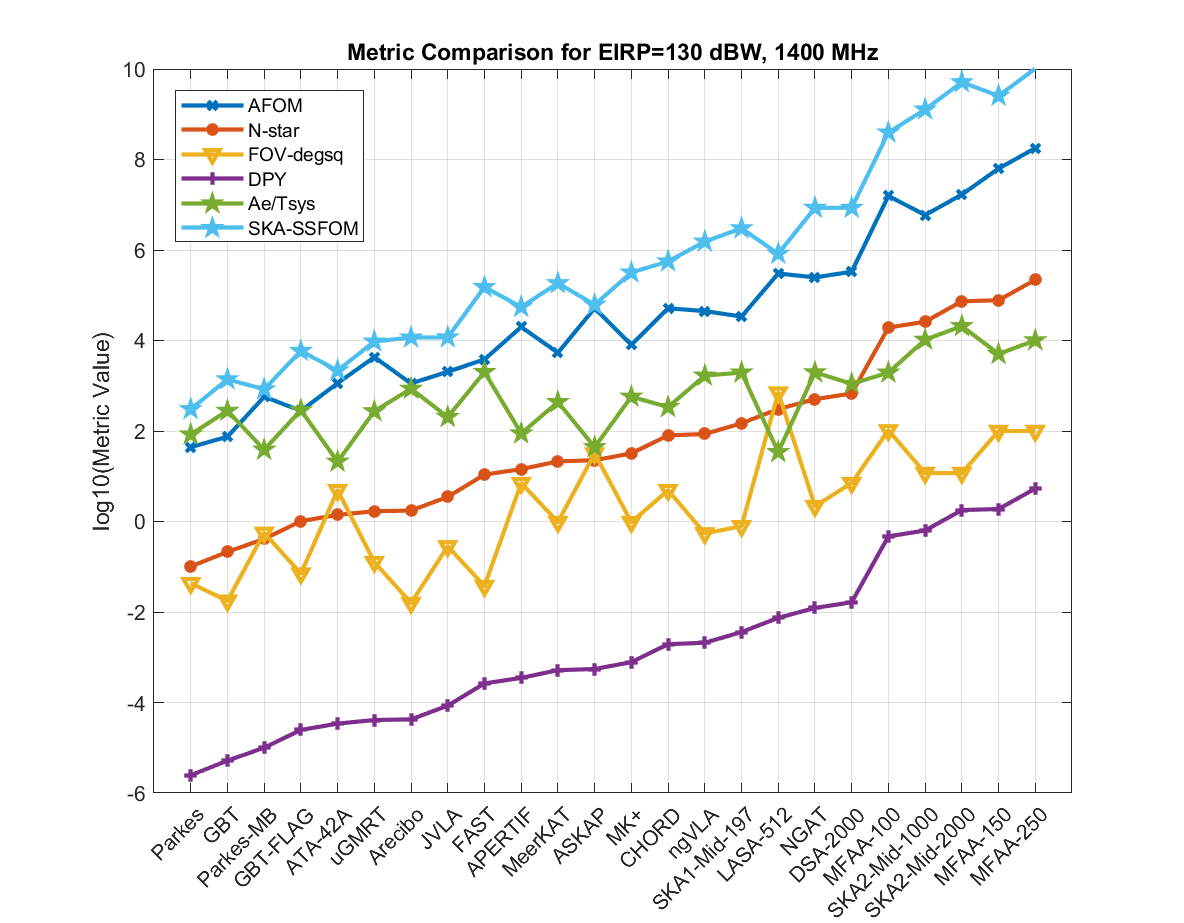}
\caption{Metric Comparison Plot for 1.4\,GHz Systems}
\label{fig:multiparam}
\end{figure}

\begin{figure}[ht!]
\epsscale{0.8}
\plotone{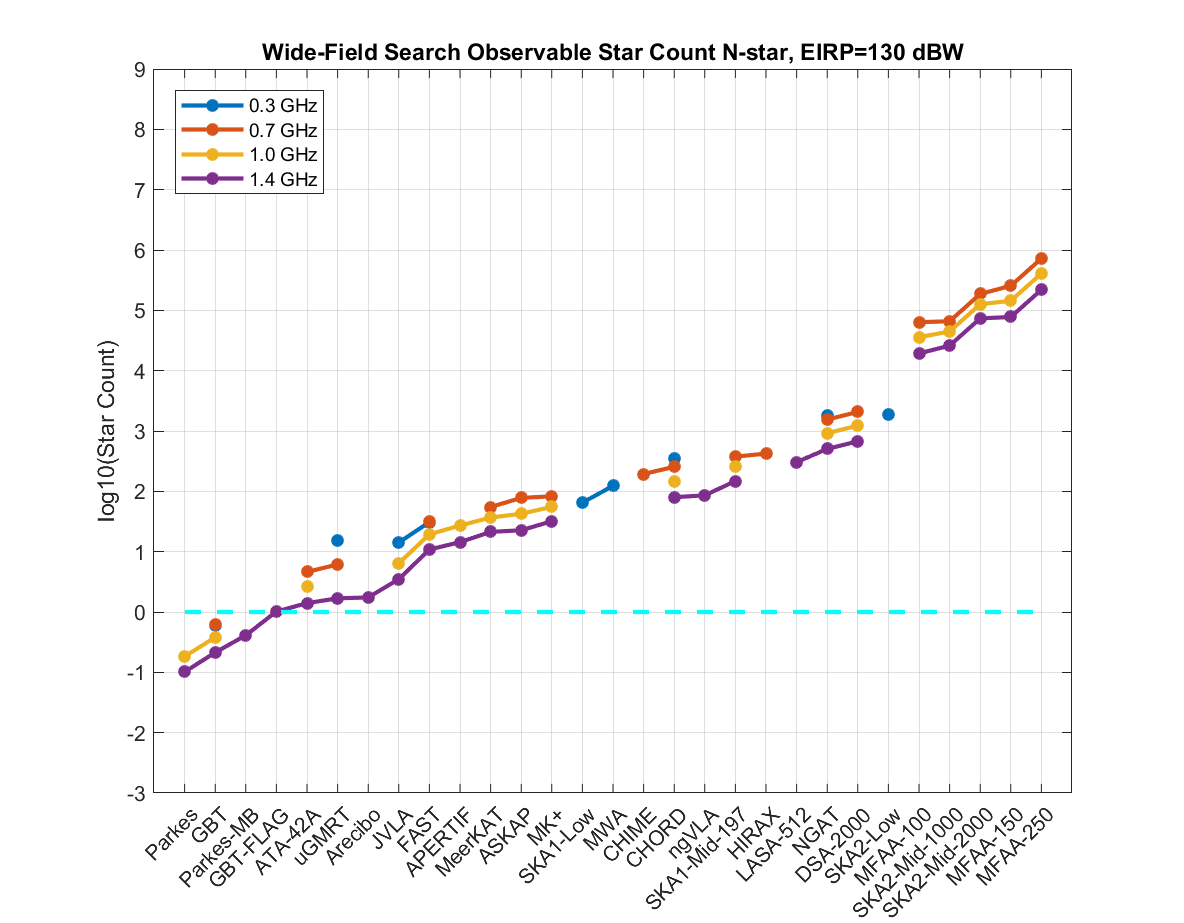}
\caption{$N_{STAR}$ Counts for Various Systems at Several Frequencies and 130\,dBW EIRP}
\label{fig:Nstar130}
\end{figure}

\begin{figure}[ht!]
\epsscale{1.15}
\plottwo{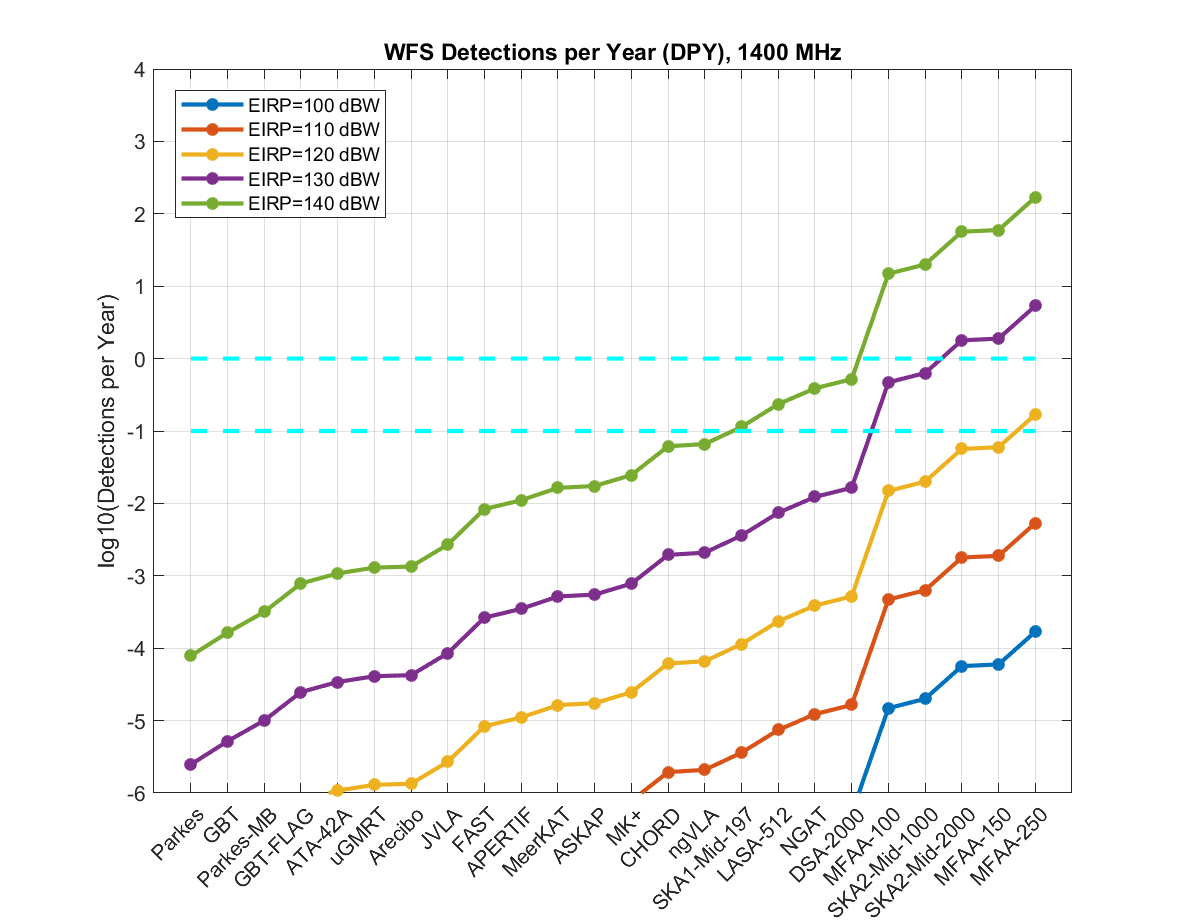}{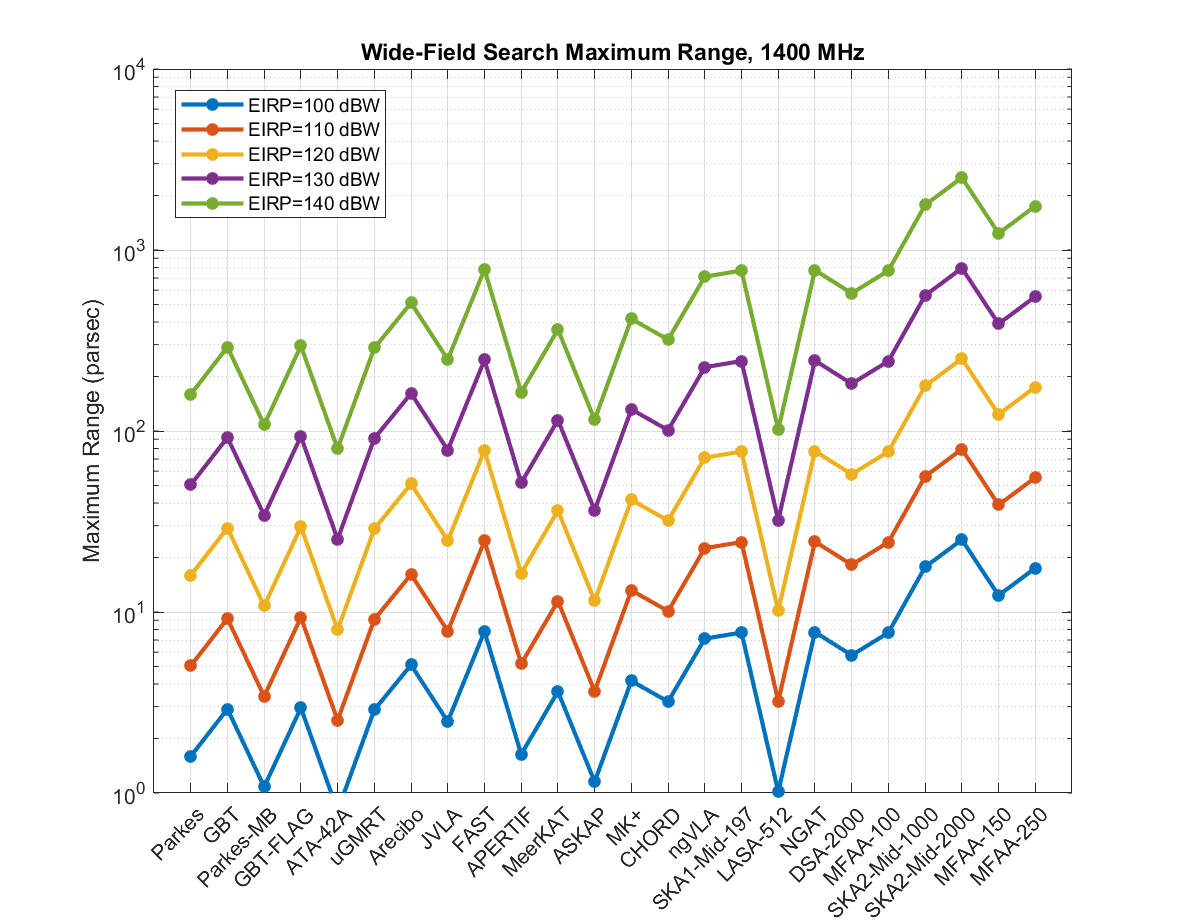}
\caption{Estimated DPY at 1.4\,GHz and $100-140$\,dBW EIRP (left) and Max Detection Range (right), $P_{civTX}=10^{-6}$}
\label{fig:DPY_EIRP_Dmax}
\end{figure}

In Table  $\ref{tbl:Table2}$, FOV, Effective Areas, AFOM, $d_{MAX}$, System Equivalent Flux Density (SEFD), $N_{STAR}$, and DPY are compared for these systems, assuming full coherent beamforming over the primary FOV.  We use 1.4\,GHz as a nominal frequency for Table $\ref{tbl:Table1}$, which most systems support, recognizing that $N_{STAR}$ and DPY may improve at lower frequencies.  For systems that do not support 1.4\,GHz, we use 700\,MHz and 300\,MHz.  The AFOM values are frequency-independent and can be directly compared between all systems.  Key parameters: 0.1 $star/pc^3$ \citep{Siemion_2015}, 130\,dB EIRP, 3\,Hz $\Delta \nu$ (0.33\,s $T_{FFT}$), 300\,s $\tau$ ($N_{avg}$=900), and DT=10. With a different star density, the $N_{STAR}$ and DPY values will change proportionally, but the relative rankings will remain unchanged.  For DPY, we also assume: $T_{TX} = 10$\,minutes $=600 $\,s, $T_{TXdwell} = 12$\,min (5 per hour, 120 per day), $P_{civTX}=10^{-6}$, $N_{RXsite}=2$, $T_{year}=365$\,day and $T_{TXscan}=30$\,day, yielding $N_{tpd} =1.0$ and $N_{STAR1} =41,100$ per equation (\ref{eq:eqn7}). 

Figure $\ref{fig:multiparam}$  presents a comparison of several of the metrics in Table $\ref{tbl:Table2}$.   Some observations:
\begin{itemize}
 \item{It may be seen that AFOM, $N_{STAR}$ and DPY all track each other, as expected.  There is a spread of 6 orders of magnitude in these quantities.  }
 \item{Sensitivity (as indicated by $A_e/T_{sys}$ or SEFD=$2k_B T_{sys}/A_e$) is important, but high sensitivity cannot make up for limited FOV as we see for the large single-pixel dishes which offer the lowest DPY values.  Systems with simultaneous high sensitivity and high FOV offer the best DPY performance.   }
 \item{FOV tracks closely with AFOM, which ranges from .002\,deg$^2$ (GBT) to 707\,deg$^2$ (LASA).}
 \item{AFOM is almost monotonic with $N_{STAR}$ counts, with slight variations due to $T_{sys}$ differences (note that $T_{sys}$ is grouped under FFOM).}
 \item{The dishes with FPAs or PAFs (APERTIF, ASKAP, FAST) do much better than their peers due to FOV multiplication (widening the search cone) and the corresponding increase in AFOM.}
 \item{LASA-512 has relatively low total area (1370\,m$^2$), but has a high AFOM due to its large FOV.}
 \item{The SKA SSFOM (survey speed figure of merit, discussed in Section \ref{ssec:SurveySpeedMetrics}) correlates fairly closely to DPY, as it a sensitivity-weighted FOV metric.}
\end{itemize}

Figure $\ref{fig:Nstar130}$ shows $N_{STAR}$ values for various systems for 130\,dBW EIRP over several frequencies.  For frequencies below 1.4\,GHz, $N_{STAR}$ increases up to 6.5 times due to the increase in FFOM. $N_{STAR}$ ranges from 0.1 to nearly 1 million, with many of the large single-pixel dishes doing worst.  For certain systems, our assumption $N_{STAR}\gg1$ is violated, and targeted search will improve $N_{STAR}$ slightly, but we will see that DPY is very poor regardless.   

Figure $\ref{fig:DPY_EIRP_Dmax}$ (left) plots the expected DPY for the 1.4\,GHz systems, with EIRP varying over $100-140$\,dBW. Lines of DPY at 1 and .1 detection/year are also shown to indicate a ``target range".  As EIRP increases, more systems approach this target range.  We see that due to the EIRP$^{3/2}$ relationship in equation (\ref{eq:eqn15}), DPY increases by 1.5 orders of magnitude ($30\times$) for every 10\,dB ($10\times$) increase in EIRP.

Figure $\ref{fig:DPY_EIRP_Dmax}$ (right) shows the corresponding detection range $d_{MAX}$ values.  For 130\,dB EIRP at 1.4\,GHz, $d_{MAX}$ ranges over $25-793$\,pc ($82-2586$ light years). Sensitivity alone does not assure higher relative levels of DPY in the model.  Many systems have high sensitivity and $d_{MAX}$ values but relatively poor DPY.   The SKA2-Mid and MFAA variants have sensitivities that are higher than the biggest dishes (FAST or Arecibo), but their DPYs are much greater, mostly the result of larger FOVs.  

\begin{figure}[ht!]
\epsscale{1.15}
\plottwo{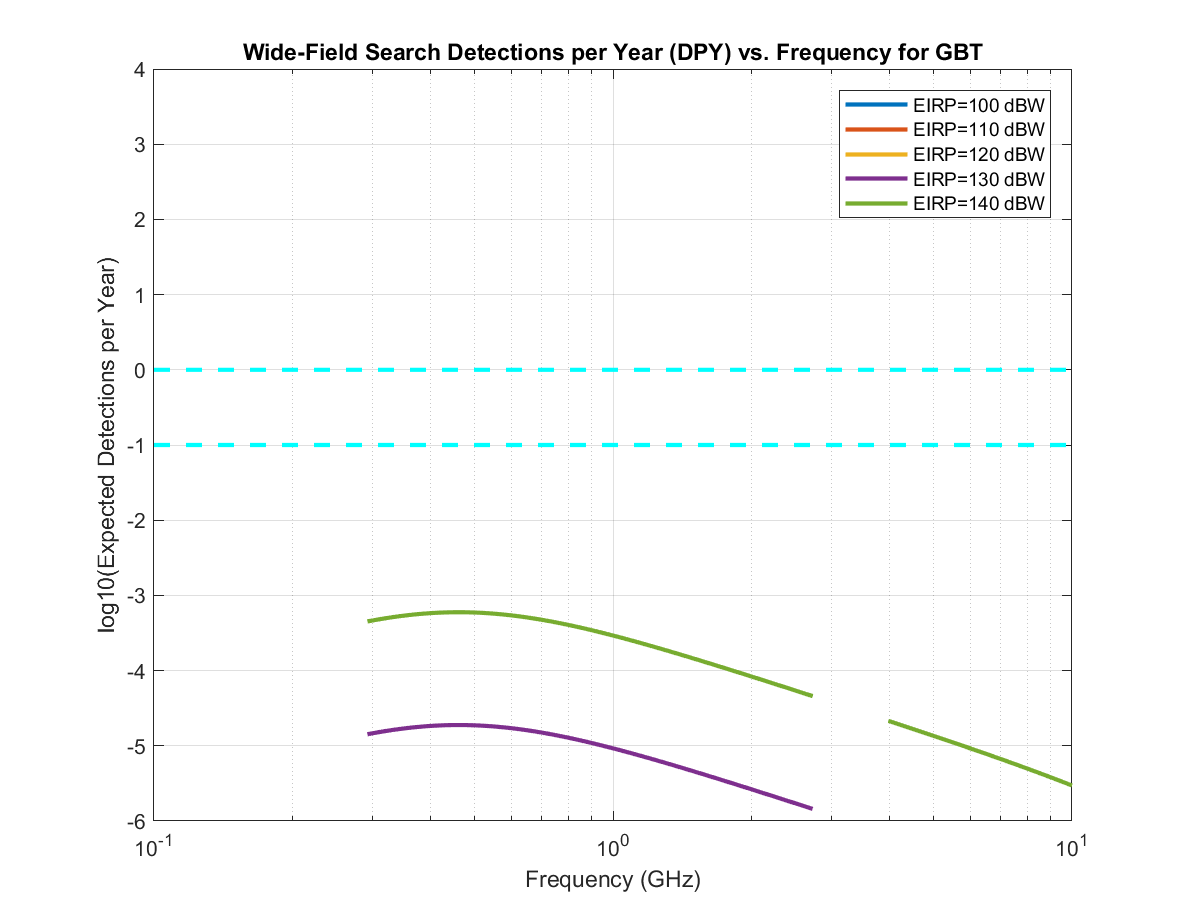}{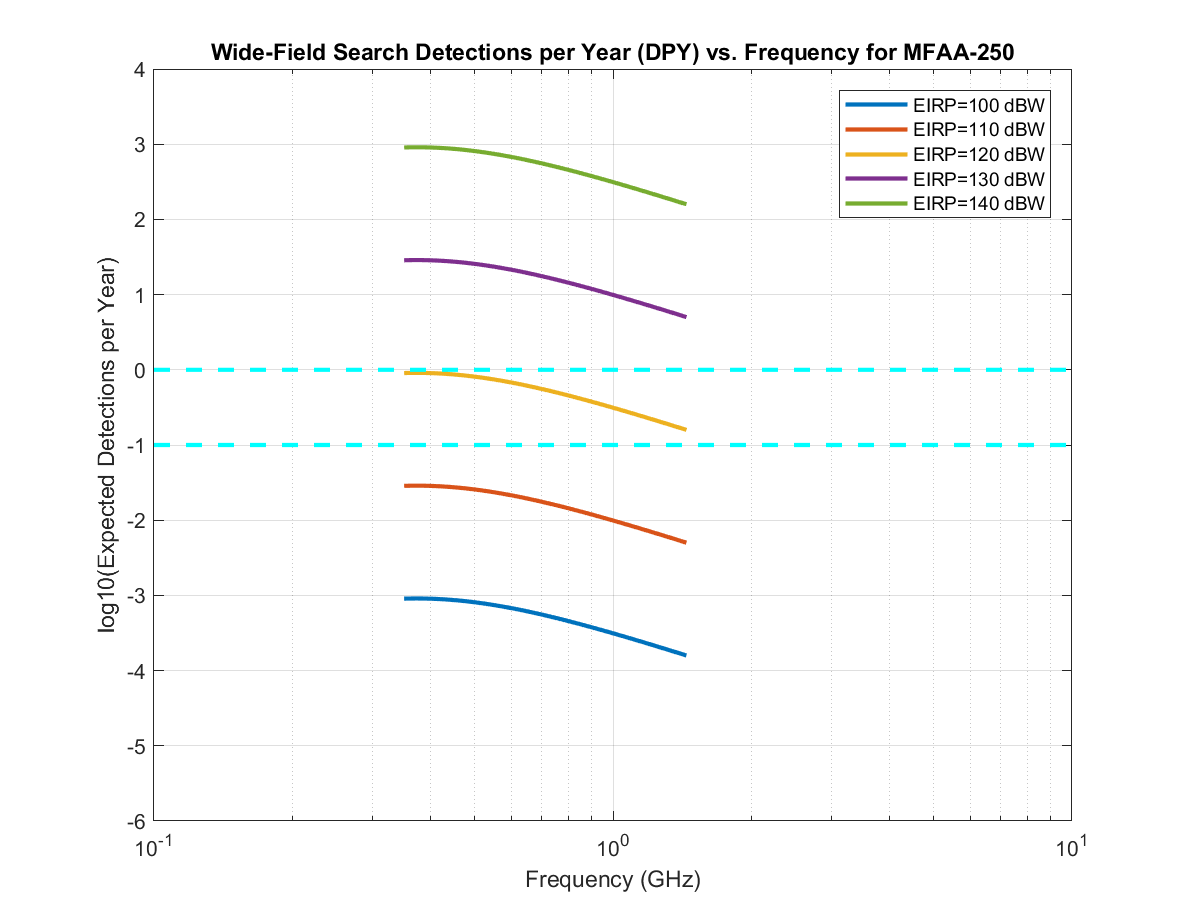}
\caption{Estimated Detections per Year (DPY) vs. Frequency for GBT (left) and MFAA-250 (right), $P_{civTX}=10^{-6}$}
\label{fig:GBT_MFAA}
\end{figure}

Figure $\ref{fig:GBT_MFAA}$ gives expected DPY versus frequency for GBT (left) and MFAA-250 (right) over $100-140$\,dBW EIRPs.  Dashed lines for DPYs of 1.0 and 0.1 are shown as a target range,   It may be seen that the DPY curves have a similar shape to the FFOM curves in Figure $\ref{fig:FFOMExample}$.  There is a factor of 30 (1.5 orders of magnitude) difference for every 10 dB change in EIRP, as expected.  The target DPY range is met for the MFAA-250 for $\sim 115$\,dBW EIRP, while GBT falls far short, where roughly 160\,dBW would be required.  

\begin{figure}[ht!]
\epsscale{0.8}
\plotone{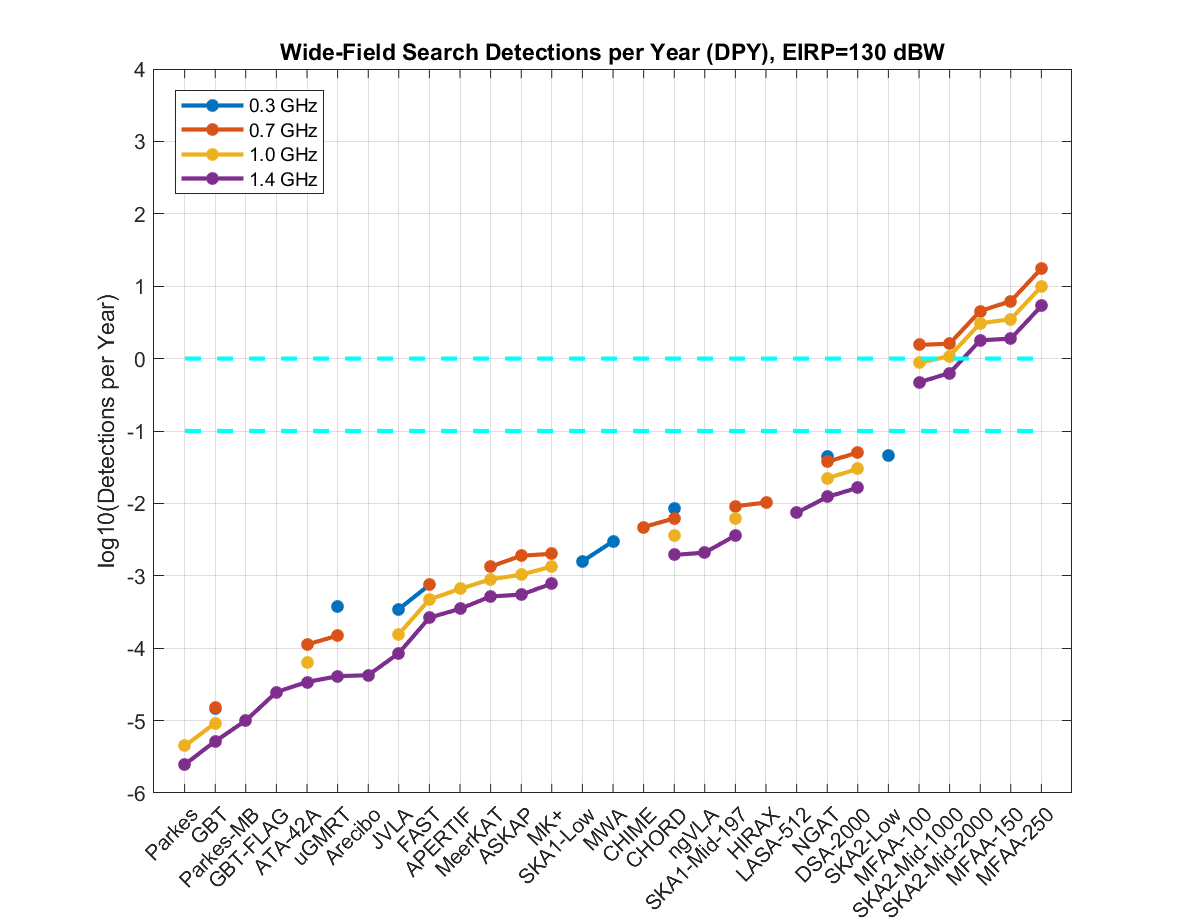}
\caption{DPY for Various Systems at Several Frequencies, 130 dBW EIRP and $P_{civTX}=10^{-6}$}
\label{fig:DPY_freq}
\end{figure}

Figure $\ref{fig:DPY_freq}$ shows the expected DPY for various systems at $P_{civTX}=10^{-6}$ and 130\,dBW EIRP over several frequencies.  Not surprisingly, it looks similar to the $N_{STAR}$ plot in Figure $\ref{fig:Nstar130}$, with increased DPY for frequencies below 1.4\,GHz.  It may be seen that most systems fall short of the .1-1 DPY target range except the SKA2-Low, SKA2-Mid dish and MFAA variants.   

\begin{figure}[ht!]
\epsscale{0.8}
\plotone{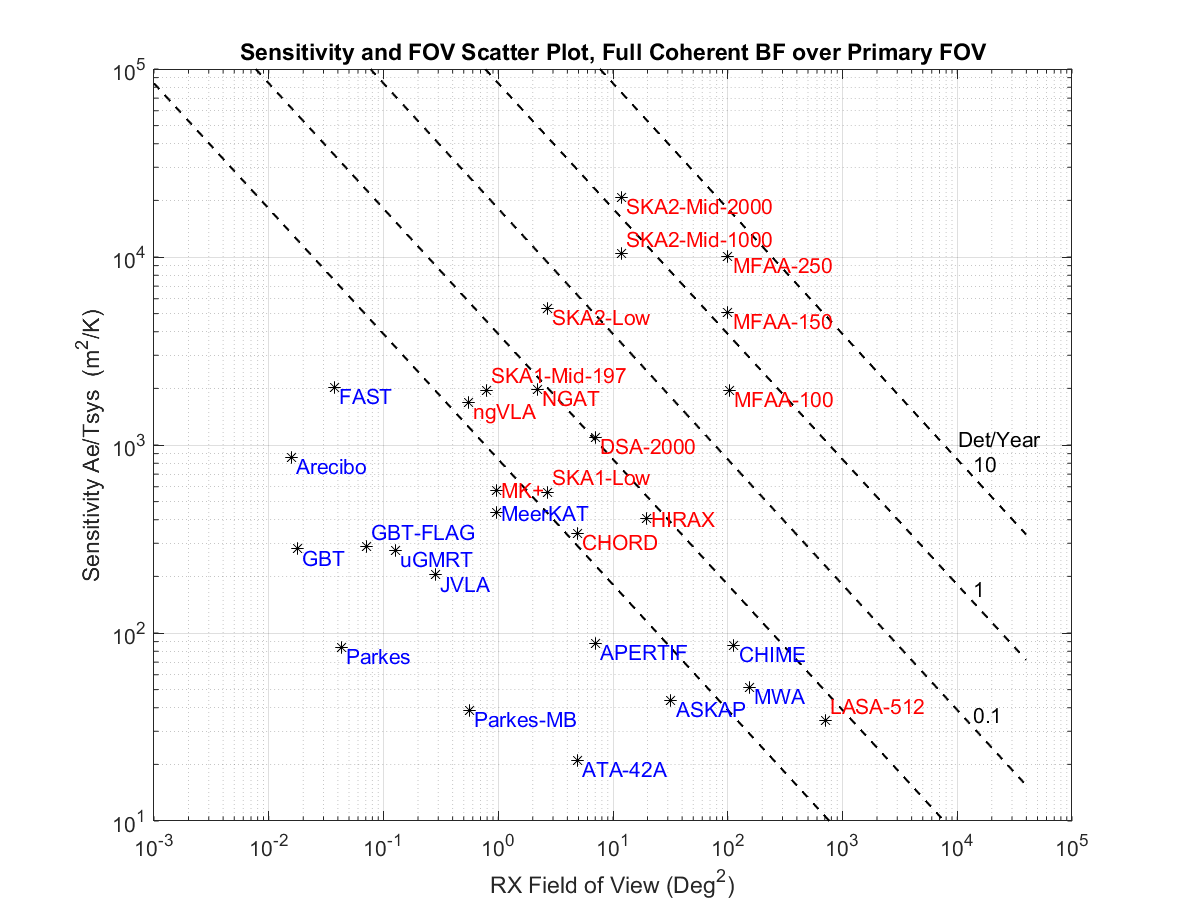}
\caption{Sensitivity and FOV, assuming Full Coherent BF, 130 dBW EIRP and $P_{civTX}=10^{-6}$, for existing systems (blue text) and potential future systems (red text).}
\label{fig:SensFOV_BF}
\end{figure}

\begin{figure}[ht!]
\epsscale{0.8}
\plotone{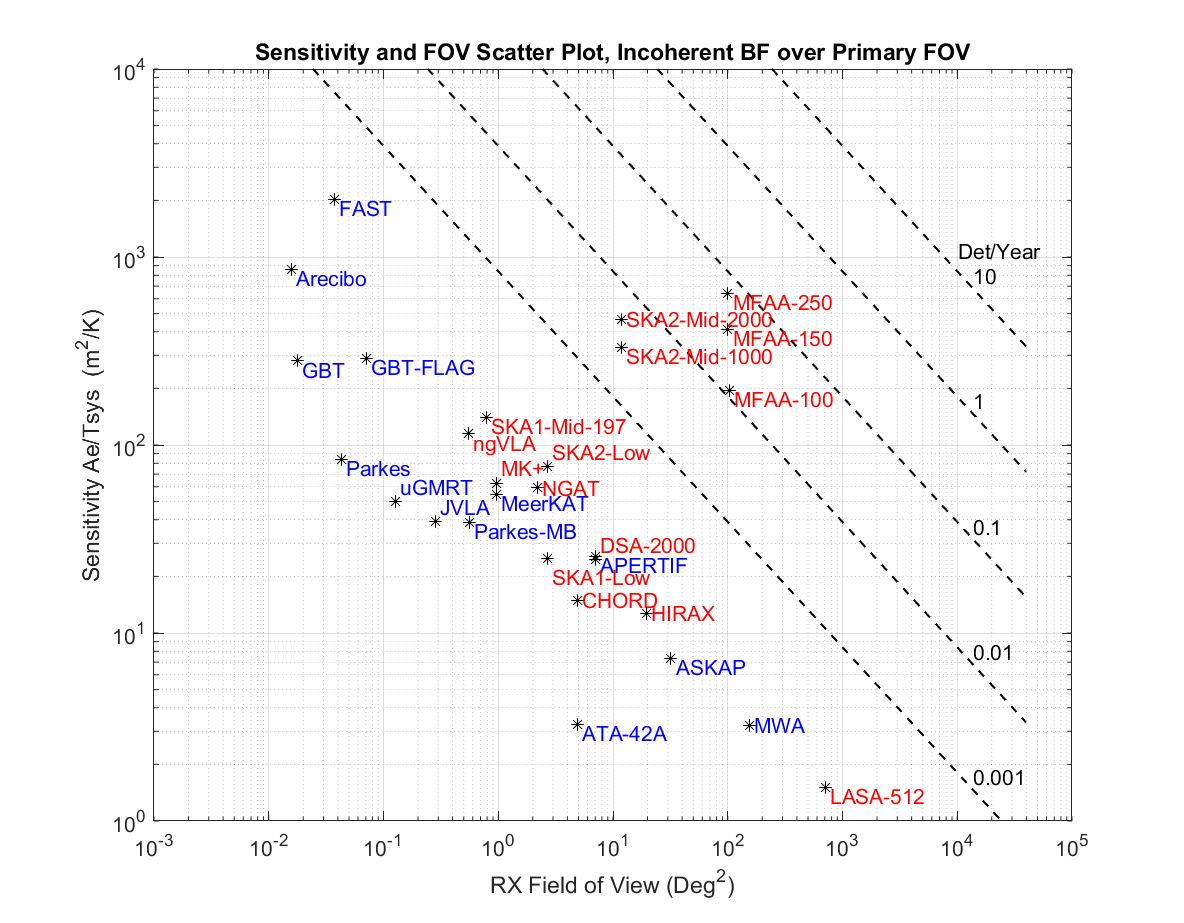}
\caption{Sensitivity and FOV, assuming Incoherent BF, 130 dBW EIRP and $P_{civTX}=10^{-6}$, for existing systems (blue text) and potential future systems (red text).  Note the significant reduction in DPY performance.}
\label{fig:SensFOV_IBF}
\end{figure}

\begin{figure}[ht!]
\epsscale{0.8}
\plotone{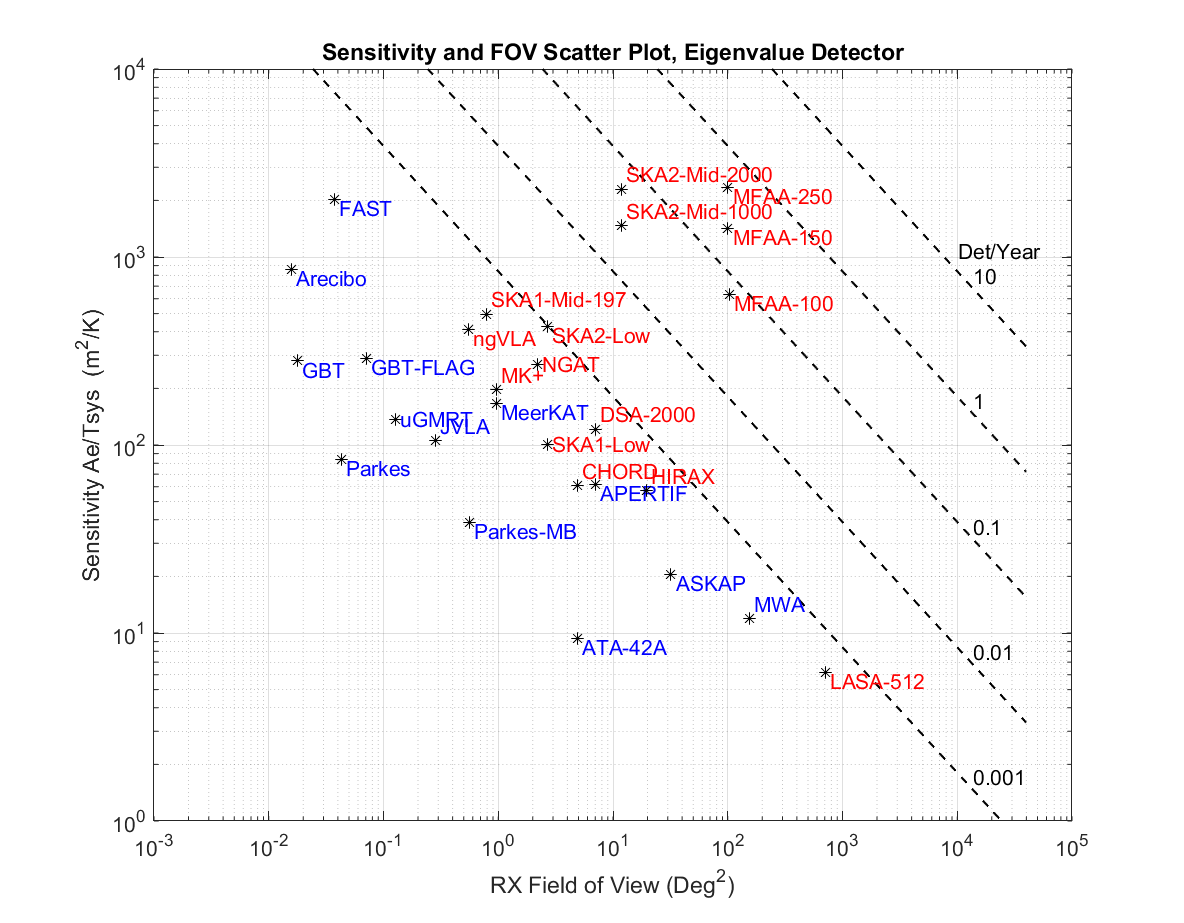}
\caption{Sensitivity and FOV, assuming Eigenvalue Detector, 130 dBW EIRP and $P_{civTX}=10^{-6}$, for existing systems (blue text) and potential future systems (red text).  The DPY has improved by an order of magnitude over IBF for many systems.}
\label{fig:SensFOV_EIG}
\end{figure}

Figure \ref{fig:SensFOV_BF} plots the sensitivity $A_e/T_{sys}$ and $\Omega_{FOV}$ for the systems and nominal frequencies in Table \ref{tbl:Table2}, assuming fully coherent beamforming, and EIRP$=130$\,dBW. Curves for DPY values of 10, 1, .1, .01 and .001 detections/year are also shown, using the relationship 

\begin{equation}  \label{eq:eqn18}
A_e/T_{sys} = \frac{8 \pi k_B}{EIRP \: AvgFOM} \: 
{[ \frac{3  \: DPY \: N_{STAR1}}{\rho_{STAR} \: \Omega_{FOV} }]}^{2/3}
\end{equation}

\noindent which can be easily derived from equations (\ref{eq:eqn1}), (\ref{eq:eqn6a}), (\ref{eq:eqn8}), and (\ref{eq:eqn12}).  

The plot format in Figure \ref{fig:SensFOV_BF} allows us to visualize the trade between sensitivity (and $d_{MAX}$) and FOV. To achieve a factor of 10 increase in DPY, we can either increase sensitivity by a factor of $10^{2/3}\approx5$ at fixed FOV (moving up the vertical axis), or increase $\Omega_{FOV}$ by a factor of 10 at fixed $A_e/T_{sys}$ (moving to the right).  For dish arrays with a fixed diameter, the main options are to increase the number of dishes and (possibly) PAF beams, generally expensive options. For aperture arrays with a large intrinsic FOV, increasing FOV may be the better option, as this often can be done by expanding computation without changing the existing RF infrastructure.  Over time, Moore's Law and improved computational architectures should enable more capability at low incremental cost.  This strategy has long been recognized by radio astronomers \citep{Garrett_2013,Siemion_SETI_Developments_2011}.  

Figure \ref{fig:SensFOV_IBF} is equivalent to Figure \ref{fig:SensFOV_BF}, only with IBF done in place of coherent BF\footnote{CHIME is omitted, as it already does fully coherent beamforming.}.  One may see that the advantage of large numbers of apertures is significantly reduced, and DPY drops as much as 2 orders of magnitude.  Full BF may be difficult or impossible with a finite power or computing budget, but resorting to incoherent BF represents a significant compromise to a system's potential performance.

Figure \ref{fig:SensFOV_EIG} shows a similar plot with array gains for an eigenvalue detector which offers $AG\sim1.7\:{N_{AP}}^{.64}$ $6.4\:log10(N_{AP})+2.3$ dB) per simulations in  \citet{Houston_SETI_Detect_URSI_2021}.  The eigenvalue detector avoids beamforming by computing a sample covariance matrix and a maximum eigenvalue by the power method, and may still be feasible computationally (\citet{Ellingson_EigenBF_2003}).  The DPY values are much improved over IBF but still below BF.

Regarding the overall comparison of systems, we see that future SKA2 variants have significant SETI performance, which is  is quite encouraging.  In particular, as noted above, the aperture array systems (MWA, SKA1-Low, SKA2-Low, and MFAA) are generally limited by beamforming capacity, rather than antenna hardware, and could achieve higher $\Omega _{FOV}$, $N_{STAR}$, and DPY with more signal processing.

\section{Search Figures of Merit} \label{sec:SearchFOMs}

Metrics are needed to evaluate search efforts for effectiveness.  Generally speaking, a SETI program can be measured in terms of the number of star-trials observed.  The most basic ETI Search FOM is just that:

\begin{equation}  \label{eq:eqn20}
Search\:FOM\:1=SFOM1(\nu,EIRP)= N_{obs}\:N_{STAR}(\nu,EIRP) 
\end{equation}

\noindent This applies to both wide-field searches and targeted searches with $N_{STAR} =1$, but as stated earlier, we assume $N_{STAR}\gg1$ even for targeted searches.  The number of time observations $N_{obs}$ is the number of   observations from all sites. While our model defines $N_{obs}\!=\!N_{RXsite}\,N_{TXdwell}\,N_{tpd}$, we can generally approximate this as $N_{obs} = N_{RXsite} T_{obs}/\tau$.\footnote{$N_{obs} =  N_{RXsite}\,N_{TXdwell}\,N_{tpd} = N_{RXsite} N_{tpd} T_{obs}/T_{TXdwell} = N_{RXsite}  [N_{tpd}(\delta_{TX}/T_{TX})\tau] [T_{obs}/\tau]$.  The quantity $[N_{tpd}(\delta_{TX}/T_{TX})\tau]$ is unknown but should be on the order of unity so it may generally be ignored.}  Per (\ref{eq:eqn642}), SFOM1 is proportional to the expected number of detections in the time interval $T_{obs}$.

If one considers a model where the transmit frequency can be uniformly distributed over a range of frequencies, each frequency bin would need to be evaluated separately.  We can introduce a frequency- and bandwidth-weighted metric as follows:

\begin{equation}  \label{eq:eqn21}
Search\:FOM\:2=SFOM2(EIRP)= \sum_i\:N_{obs}\:N_{STAR}(\nu_i,EIRP) \:\Delta \nu
\end{equation}

\noindent The units would be star observations-Hz, or simply Hz.   

We can also scale SFOM1 per (\ref{eq:eqn642}) to produce a metric corresponding to the expected number of detections:

\begin{equation}  \label{eq:eqn21b}
Search\:FOM\:3=SFOM3(\nu,EIRP,P_{civTX},N_{TXscan})= \frac{P_{civTX}}{N_{TXscan}}\,N_{obs}\:N_{STAR}(\nu,EIRP) = \overline{N_{Det}}
\end{equation}

\noindent We need to assume values for $P_{civTX}$ and $N_{TXscan}=T_{TXdwell}/T_{TXscan}$ as well as EIRP, but SFOM3 is easily understood.

Some comments:

\begin{itemize}
 \item{In  (\ref{eq:eqn21}), the frequency bin width $\Delta \nu$ is associated with each summation point, so total bandwidth coverage is taken into account.  }
 \item{We have retained EIRP as a variable.  \citet{Enriquez_2017}, \citet{Wright_2018AJ} and others assume an Arecibo-level 130\,dBW EIRP as a minimum for discussion, which we also use, but other values in the range of $100-120$\,dBW may be entertained (see Section \ref{sec:SystemRequirements}).}
 \item{We have not factored in the number of polarizations (as is sometimes done) but have assumed 2 polarizations with Stokes I summation as in most modern receivers.  }
 \item{We could base the metric on the search cone volume $V_{cone}(\nu_i,EIRP)$ and avoid possible ambiguity regarding star density $\rho_{STAR}$.}
\end{itemize}

The SFOMs are useful to compare current efforts against past efforts, and to determine the “haystack fraction” of local space actually searched, but should also guide future search efforts, particularly with regard to the efficacy of search frequency coverage.  In  (\ref{eq:eqn21}), there is a frequency dependence according to the FFOM  (\ref{eq:eqn8}) that allows us to evaluate the SFOM2 over a proposed frequency band.  Given a cost budget, high frequencies that contribute minimally to the SFOM2 may be discouraged. See additional discussion in appendix  \ref{sec:LimitedComp}.

While many past systems had limited bandwidth that varied from system to system, modern systems cover wide frequency ranges, so this is less important for “scoring”.  SFOM1 and SFOM3 are both functions of frequency and are in units of ``star-observations'' and ``expected detections'', respectively, which should be intuitively easier to grasp. 

Not surprisingly, the SFOMs above are similar to others described by \citet{Wright_2018AJ}.  The primary advantage of SFOM1 and SFOM3 is that they relate directly to the probability of a detection occurring within an observation interval, given a set of assumed parameters. 

\subsection{Comparison to the Drake FOM}

The formulation in the SFOMs above in terms of $N_{STAR}$ is entirely consistent with Drake’s original search FOM \citep{Drake_1984},which applies to wide-field search.  First, define the minimum flux $\phi_{MIN}$  $(W/m^2)$ that may be detected, associated with a source of a given EIRP at a distance $d_{MAX}$, in terms of $N_{STAR}$ from Equation (\ref{eq:eqn1}):

\begin{equation}  \label{eq:eqn22}
\phi_{MIN}= \frac{EIRP}{4\pi\:{d_{MAX}}^2}
= \frac{EIRP}{4\pi}\:[\frac{\rho_{STAR}\:\Omega_{FOV}}{3\:N_{STAR}}]^{2/3}
\end{equation}

Drake’s figure of merit for a search program is:

\begin{equation}  \label{eq:eqn23}
DFM= \frac{\Omega _{Total}\: BW_{Total}}{{\phi_{MIN}}^{3/2}}   	
\end{equation}

where 

{\addtolength{\leftskip}{10 mm}
\setlength{\parindent}{0 mm}

$ $

$\Omega_{Total} = N_{obs}\:\Omega_{FOV} = $Total solid angle covered in a survey

$ $

$BW_{Total}$ = Total bandwidth covered 

$ $

}

Upon substituting the above, we obtain:

\begin{equation}  \label{eq:eqn24}
DFM(\nu) = \frac{ 24\:\pi^{3/2} }{ \rho_{STAR}\:{EIRP}^{3/2} }
\:N_{obs}\:N_{STAR}(\nu,EIRP)\:BW_{Total}
\end{equation}

Therefore, the DFM is a scaled version of SFOM2 above.  DFM is normalized for EIRP and $\rho_{STAR}$, and the bandwidth is simply multiplied by $N_{STAR}(\nu)$ rather than integrating $N_{STAR}(\nu)$ over frequency.  The quantity $\Omega_{Total}/{\phi_{MIN}}^{3/2}$, or the DFM with $BW_{Total}$ omitted, is a scaled version of SFOM1.  The DFM assumes the same uniform star density concept that is behind the SFOMs above. 

\subsection{Comparison to CWTFM}

\citet{Enriquez_2017} define a Continuous Waveform Transmitter Figure of Merit (CWTFM)

\begin{equation}  \label{eq:eqn24a}
CWTFM = \eta \frac{ EIRP_{MIN} }{ N_{STAR} \: N_{obs} } \:\frac{\nu_{mid}}{ BW_{total} } = \eta \frac{ EIRP_{MIN} }{ N_{STAR} \: N_{obs} } \: \nu_{rel} 
\end{equation}

\noindent where $\nu_{mid}$ is the band center, $\nu_{rel}=BW_{total}/ \nu_{mid}$ is the fractional total bandwidth, and $\eta=EIRP_0 \: \nu_{rel}$ is a normalizing constant with representative values chosen as $EIRP_0=10^{13}$\,W (Arecibo) and $\nu_{rel}=0.5$. The value $EIRP_{MIN}=4 \pi {d_{TgtStar}}^2 \phi_{MIN}$ is the minimum EIRP for a SETI transmitter at the targeted star.  A ``Transmitter Rate'' is also defined as 

\begin{equation}  \label{eq:eqn24b}
Transmitter\:Rate = TR =  1 / (N_{STAR} \: N_{obs} \: \nu_{rel}) \: .
\end{equation}

\noindent so $TR = 1 / (SFOM1 \: \nu_{rel})$.  Transmitter Rate scores fractional bandwidth instead of total bandwidth searched.  TR is defined as an inverse quantity, which might be counter-intuitive (lower TR implies a better survey).  TR has been plotted against $EIRP_{MIN}$ for many surveys \citep{Enriquez_2017,Price_2018,Wlodarczyk_Sroka_2020,Li_2020,Siemion_2015} and appears more widely adopted than CWTFM itself.  For past comparisons of TR, most surveys were scored as $N_{STAR}=1$, and $d_{TgtStar}$ and $EIRP_{MIN}$ were well defined.  With $N_{STAR}\gg1$ in the FOV for most surveys \citep{Wlodarczyk_Sroka_2020}, TR and CWTFM become problematic because  $N_{STAR}(\nu,EIRP)$ is a function of frequency which is then multiplied by a fractional bandwidth.  CWTFM and TR appear to be suitable for comparing surveys with either 1) single stars or 2) a single frequency.  It might be possible to extend them to be more generally applicable, but at this point we prefer SFOM1 or SFOM3 because they quantify a survey's star-observations, which will be proportional to the probability of achieving an ET detection. 

\subsection{Comparison to Haystack Fraction Metrics} \label{ssec:HaystackFractionMetrics}

\citet{Wright_2018AJ} developed a set of ET search metrics called ``Haystack'' functions, which are meant to evaluate SETI effectiveness over the complete range of possible TX characteristics.  The haystack volume involves 9 dimensions. Haystack functions are defined for a survey and the ideal case which spans the "total" haystack volume.  The ratio of the survey and total haystack functions is the ``haystack ratio''.  Haystack ratios over many SETI efforts can be summed to obtain a measure of how much SETI search has been done to date. 

An evaluation of Wright's haystack functions appears in Appendix \ref{sec:HaystackFraction}. While the effort is on the right track, we identify issues and suggest improvements in the following areas:

\begin{itemize}
\item{The method of integration over all dimensions creates a result which is hard to interpret and may be unreliable.  Parts of the volume that are included may have zero or very low probability.  A weighting based on \textit{a priori} probability densities for the TX parameters creates results which are intuitively more sensible, with the added benefit that much of the haystack integration is an expected value operation. The net haystack is a volume integration of a desired FOM quantity (sensitivity or other function) where the FOM quantity is averaged over the range of possible TX parameters. This forces us to hypothesize \textit{a priori} probability densities for the TX parameters, which may take many forms depending on assumptions.  However, without these pdfs, we make an implicit assumption that the probabilities of all voxels in the N-dimensional integration are equal, which we know is not the case.}
\item{The sensitivity FOM which Wright uses in the integration has shortcomings. We suggest: 1) integrating star density over a detection volume, which yields $N_{STAR}$; 2) summing over observations, which yields star-observations like in SFOM1; and 3) scaling to obtain a measure similar to SFOM3 or $\overline{N_{Det}}$. The suggested FOM integrand is star density times the probability of detection, averaged over the full set of TX parameters.}
\end{itemize}

The revised haystack volume is an estimate of the expected number of detections in a survey effort, averaged over the full range of TX parameters.  If we choose a single detection as a reference value, the survey haystack fraction is equal to the survey haystack volume.  The end result is completely consistent with the DPY analysis presented to this point. More investigation is needed, but we believe this is the right approach for search metrics going forward.  Refer to Appendix \ref{sec:HaystackFraction} for more details.

\subsection{Comparison to Survey Speed Metrics} \label{ssec:SurveySpeedMetrics}

In radio astronomy, search speed metrics are defined as a measure of FOV adjusted for sensitivity, since the survey speed $\propto  \Omega_{FOV}/\tau$.  The SKA program uses the following \citep{Bunton_2003_SKA_SSFOM}:

\begin{equation}  \label{eq:eqn25}
\textit{SKA{\textendash}SSFOM}= \Omega_{FOV} \: (\frac{A_e}{T_{sys}})^{2} 
\end{equation}

For broad-band sources, in order to keep a constant minimum detectable flux density, the integration time $\tau$ scales as  $(T_{sys}/A_e)^2$.  SKA-SSFOM needs to scale as $1/\tau$, hence the $(A_e/T_{sys})^{2}$ factor.  If $N_{FOV}=1$ and $AG=N_{AP}$, we can easily show that 

\begin{equation}  \label{eq:eqn26}
\textit{SKA{\textendash}SSFOM}= \frac{ \lambda^2 }{ A_{eAP} } \:
(\frac{ A_{eAP} N_{AP} }{ T_{sys} } )^2 \:= (FFOM)\:(AFOM)\:(\frac{ A_e }{ T_{sys} } )^{1/2}
\end{equation}

Therefore, while the SKA-SSFOM in Figure $\ref{fig:multiparam}$  tracks $N_{STAR}$, DPY and AFOM reasonably well, there is a mismatch in slope brought about by the extra  $(A_e/T_{sys})^{1/2}$  factor.  This mismatch is noted in the pulsar/FRB literature as well \citep{Macquart_2011}.

Another survey speed metric is described by \citet{Enriquez_2017}:  

\begin{equation}  \label{eq:eqn27}
\textit{Enriquez{\textendash}SSFM} = \frac{ BW_{Total} }{(SEFD)^2 \: \Delta \nu}  
= \frac{ BW_{Total} }{(2k_B)^2 \: \Delta \nu}  \:(\frac{ A_e }{ T_{sys} } )^2
\propto \frac{ BW_{Total} }{\tau}
\end{equation}

Noting $SEFD=2 k_B T_{sys}/A_e$, the Enriquez-SSFM appears to be a sensitivity measure only, and does not explicitly take FOV into account.  

We can define a more suitable and generally-applicable speed metric, as follows:

\begin{equation}  \label{eq:eqn28}
SSFOM = \frac{N_{RXsite} \: N_{STAR} }{ \tau } \propto DPY
\end{equation}

For each site, maximizing $N_{STAR}/\tau$ should be sufficient to guarantee survey speed and DPY at the same time.  We also note that

\begin{equation}  \label{eq:eqn29}
SSFOM = \frac{N_{RXsite} }{ \tau } \: \frac{1}{3}\: \rho_{STAR} \: \Omega_{FOV} \: {d_{MAX}}^3
 \propto \Omega_{FOV} \: (\frac{ A_e }{ T_{sys} } )^{3/2}
 = (FFOM)\:(AFOM)
\end{equation}

We obtain the 3/2 power relationship to sensitivity $A_e/T_{sys}$ as might be expected.  We should also note: \textbf{Since SSFOM in  (\ref{eq:eqn28}) is proportional to DPY, SSFOM is redundant.  It is sufficient to focus on DPY alone.}

\section{Expected Time to First Detection of a SETI Program} \label{sec:TimeToFirstDetection2}

The required search time to achieve detections is often neglected in SETI analyses, partly because many parameters like $P_{civTX}$ are completely unknown.  With an estimate of DPY, we can evaluate the time to a first detection.  Assuming a large number of independent potential sources, the number of detections over a given time interval should follow a Poisson arrival process.  Over an observation period $T_{obs}$ (expressed in years)\footnote{Note that we have defined DPY to be a detection rate expressed in units ``detections/year''.  We can define an equivalent detection rate $R_{Det}$=DPY/$T_{year}$ where $R_{Det}$ is is expressed in arbitrary time units, e.g. detections/second.  Most formulas above involve ratios of time quantities, so the choice of time units doesn't matter. We assume any time quantities in formulae involving DPY or $N_{STAR1}$ to be expressed in years, with $T_{year}$=1.}, we would expect

\begin{equation}  \label{eq:eqn31}
P(N_{Det}=k)=(DPY \: T_{obs} )^k \: e^{-DPY \:T_{obs} }/k!  \;,
\end{equation}

\noindent
where $P(N_{Det}=k)$ is the probability of exactly k detections in the interval $T_{obs}$. Noting that the probability of at least one detection is just one minus the probability of zero detections

\begin{equation}  \label{eq:eqn33}
P(N_{Det} \ge 1) = 1-P(N_{Det}=0) = 1-e^{-DPY \:T_{obs}} \;,
\end{equation}

\noindent
we can choose a constant $\alpha$ such that $DPY \: T_{obs} = \alpha$ makes $P(N_{Det} \ge 1)$ a high value.  From (\ref{eq:eqn6a}) and (\ref{eq:eqn6b}), we then obtain: 

\begin{equation}  \label{eq:eqn34}
T_{obs-min} = \frac{\alpha}{DPY} 
= \frac{\alpha\, N_{STAR1}}{N_{STAR}}
= \frac{\alpha \: T_{TXscan} }{  P_{civTX} \: N_{tpd} \:  N_{RXsite} \: N_{STAR} }
\end{equation}

\noindent where $T_{obs-min}$ is the required observation time in years to achieve one or more detections with high probability.  This applies to wide-field searches ($N_{tpd}=T_{TX}/\tau-1$), as well as slow-scan searches and targeted searches ($N_{tpd}=1$), provided the observation time per pointing is much less than $T_{TXscan}$.  For simplicity let us choose $\alpha$=1, which implies  $T_{obs-min}$=1/DPY.\footnote{For the Poisson distribution the expected number of detections over $T_{obs}$ is E(k)=$DPY\:T_{obs}=\overline{N_{Det}}$, so $\alpha$=$DPY\:T_{obs}$=1 corresponds to one expected detection.  If $T_{obs-min}$=1/DPY, we need to observe for 10 years if DPY=.1.  For $\alpha$=1, $P(N_{Det} \ge 1) = 63\%$.  For more confidence we can choose a higher value of $\alpha$ (e.g. $\alpha$=2.3 yields $P(N_{Det} \ge 1) = 90\%$), and $T_{obs-min}$=$\alpha$/DPY. }

For example, for WFS with with 100 observable stars, and $N_{STAR1}$=41,100 per (\ref{eq:eqn6a})\footnote{As before, we assume $\tau =5$\,min, $T_{TX}=10$\,min, $T_{TXdwell}=12$\,minutes, $T_{TXscan} =30$\,d $=0.0822$\,yr, $\delta_{TX}=83\%$ ,  $N_{RXsite} = 2$,  and  $P_{civTX} = 10^{-6}$, so $T_{TX}/\tau =2$ and $N_{tpd} =1.0$. } we get 

\begin{eqnarray}
T_{obs-min}=  \frac{1 \: (41100) }{100} = 411 \:{\rm years!}  \nonumber
\end{eqnarray}

\noindent
With 10,000 stars, $T_{obs-min}$ reduces to 4.1 years. Clearly $N_{STAR}$ must be large to counteract a small value of $P_{civTX}$.

\noindent
Alternatively, if we have an allowable observation time $T_{obs}$, we may estimate the number of observable stars required to achieve a single detection with high probability:

\begin{equation}  \label{eq:eqn36}
N_{STAR} \ge  \frac{\alpha \, N_{STAR1}}{T_{obs} }
= \frac{\alpha \: T_{TXscan} }{  P_{civTX} \: N_{tpd} \:  N_{RXsite} \: T_{obs} } \;.
\end{equation}

\noindent
For example, for the same parameters above and $T_{obs}$ limited to 10 years, we get 

\begin{eqnarray}
N_{STAR} \ge  \frac{1 \: (41100)}{10} = 4110 \;.    \nonumber
\end{eqnarray}

\noindent
This is a big number compared to most of the $N_{STAR}$ values in Table $\ref{tbl:Table2}$, but this could potentially be met by the SKA2/MFAA systems.

Stated still another way, if we have a known $N_{STAR}$ value, the minimum $P_{civTX}$ value which would result in a detection would be

\begin{equation}  \label{eq:eqn37}
 P_{civTX} \ge  \frac{\alpha \: T_{TXscan} }{  N_{tpd} \:  N_{RXsite} \: N_{STAR} \: T_{obs} } \;.
\end{equation}

\noindent
More generally, the $P_{civTX}\,N_{STAR}$ product, or the average number of active transmitters per observation $N_{TX}$, needs to be

\begin{equation}  \label{eq:eqn38}
 N_{TX} = P_{civTX} \, N_{STAR} \ge  \frac{\alpha \: T_{TXscan} }{  N_{tpd} \:  N_{RXsite} \: T_{obs} } \;.
\end{equation}

\noindent
With the representative numbers above for 10 years observation, $P_{civTX} \: N_{STAR} \ge  1 (0.0822)/(1 \: 2 \: 10 )  \approx .004$.   An $N_{STAR}$ value of 1 will result in a probable detection only if $P_{civTX} \ge .004$, but if $N_{STAR} =100,000$ a detection should occur for $P_{civTX} \ge 4\times10^{-8}$.   Clearly this is a strong motivator to get $N_{STAR}$ as high as possible.

Finally, if we have observed over $T_{obs}$, we can show that the achieved fraction of required search for a probable detection will be

\begin{equation}  \label{eq:eqn39}
\Phi_{Det} = Search\:Fraction =  \frac{ T_{obs} }{ T_{obs-min} }
= \frac{  T_{obs} \, DPY} {\alpha}
= \frac{  T_{obs} \, N_{STAR}} {\alpha \, N_{STAR1}}
= \frac{  P_{civTX}}{ \alpha \:  N_{TXscan} } \: N_{obs}  N_{STAR}
= \frac{SFOM3}{\alpha}\;.
\end{equation}

\noindent If we search for $\alpha$/DPY years, the search fraction will be 100\% with confidence 1-$e^{-\alpha}$. The search fraction is equivalent to SFOM3 defined in (\ref{eq:eqn21b}) with $\alpha$=1.  Instead of increasing the number of observations $N_{obs}$ over all receive sites by many orders of magnitude, a better tactic should be to elevate $N_{STAR}$.

Consider 2 years of observation with the previous parameters.  The search fraction will be

\begin{eqnarray}
\Phi_{Det} = \frac{ 2 }{1 \: (41100)} \: N_{STAR} = 4.9\times10^{-5} \: N_{STAR}  \;,    \nonumber
\end{eqnarray}

or $4.9\times10^{-5}$ for $N_{STAR}$ = 1 and 0.49 for $N_{STAR}$ = 10,000, a huge difference.

\section{Expected Time to First Detection of An Existing ET Source for a Fast Wide-Field Search} \label{sec:TimeToFirstDetection1}

Consider next a case where a source is actively transmitting within our maximum detection range $d_{MAX}$ with a scan period of $T_{TXscan}$.  We would like assurances that it can be detected in a reasonable period of time in a Fast-Scan WFS.  Let us define and evaluate the metric $T_{Det1}$:

{\addtolength{\leftskip}{10 mm}
\setlength{\parindent}{0 mm}

$ $

$T_{Det1}(d)$ = The expected time to a first ET detection given a transmitter at range d

$ $

=$\; \sum_{j=1}^\infty$ E($T_{Det}\: \vert$ First Detection in TX cycle j with TX at range d ) *

$\;\;\;\;\;\;\;\;\;$ p(First Detection in TX cycle j with TX at range d)

$ $

$= \; \sum_{j=1}^\infty \: [(j-\frac{1}{2})\:T_{TXscan} ) ]$ p(Detection in TX cycle j) p(No Detections in TX cycles 1..j-1)

$ $

$= \; \sum_{j=1}^\infty \: (j-\frac{1}{2})\:T_{TXscan} \: p_{DTXC}(d) \: (1-p_{DTXC}(d))^{j-1}$

$ $

}

where

{\addtolength{\leftskip}{10 mm}
\setlength{\parindent}{0 mm}

$ $

$p_{DTXC}(d)$ = p(Detection in TX cycle j $\vert$ TX at range d)

\begin{equation}
  = P_D(d)\:N_{tpd}/N_{RXscan}=\begin{cases}
     \frac{T_{TX}/\tau - 1 }{ N_{RXscan} } & d\:<\:d_{MAX} \\
     0 & \text{Otherwise}.
  \end{cases}
\end{equation}

$ $

}

Noting that  $\sum_{j=1}^\infty \:j\: p\:(1-p)^{j-1} = 1/p$  and $\sum_{j=1}^\infty \: p \:(1-p)^{j-1} = 1$ , we get for $d < d_{MAX}$: 

\begin{eqnarray}
T_{Det1} = T_{TXscan} (\frac{1}{p_{DTXC}} - \frac{1}{2}) 
= T_{TXscan} (\frac{N_{RXscan}}{T_{TX}/\tau-1}-\frac{1}{2})
\approx \frac{\tau \: T_{TXscan} \: N_{RXscan}}{T_{TX}-\tau}  \nonumber
\end{eqnarray}

or

\begin{equation}  \label{eq:eqn30}
T_{Det1} \approx \frac{T_{TXscan} \: T_{RXscan}}{T_{TX}-\tau}
\approx \frac{\tau}{T_{TX}-\tau} \: \frac{4\pi \:T_{TXscan}}{\Omega_{FOV} \: N_{RXsite}}
\end{equation}

$T_{Det1}$ is therefore proportional to TX scan time and inversely proportional to $\Omega_{FOV}$.  The time to a first detection of an existing source will be increased as scan times increase in both transmit and receive.  Large antenna gains result in small values of $\Omega_{TX}$ and $\Omega_{FOV}$ which can cause $T_{Det1}$ to grow very large.  $\textit{It is a multiplicative, rather than additive, effect.}$  Given that small $\Omega_{TX}$ is needed to raise EIRP to cover interstellar distances, wide field of view in the receiver is critical to reducing the time to a first detection.

\begin{figure}[ht!]
\epsscale{0.8}
\plotone{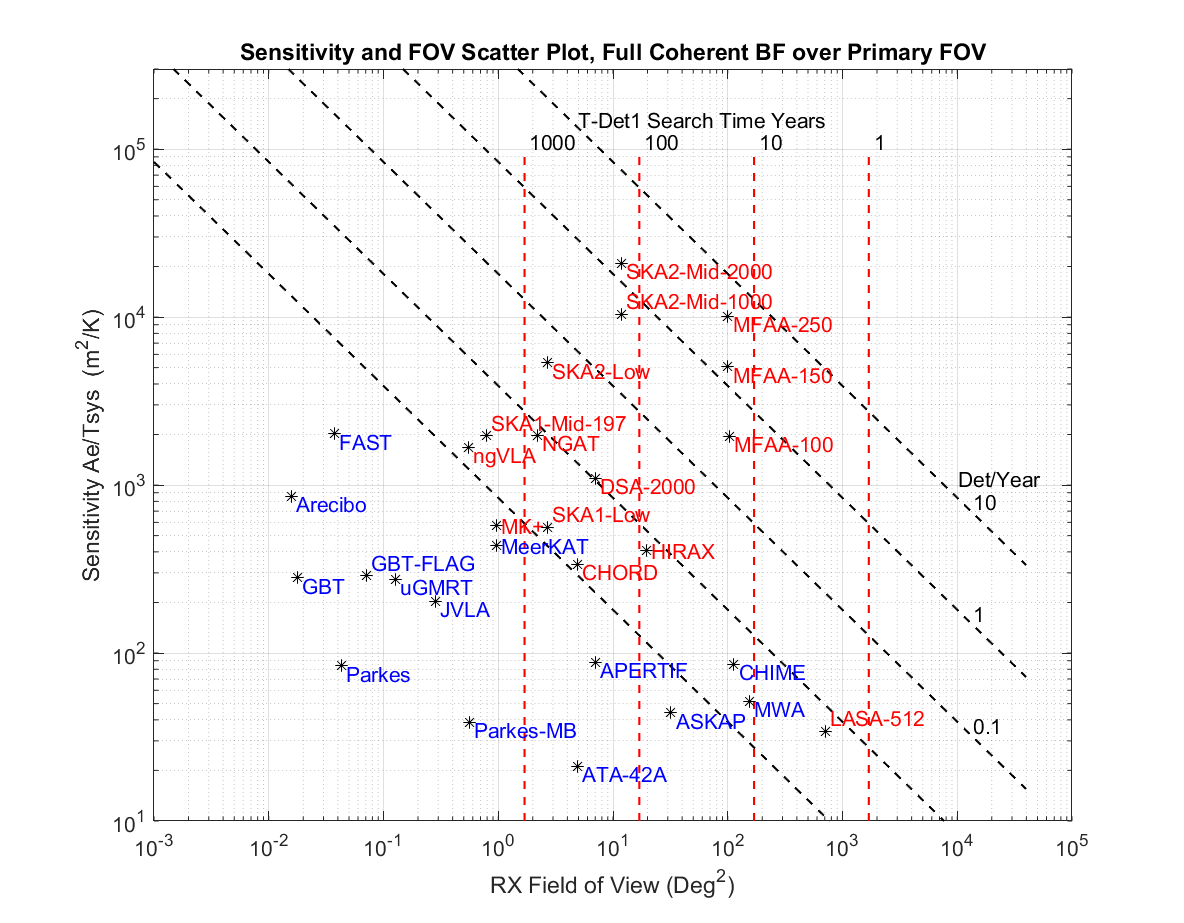}
\caption{Sensitivity and FOV with Search Time $T_{Det1}$, Full Coherent BF, 130 dBW EIRP and $P_{civTX}=10^{-6}$}
\label{fig:SensFOV_BF_TDet1}
\end{figure}

Figure \ref{fig:SensFOV_BF_TDet1} depicts Figure \ref{fig:SensFOV_BF} with lines of contant $T_{Det1}$ added. We see that $T_{Det1}$ times will be impractically long for many systems.  The systems identified in Tables  $\ref{tbl:Table1}$ and $\ref{tbl:Table2}$ would require from 1.9\,yr (LASA) to 590,000\,yr (Arecibo) to a first detection at 1.4\,GHz.   The FOV required to bring the first detection time to within 10\,yr or less may be surprisingly large.  For $T_{Det1} =10$\,yr, a 170\,deg$^2$ FOV is needed, or more than 4 times the FOV of ASKAP. For $T_{Det1} =2$\,yr, 848\,deg$^2$ is needed.  Assuming $T_{TXscan} =30$\,d, $T_{TX} =10$\,min and $\tau =5$\,min, we need RX scan times over a hemisphere on the order of 10\,h to keep $T_{Det1}<10$\,yr.  Clearly ultra-wide FOV is critical for timely results, and most candidate systems are inadequate.

The ``Lone Wolf" search time $T_{Det1}$ above only applies to fast WFS systems which scan the celestial sphere rapidly and systematically.  A slow WFS system, or SETI system based on commensal observations, will in general not be able to detect specific transmitters at arbitrary directions. However, we still should be able to achieve detections based on statistical arguments if $N_{STAR}$ is large.

\section{System Requirements} \label{sec:SystemRequirements}

The above analysis provides insight into the requirements for a practical SETI system.  First, we need to understand EIRP values in common Earth-bound practice.  Table $\ref{tbl:Table4}$ gives representative values for several types of RF transmitters.  It may be seen that EIRP values for common continuous transmitters (100$\%$ duty cycle) range from around 30\,dBW for GPS up to 62\,dBW for UHF Digital TV.  Radars have much larger EIRPs (e.g. 94\,dBW for air traffic control) but these by nature will have a low duty cycle.  For long SETI integrations the EIRP should be derated by $10\,{\rm log}_{10}$(duty-cycle) to a continuous average power, so the ATC radar with a peak EIRP of 94\,dBW at .1$\%$ duty cycle has an effective average EIRP of $94 + 10\,{\rm log}_{10}(.001) = 64$\,dBW, which is consistent with a UHF DTV station.  The former Arecibo planetary radar is a notable exception, which achieved 133\,dBW EIRP at approximately 50\% duty cycle from $P_{TX} =1$\,MW.  Note also that the Arecibo radar covered an extremely small solid angle $\Omega_{TX}$.


\begin{deluxetable*}{|c|c|c|c|c|l|}
\tablecaption{Representative EIRP Values from Terrestrial/Satellite Sources}
\tablewidth{0pt}
\label{tbl:Table4}
\tablehead{
\colhead{ } & \colhead{ }       & \colhead{ }        & \colhead{\textbf{Typical}} & \colhead{ }  & \colhead{ }  \\[-2mm]
\colhead{ } & \colhead{\textbf{EIRP}} & \colhead{\textbf{Duty}}  & \colhead{\textbf{Antenna}} & \colhead{\textbf{Nominal}} & \colhead{ } \\[-2mm]
\colhead{ } & \colhead{\textbf{dBW}} & \colhead{\textbf{Cycle}} & \colhead{\textbf{Gain dB}} & \colhead{\textbf{Frequency}} & \colhead{\textbf{Notes}}
}
\startdata
GPS & 26-34 & 100$\%$ & $13-15$ & 1575\,MHz & Actual EIRP up to 8 dB above spec \\ \hline 
FM Radio & 52 & 100$\%$ & $7-10$ & 100\,MHz & Dipole, toroidal pattern toward horizon \\ \hline 
DTV Ch 2-6 & 44 & 100$\%$ & $7-10$ & $54-88$\,MHz & $''$, VHF Low Band \\ \hline
DTV Ch 7-13 & 47 & 100$\%$ & $7-10$ & $174-216$\,MHz & $''$,  VHF High Band \\ \hline
DTV Ch 14-51 & 59-62 & 100$\%$ & $7-10$ & $470-698$\,MHz & $''$,  UHF \\ \hline
ASR-9 ATC Radar & 94 & 0.1$\%$ & 34 & $2.7-2.9$\,GHz & Air Traffic Control,  Ref US FAA \\ \hline
Intelsat Uplink & 58-87 & High & 48-58 & 6\,GHz & Earth Station, C-Band \\ \hline
Intelsat Downlink & 36-46 & High &  & 4\,GHz & Satellite, C-Band \\ \hline
Intelsat Downlink & 50-56 & High &  & 11\,GHz & Satellite, Ku-Band \\ \hline
Arecibo Radar & 125 & 6$\%$ & 62 & 430\,MHz & 305-m spherical dish \\ \hline
Arecibo Radar & 133 & $\sim50\%$ & 73 & 2380\,MHz & 305-m spherical dish \\ \hline
\enddata
\end{deluxetable*}

Table $\ref{tbl:Table5}$ shows the EIRP values resulting from continuous transmission at different input power levels and antenna gains.  Note that values in decibels are $P_{dBW} = 10\,{\rm log}_{10}$($P_{W}$).  We assume $T_{TXdwell}$ is set to 12\,min and beam widths are adjusted to cover a full sphere in 1\,d, 1 week, 1 month, and 1\,yr.  At 1\,MW power levels, similar to Arecibo, we get EIRPs of 84, 92, 99, and 109\,dBW respectively.  In reference to current technology, generating 1\,MW average RF power requires a single 2\,MW Diesel power module in a standard 40-foot shipping container.  Large power plants have capacities on the order of $1-10$\,GW, while the Three Gorges Dam (the world's largest) averaged 11.6\,GW in 2018\footnote{For example, see \url{https://en.wikipedia.org/wiki/List_of_largest_power_stations}}.  The EIRPs rise proportionally as average power rises to 10, 100 and 1000\,MW.  One can see that, based on our technology levels and desire for reasonable detection time scales, it would be very challenging to achieve even 120\,dBW EIRP, requiring approximately 125\,MW RF power to achieve a hemisphere transmit cycle time of 1 month at 2 sites.


\begin{deluxetable*}{|c|c|c|c|c|c|}
\tablecaption{Example of Trade Between EIRP and Transmit Cycle Time}
\tablewidth{0pt}
\label{tbl:Table5}
\tablehead{
\colhead{ } & \colhead{ } & \colhead{ } & \colhead{ } & \colhead{ } & \colhead{ }}
\startdata
\textbf{Hemisphere Scan Time $\rightarrow$}  & $T_{TXscan}$ & 1 Day & 1 Week & 1 Month & 1 Year \\ \hline 
\textbf{Required Dwell Cycles $\rightarrow$}  & $N_{TXscan}$ & 120	& 840 & 3600 & 43800  \\ \hline
\textbf{Antenna Gain dB $\rightarrow$} & $G_{TX}$ & 24 & 32 & 39 & 49 \\ \hline
\textbf{Average RF Power}  & \textbf{dBW}  &  \multicolumn{4}{|c|}{\textbf{EIRP dBW}}  \\ \hline 
1 MW & 60 & 84 & 92 & 99 & 109	 \\ \hline 
10 MW & 70 & 94 & 102 & 109 & 119  \\ \hline 
100 MW & 80 & 104 & 112 & 119 & 129	 \\ \hline 
1 GW & 90 & 114 & 122 & 129 & 139  \\ \hline 
\enddata
\tablecomments{Assumes TX Dwell Time =12 Minutes and 2 TX Sites}
\end{deluxetable*}


\begin{figure}[ht!]
\epsscale{1.0}
\plotone{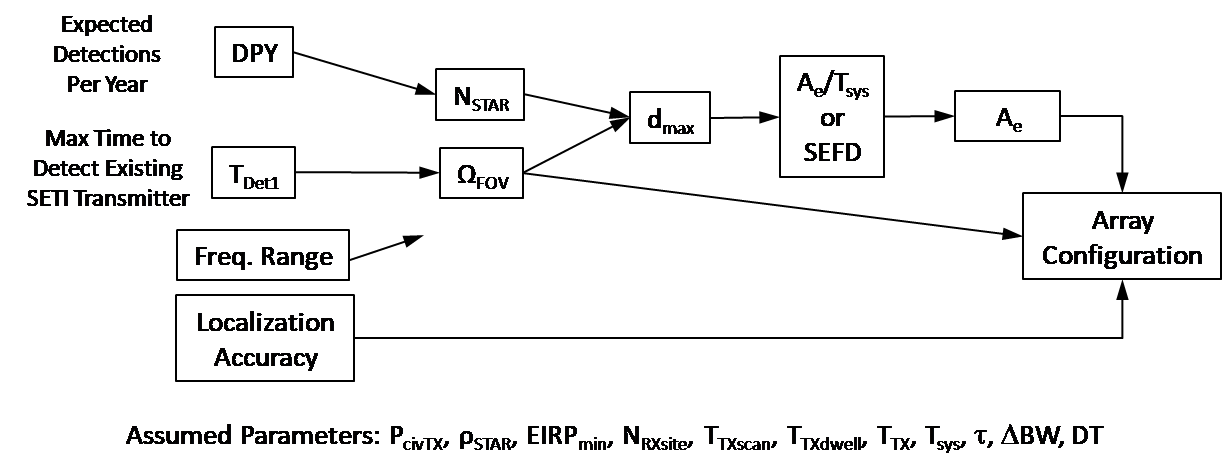}
\caption{Possible Requirements Flow-Down for a SETI Receiving System}
\label{fig:Reqts_flow}
\end{figure}

Clearly this discussion relates to our perceptions of what is easy and hard according to current technology, and may not apply to more advanced ET societies.  However, a conservative design approach will not demand EIRPs that are many orders of magnitude higher than what we may consider feasible or reasonable.  This leads to a potential list of requirements for a SETI system in Table $\ref{tbl:Table6}$.  These requirements are hypothetical and presented for discussion and further refinement.  Any SETI program should have a list of objectives like those in Table $\ref{tbl:Table6}$ to focus resources, achieve goals and understand limitations.

The requirements in Table $\ref{tbl:Table6}$ are given as primary requirements, derived or implied secondary requirements, and assumed parameters\footnote{Per the discussion of haystacks in Appendix \ref{sec:HaystackFraction}: To fully adopt the haystack concept, many of the assumed TX parameters will have ranges rather than point values. A multidimensional prior pdf of these parameters will need to be specified, and the DPY requirement will be an expected value of DPY over all transmit parameters.}.  A conceptual flow-down of primary to secondary requirements is illustrated in Figure $\ref{fig:Reqts_flow}$.  Based on a desired DPY and detection time for a given SETI transmitter, we end up with required sensitivities and then total effective area.  It is interesting to note that $T_{Det1}$ drives the field of view, but more FOV can drive down the required effective area.


\begin{deluxetable*}{|c|c|p{40mm}|c|c|c|p{50mm}|}
\tablecaption{\textbf{Hypothetical} SETI System Requirements}
\tablewidth{0pt}
\label{tbl:Table6}
\tablehead{
\colhead{ } & \colhead{\textbf{Name}}  & \colhead{\textbf{Description}} & \colhead{\textbf{Nominal}} & \colhead{\textbf{Goal}}  & \colhead{\textbf{Units}}   & \colhead{\textbf{Comments}} 
}
\startdata
 &  & \textbf{Primary Requirements} &  &  &  & \\[-3mm] \hline
1 & DPY & Expected Detections per \newline Year of Observation & 1 & 10 &  & For all frequencies and freq. rates within ranges \\ \hline 
2 & $T_{Det1}$ & Expected time to Detection given one TX present & 10 & 2 & yr & SETI TX within detection range \\ \hline
3 & $\nu_{min}-\nu_{max}$ & Frequency range & \textbf{$0.3-1.7$} & $0.1-2$ & GHz & Low UHF to water hole \\ \hline
4 & $\dot{\nu}_{norm}=\dot{\nu}/\nu$ & Frequency rate range & $\pm10$ & $\pm200$ & nHz & Fraction of bin frequency \\ \hline
5 & $\sigma_{Loc}$ & Localization Accuracy & TBD &  & arcsec & Assuming SNR$ = $DT$+3$\,dB \\ \hline
&  & \textbf{Derived Requirements} &  &  &  &  \\[-3mm] \hline
6 & $N_{STAR}$ & Number of Observable Stars within Detection Cone & 41100 & 411000 &  & \\ \hline
7 & $\Omega_{FOV}$ & RX Search Field of View & 170 & 848 & deg$^2$ & Instantaneous FOV, hemispherical reach \\ \hline
8 & $d_{MAX}$ & Maximum Detection Range & 288 & 363 & parsec & \\[-3mm] \hline
9 & $A_e/T_{sys}$ & Sensitivity  & 2740 & 4349 & m$^2/$K & \\[-3mm] \hline
10 & SEFD & System Equivalent Flux Density & 1.0 & 0.63 & Jy & \\ \hline
11 & $A_e$ & Effective Total Area & 95900 & 152000 & m$^2$ & \\[-3mm] \hline
 &  & \textbf{Assumed Parameters} &  &  &  & \\[-3mm] \hline
12 & $f_0$ & Nominal Frequency & 1.4 &  & GHz & \\[-3mm] \hline
13 & $T_{sys}$ & System Temperature & 35 & & K & At 1.4\,GHz \\ \hline
14 & $P_{civTX}$  & Probability of active TX \newline within a star system & $10^{-6}$ &  &  & Reference value \\ \hline
15 & $\rho_{STAR}$  & Average Star Density & 0.1 &  & star/pc$^3$ & \\[-3mm] \hline
16 & $EIRP_{min}$ & Minimum Detectable EIRP & 130 &  & dBW & 120 dBW desirable \\ \hline
17 & $N_{RXsite}$ & Number of Receiver Sites & 2 &  &  & Sites providing $4\pi$ overall coverage \\ \hline
18 & $T_{TXscan}$  & TX Hemisphere Scan Time  & 30 &  & Days & Time for TX to cover full sphere \newline assuming 2 transmitters \\ \hline
19 & $T_{TXdwell}$  & TX dwell time  & 12 &  & min & Time at a given TX pointing \newline direction \\ \hline
20 & $T_{TX}$  & TX transmit time  & 10 &  & min & Transmit time within TX dwell \newline period \\ \hline
21 & $\tau$  & RX total integration time  & 5 &  & min & \\[-3mm] \hline
22 & $\Delta \nu$ & RX coherent bandwidth & 3.0 &  & Hz & = $1/T_{FFT}$, $T_{FFT} = 0.33$\,s \\ \hline
23 & DT & Detection Threshold & 10 &  &  & SNR after non-coherent integration \\ \hline
\enddata
\end{deluxetable*}

Examining the requirements table, an MFAA-like system with a larger FOV but relatively modest area (equivalent to a 140-station MFAA with $D_{AP} =32$\,m) should be able to meet most of the nominal requirements.  Perhaps most surprising is the difficulty in meeting $T_{Det1}$ and FOV (items 2 and 7), where all systems beside MFAA are orders of magnitude short.  In fact, it is doubtful that any dish system could meet the FOV requirement without greatly scaled-up PAF technology.  Aperture arrays as currently envisioned become impractical above L-Band, so dishes may be the only option for higher frequencies.  SETI above L-Band would need to relax FOV and depend on large sensitivity $A_{e}/T_{sys}$ to achieve desired DPY values.

\clearpage
\section{Conclusions} \label{sec:Conclusions}

A SETI system design methodology which optimizes the expected number of ET detections per year was described.  To summarize:
\begin{itemize}
\item{Expressions were derived for the detection rate, expressed as the expected number of detections per year (DPY).  The detection rate depends critically on the number of observable stars $N_{STAR}$, and requires receivers to have both high sensitivity and field of view.  Targeted searches don't need to be treated separately, as they are a special case of slow-scan wide-field searches. The expected number of detections in an observation period $\overline{N_{Det}}$ was similarly determined, and found to be proportional to star-observations.}
\item{DPY performance was examined in terms of frequency, array and averaging metrics.  }
\begin{itemize}
\item{It was found that DPY improves with lower frequencies due to the effect of wider FOV, so that UHF frequencies ($300-1000$\,MHz) below the $1-10$\,GHz terrestrial microwave window may be more attractive for ET detections per unit observation time.  This result is opposite to the usual desire to simply maximize SNR and detection range, which favors large apertures and higher frequencies.  }
\item{It was shown that (for coherent beamforming) many small dishes are much more favorable than a single large dish with the same total area, and a single cluster is favored over Fly's Eye configurations.}
\end{itemize}
\item{A large number of radio telescope platforms that might potentially be used for SETI were compared for anticipated $N_{STAR}$ and DPY performance.  In general, interferometric receive arrays greatly outperform large single dishes due to their large FOV. }
\item{The DPY will be a function of both sensitivity and FOV.  To increase DPY it is possible to increase sensitivity, FOV, or both.  With appropriate cost functions, DPY can be optimized. }
\item{Several search figures of merit were defined based on the the number of star-observations, $N_{obs}\,N_{STAR}$. These are consistent with Drake's FOM and others, and can accommodate both wide-field and targeted searches.  Search speed FOMs were also discussed and found to be equivalent to DPY.}
\item{The most comprehensive search FOM, Wright's haystack fraction, was examined in depth in Appendix \ref{sec:HaystackFraction}, and found to have shortcomings.  A revised haystack fraction was proposed, which measures expected number of detections averaged over the ensemble of possible TX parameters.}
\item{Expressions were derived for the time to detect a specific ``Lone Wolf" SETI transmitter, and the expected time to a first detection based on a probabilistic model.  The required observation times may be surprisingly long without large FOVs.}
\item{When Equivalent Isotropic Radiated Power (EIRP) levels needed for detection at interstellar distances and realistic average power levels based on our technology are compared, we conclude that significant transmit antenna gain is required.  This in turn means that many transmissions and significant scan cycle time are needed to completely scan the sky.  SETI strategies assuming near-omnidirectional transmissions are not realistic.  To achieve detections in a reasonable time frame, receiver FOV must be a significant fraction of the sky.  }
\item{A hypothetical set of SETI system requirements was derived based on DPY and detection time requirements.  Required total effective area (nominally 96000\,m$^2$) is within the realm of existing or planned systems, but the required FOV (nominally 170\,deg$^2$) is quite high by current standards, except at low frequencies.}
\item{Strategies for maximizing detection rate with limited computation were discussed in Appendix \ref{sec:LimitedComp}.  Each frequency/drift rate bin contributes to $\overline{N_{Det}}$ in a predictable way for given FFOM and \textit{a priori} frequency/drift rate distributions.  We can optimize the set of bins to be evaluated within a computational budget. }
\end{itemize}

Approaches to improve DPY are consistent with radio astronomy trends toward ever-increasing sensitivity, ultra-wide bandwidth, and all-sky reception.  The trend toward aperture arrays at both low- and mid-frequencies is notable \citep{Garrett_2017}, and desirable because of the very large FOV of the individual array elements. If we assume for discussion that AA antennas can be designed to cover scan angles of 45$\degr$ or more from zenith \citep{Torchinsky_2017_MFAA}, this implies element coverage of at least 6000\,deg$^2$ FOV.  With greater beamforming resources, a nominal 100\,deg$^2$ FOV could be expanded to thousands of square degrees. This would greatly increase star counts and DPY, greatly reduce the detection times to practical values, and potentially allow full-time observation of certain patches of sky.  Scaling up FOV with dishes is much more difficult.

Challenges for future RF SETI systems include the following:

\begin{itemize}
\item{The signal processing requirements for both beamforming and spectral analysis have not been considered and are probably the limiting factor with current technology.  Coherent beamforming over the primary FOV is needed to achieve DPY performance, but BF processing is extremely compute-intensive, particularly for interferometric arrays.  Compromises in computation and array gain are required, and need to be investigated further.  Dense arrays (e.g. CHIME, CHORD, HIRAX or NGAT) will reduce $D_{array}$ greatly, and potentially allow FFT beamforming (e.g. \citealt{CHIME_Ng_2017}, \citealt{CHIME_Amiri_2018}, and also \citealt{VanArdenne_2009}, \citealt{BijDeVaate_2020} for AA station beamforming), which will decrease computation dramatically.  It may be more fruitful to perform commensal observations on dense arrays if they can be scaled to higher sensitivity levels. Dense arrays are a natural fit for pulsar and FRB explorations, suggesting that transient search and SETI may be a good commensal pairing.}
\item{We did not consider the effect of radio frequency interference (RFI).   A large FOV will let in more RFI, so dynamic range is critical and RFI must be dealt with.  One positive: with large interferometric arrays that provide high sensitivity and FOV, it should be possible to monitor the direction of arrival of the RFI.  It may be possible to reject bad arrival directions, or to distinguish point-source ET signals from earth-bound signals which will probably have different characteristics (more diffuse, exhibit multipath, variable correlation between dishes, low elevation angles, etc.).   One can also reject interference from satellites by tracking them across the sky.}
\item{As a minor item, star densities $\rho_{STAR}$ as a function of spatial direction need to be updated based on recent GAIA surveys.}
\end{itemize}

The DPY analysis predicts less-than-ideal detection rates for current SETI efforts on existing systems.  However, these efforts are still valuable as trailblazers, and should continue, because:
\begin{itemize}
\item{The parameters we have assumed may be too conservative (e.g. $P_{civTX}$), so true DPY levels may be higher, and we just need to search more with available systems.  }
\item{Nearby stars could have low-power transmitters operating at low duty cycles.  Long observation times are needed to assess presence or absence of transmitters beyond what has typically been done.  This may require weeks or months per targeted star.  We need to listen longer and more systematically.}
\item{We need to apply and perfect algorithms that will combine high DPY and computational efficiency for array systems.  Most SETI has been done on single-pixel dishes, which is difficult enough over a wide bandwidth.  There are many challenges to achieving large array gains on interferometric arrays at feasible levels of computation, but the DPY payoff will be large.}
\item{As indicated above, we need to improve screening against RFI through spatial approaches based on array processing, as well as single-channel methods involving machine learning and identification of signal modulations based on cyclostationarity \citep{Hellbourg_RFI_2019,Morrison_Cyclo_2018}.}
\item{Most importantly, it is unclear whether many of the proposed high-sensitivity/high-FOV systems like SKA2 or MFAA will be built any time soon.  We need to continue using the tools that are available today.}
\end{itemize}

A special call-out should be given to two existing systems with potentially high DPY values: MWA and CHIME.  Both have large intrinsic primary FOVs and compute coherent beams over a large fraction of the FOV.  They cover low frequencies (70-300 MHz for MWA and 400-800 MHz for CHIME) that have favorable FFOM values. To realize their full potential for SETI, algorithms with Hz-level resolution bandwidths would need to be applied.

Clearly it will take time for more ultra-wide FOV systems to come online.  In the near term with existing radio telescopes, one can enhance DPY by using systems with higher star counts, and observing at UHF and L-Band frequencies at higher priority than S-Band and above.  Targeted searches should aim for $N_{STAR}$ values that are much greater than unity.  One may be able to “cherry-pick” and observe patches of sky with high star concentrations (e.g. \citet{Gajjar_AJ_2021}).  Computational upgrades to current or near-future systems that provide much larger FOV (e.g. SKA1-Low and possibly MWA) should be encouraged.  Frequency range, resolution bandwidth and beamforming upgrades should also be implemented where possible on all systems.

Commensal observations on top of sky surveys should also be a reasonable strategy, as a wide-field survey program can be doubly productive.  Ethernet interconnects can allow multiple subscribers to a telescope's raw data stream with little impact to the primary user (\citet{Hickish_Commensal_SETI_JVLA_2019}). The survey will create images or detect transient events regularly, and produce SETI detections occasionally.  Of course, most SETI detections will be RFI and true ET candidate detections will be rare.  Still, if true ET detections were to occur, it would be a significant milestone for humankind, and the effort to achieve them would be relatively modest compared to space programs or other large-scale science or engineering efforts.

\section{Acknowledgments}
Breakthrough Listen is managed by the Breakthrough Initiatives, sponsored by the \href{http://breakthroughinitiatives.org}{Breakthrough Prize Foundation}.

\clearpage

\bibliographystyle{aasjournal}
\vspace{1 cm}
\bibliography{references}

\begin{thebibliography}{}
\expandafter\ifx\csname natexlab\endcsname\relax\def\natexlab#1{#1}\fi
\providecommand{\url}[1]{\href{#1}{#1}}
\providecommand{\dodoi}[1]{doi:~\href{http://doi.org/#1}{\nolinkurl{#1}}}
\providecommand{\doeprint}[1]{\href{http://ascl.net/#1}{\nolinkurl{http://ascl.net/#1}}}
\providecommand{\doarXiv}[1]{\href{https://arxiv.org/abs/#1}{\nolinkurl{https://arxiv.org/abs/#1}}}

\bibitem[{Amiri {et~al.}(2018)Amiri, Bandura, Berger, Bhardwaj, Boyce, Boyle,
  Brar, Burhanpurkar, Chawla, \& et~al.}]{CHIME_Amiri_2018}
Amiri, M., Bandura, K., Berger, P., {et~al.} 2018, The Astrophysical Journal,
  863, 48, \dodoi{10.3847/1538-4357/aad188}

\bibitem[{Barker(2017)}]{SKA_PAF_Barker_2017}
Barker, S. 2017, {SKA Phased Array Feed Consortium Update},
  \url{https://indico.skatelescope.org/event/432/attachments/3617/4781/Barker-PAF_June2017_Final.pdf},
  SKA Organisation

\bibitem[{Bij~de Vaate {et~al.}(2020)Bij~de Vaate, de~Villiers, Davidson, \&
  van Cappellen}]{BijDeVaate_2020}
Bij~de Vaate, J.~G., de~Villiers, D. I.~L., Davidson, D.~B., \& van Cappellen,
  W.~A. 2020, Experimental Astronomy: Astrophysical Instrumentation and
  Methods, 1

\bibitem[{{Braun} {et~al.}(2019){Braun}, {Bonaldi}, {Bourke}, {Keane}, \&
  {Wagg}}]{Braun_2019}
{Braun}, R., {Bonaldi}, A., {Bourke}, T., {Keane}, E., \& {Wagg}, J. 2019,
  arXiv e-prints, arXiv:1912.12699.
\newblock \doarXiv{1912.12699}

\bibitem[{Braun \& van
  Cappellen(2006)}]{Braun_2006_SKA_memo_87_Dense_Sparse_AA}
Braun, R., \& van Cappellen, W. 2006, SKA Memo 87 - Aperture Arrays for the
  SKA: Dense or Sparse?
\newblock \url{https://https://arxiv.org/abs/astro-ph/0611160}

\bibitem[{{Bunton}(2003)}]{Bunton_2003_SKA_SSFOM}
{Bunton}, J.~D. 2003, {Figure of Merit for SKA Survey Speed}.
\newblock \url{http://hdl.handle.net/102.100.100/190222?index=1}

\bibitem[{Chippendale {et~al.}(2016)Chippendale, Beresford, Deng, Leach,
  Reynolds, Kramer, \& Tzioumis}]{Chippendale_2016}
Chippendale, A.~P., Beresford, R.~J., Deng, X., {et~al.} 2016, 2016
  International Conference on Electromagnetics in Advanced Applications
  (ICEAA), \dodoi{10.1109/iceaa.2016.7731550}

\bibitem[{{Conselice} {et~al.}(2016){Conselice}, {Wilkinson}, {Duncan}, \&
  {Mortlock}}]{Conselice_2016ApJ}
{Conselice}, C.~J., {Wilkinson}, A., {Duncan}, K., \& {Mortlock}, A. 2016,
  \apj, 830, 83, \dodoi{10.3847/0004-637X/830/2/83}

\bibitem[{Cordes \& Lazio(1991)}]{Cordes_ISM_NB_Signals_1991}
Cordes, J.~M., \& Lazio, T.~J. 1991, The Astrophysical journal, 376, 123

\bibitem[{{Dewdney}(2015)}]{SKA1_Baseline_Dewdney_2015}
{Dewdney}, P. 2015, {SKA1 SYSTEM BASELINE V2 DESCRIPTION},
  \url{https://www.skatelescope.org/wp-content/uploads/2014/03/SKA-TEL-SKO-0000308_SKA1_System_Baseline_v2_DescriptionRev01-part-1-signed.pdf},
  SKA Organisation

\bibitem[{{Drake} {et~al.}(1984){Drake}, {Wolfe}, \& {Seeger}}]{Drake_1984}
{Drake}, F., {Wolfe}, J.~H., \& {Seeger}, C.~L. 1984, NASA Technical Paper 2244

\bibitem[{{Drake}(1961)}]{Drake_1961}
{Drake}, F.~D. 1961, Physics Today, 14, 40, \dodoi{10.1063/1.3057500}

\bibitem[{{Drake}(1973)}]{Drake_1973}
---. 1973, {In Communication with Extraterrestrial Intelligence (CETI), ed. C
  Sagan},  Cambridge, MA: MIT Press

\bibitem[{{Dreher}(2004)}]{Dreher_2002}
{Dreher}, J.~W. 2004, in Bioastronomy 2002: Life Among the Stars, ed.
  R.~{Norris} \& F.~{Stootman}, Vol. 213, 467

\bibitem[{{Dreher} \& {Cullers}(1997)}]{Dreher_1997}
{Dreher}, J.~W., \& {Cullers}, D.~K. 1997, in IAU Colloq. 161: Astronomical and
  Biochemical Origins and the Search for Life in the Universe, ed. C.~{Batalli
  Cosmovici}, S.~{Bowyer}, \& D.~{Werthimer}, 711

\bibitem[{{Ellingson}(2003)}]{Ellingson_EigenBF_2003}
{Ellingson}, S.~W. 2003, in IEEE Antennas and Propagation Society International
  Symposium. Digest. Held in conjunction with: USNC/CNC/URSI North American
  Radio Sci. Meeting (Cat. No.03CH37450), Vol.~4, 196--199 vol.4,
  \dodoi{10.1109/APS.2003.1220154}

\bibitem[{{Enriquez} {et~al.}(2017){Enriquez}, {Siemion}, {Foster}, {Gajjar},
  {Hellbourg}, {Hickish}, {Isaacson}, {Price}, {Croft}, {DeBoer}, {Lebofsky},
  {MacMahon}, \& {Werthimer}}]{Enriquez_2017}
{Enriquez}, J.~E., {Siemion}, A., {Foster}, G., {et~al.} 2017, \apj, 849, 104,
  \dodoi{10.3847/1538-4357/aa8d1b}

\bibitem[{{Friis}(1946)}]{FRIIS_1946}
{Friis}, H.~T. 1946, Proceedings of the IRE, 34, 254,
  \dodoi{10.1109/JRPROC.1946.234568}

\bibitem[{Gajjar(2021)}]{Gajjar_AJ_2021}
Gajjar, V. 2021, Submitted to \aj

\bibitem[{{Garrett} {et~al.}(2017){Garrett}, {Siemion}, \& {van
  Cappellen}}]{Garrett_2017}
{Garrett}, M., {Siemion}, A., \& {van Cappellen}, W. 2017, arXiv e-prints,
  arXiv:1709.01338.
\newblock \doarXiv{1709.01338}

\bibitem[{{Garrett}(2013)}]{Garrett_2013}
{Garrett}, M.~A. 2013, in 2013 Africon, 1--5,
  \dodoi{10.1109/AFRCON.2013.6757830}

\bibitem[{Gunst {et~al.}(2020)Gunst, Faulkner, Wijnholds, Jongerius,
  Torchinsky, \& van Cappellen}]{Gunst_MFAA_2020}
Gunst, A.~W., Faulkner, A.~J., Wijnholds, S., {et~al.} 2020, Mid Frequency
  Aperture Array Architectural Design Document.
\newblock \doarXiv{2008.04583}

\bibitem[{{Gupta} {et~al.}(2017){Gupta}, {Ajithkumar}, {Kale}, {Nayak},
  {Sabhapathy}, {Sureshkumar}, {Swami}, {Chengalur}, {Ghosh},
  {Ishwara-Chandra}, {Joshi}, {Kanekar}, {Lal}, \& {Roy}}]{GMRT_Gupta_2017}
{Gupta}, Y., {Ajithkumar}, B., {Kale}, H.~S., {et~al.} 2017, Current Science,
  113, 707

\bibitem[{{Hagen}(2001)}]{Arecibo_430MHz_Manual_2001}
{Hagen}, J. 2001, Arecibo 430 MHz Radar System Operation and Maintenance
  Manual, NAIC Arecibo Observatory.
\newblock
  \url{https://www.naic.edu/~phil/hardware/xmiter430/430tx_manual_hagen.pdf}

\bibitem[{Hallinan {et~al.}(2019)Hallinan, Ravi, Weinreb, Kocz, Huang, Woody,
  Lamb, D'Addario, Catha, Shi, Law, Kulkarni, Phinney, Eastwood, Bouman,
  McLaughlin, Ransom, Siemens, Cordes, Lynch, Kaplan, Chatterjee, Lazio,
  Brazier, Bhatnagar, Myers, Walter, \& Gaensler}]{Hallinan_DSA2000_2019}
Hallinan, G., Ravi, V., Weinreb, S., {et~al.} 2019, The DSA-2000 -- A Radio
  Survey Camera.
\newblock \doarXiv{1907.07648}

\bibitem[{Harris \& Haines(2011)}]{Harris_Haines_PFB_2011}
Harris, C., \& Haines, K. 2011, Publications of the Astronomical Society of
  Australia, 28, 317–322, \dodoi{10.1071/AS11032}

\bibitem[{{Hellbourg} \& {Morrison}(2019)}]{Hellbourg_RFI_2019}
{Hellbourg}, G., \& {Morrison}, I. 2019, in 2019 RFI Workshop - Coexisting with
  Radio Frequency Interference (RFI), 1--5,
  \dodoi{10.23919/RFI48793.2019.9111786}

\bibitem[{Hickish {et~al.}(2019)Hickish, Beasley, Bower, Burke-Spolaor, Croft,
  DeBoer, Demorest, Diamond, Gajjar, Law, Lazio, Manley, Paragi, Ransom, \&
  Siemion}]{Hickish_Commensal_SETI_JVLA_2019}
Hickish, J., Beasley, T., Bower, G., {et~al.} 2019, Commensal, Multi-user
  Observations with an Ethernet-based Jansky Very Large Array.
\newblock \doarXiv{1907.05263}

\bibitem[{Hobbs {et~al.}(2020)Hobbs, Manchester, Dunning, Jameson, Roberts,
  George, Green, Tuthill, Toomey, Kaczmarek, \& et~al.}]{Hobbs_2020}
Hobbs, G., Manchester, R.~N., Dunning, A., {et~al.} 2020, Publications of the
  Astronomical Society of Australia, 37, \dodoi{10.1017/pasa.2020.2}

\bibitem[{{Houston}(2021)}]{Houston_SETI_Detect_URSI_2021}
{Houston}, K. 2021, A Novel Detector for SETI on Radio Telescope Arrays, URSI
  GASS 2021

\bibitem[{{Hurter} \& {Kotzé}(2020)}]{MeerKAT_Ext_2020_SOW}
{Hurter}, P.~H., \& {Kotzé}, J. 2020, {MEERKAT EXTENSION INFRASTRUCTURE SCOPE
  OF WORK},
  \url{https://www.sarao.ac.za/wp-content/uploads/2020/02/SSA4003-0007-004_RevF_MeerKATExtInfraSOW.pdf},
  South African Radio Astronomy Observatory

\bibitem[{Intelsat(2007)}]{INTELSAT_16QAM_2005}
Intelsat. 2007, Performance Characteristics for Digital Carriers Using 16QAM
  Modulation, Tech. rep., Intelsat.
\newblock
  \url{https://www.intelsat.com/wp-content/uploads/2020/08/iess-316e.pdf}

\bibitem[{Jonas(2018)}]{MeerKAT_Jonas_2016}
Jonas, J. 2018, in Proceedings of MeerKAT Science: On the Pathway to the SKA
  {\textemdash} PoS(MeerKAT2016), Vol. 277, 001, \dodoi{10.22323/1.277.0001}

\bibitem[{{Kerins}(2021)}]{Kerins_2021}
{Kerins}, E. 2021, \aj, 161, 39, \dodoi{10.3847/1538-3881/abcc5f}

\bibitem[{{Lehmensiek} \& {Theron}(2014)}]{MeerKAT_Lehmensiek_2014}
{Lehmensiek}, R., \& {Theron}, I.~P. 2014, in 2014 XXXIth URSI General Assembly
  and Scientific Symposium (URSI GASS), 1--4,
  \dodoi{10.1109/URSIGASS.2014.6930057}

\bibitem[{Li {et~al.}(2020)Li, Gajjar, Wang, Siemion, Zhang, Zhang, Yue, Zhu,
  Jin, Li, Berger, Brzycki, Cobb, Croft, Czech, DeBoer, DeMarines, Drew,
  Enriquez, Gizani, Korpela, Isaacson, Lebofsky, Lacki, MacMahon, Nanez, Niu,
  Pei, Price, Werthimer, Worden, Zhang, Zhang, \& Collaboration}]{Li_2020}
Li, D., Gajjar, V., Wang, P., {et~al.} 2020, Opportunities to Search for
  Extra-Terrestrial Intelligence with the Five-hundred-meter Aperture Spherical
  radio Telescope.
\newblock \doarXiv{2003.09639}

\bibitem[{{Lynch} {et~al.}(2018){Lynch}, {Lorimer}, {Ellingson}, {Bandura}, \&
  {McLaughlin}}]{Lynch_2018}
{Lynch}, R., {Lorimer}, D., {Ellingson}, S., {Bandura}, K., \& {McLaughlin}, M.
  2018, in {Towards an All-Sky Radio Telescope for SETI}.
\newblock
  \url{http://www.jodrellbank.manchester.ac.uk/media/eps/jodrell-bank-centre-for-astrophysics/news-and-events/2018/wide-field-seti/whitepapers/RLynch_white_paper.pdf}

\bibitem[{{Macquart}(2011)}]{Macquart_2011}
{Macquart}, J.-P. 2011, The Astrophysical Journal, 734, 20,
  \dodoi{10.1088/0004-637x/734/1/20}

\bibitem[{{Macquart}(2014)}]{Macquart_2014}
---. 2014, \pasa, 31, e031, \dodoi{10.1017/pasa.2014.27}

\bibitem[{McConnell {et~al.}(2020)McConnell, Hale, Lenc, Banfield, Heald,
  Hotan, Leung, Moss, Murphy, O'Brien, Pritchard, Raja, Sadler, Stewart,
  Thomson, Whiting, Allison, Amy, Anderson, Ball, Bannister, Bell, Bock,
  Bolton, Bunton, Chippendale, Collier, Cooray, Cornwell, Diamond, Edwards,
  Gupta, Hayman, Heywood, Jackson, Koribalski, Lee-Waddell, McClure-Griffiths,
  Ng, Norris, Phillips, Reynolds, Roxby, Schinckel, Shields, Tremblay,
  Tzioumis, Voronkov, \& Westmeier}]{ASKAP_McConnell_2020_Survey}
McConnell, D., Hale, C.~L., Lenc, E., {et~al.} 2020, PUBLICATIONS OF THE
  ASTRONOMICAL SOCIETY OF AUSTRALIA, 37.
\newblock
  \url{https://libproxy.berkeley.edu/login?qurl=https%3a%2f%2fsearch.ebscohost.com%2flogin.aspx%3fdirect%3dtrue%26db%3dedswsc%26AN%3d000594414600001%26site%3deds-live}

\bibitem[{McPherson {et~al.}(2018)McPherson, McMullin, Stevenson, Dewdney,
  Casson, Stringhetti, Deegan, Hekman, Austin, Harman, Gibbs, Es, Labate,
  Swart, Ciaizzo, Hayden, \& Turner}]{McPherson_SKA1_2018}
McPherson, A.~M., McMullin, J., Stevenson, T., {et~al.} 2018, in Ground-based
  and Airborne Telescopes VII, ed. H.~K. Marshall \& J.~Spyromilio, Vol. 10700,
  International Society for Optics and Photonics (SPIE), 281 -- 311,
  \dodoi{10.1117/12.2312438}

\bibitem[{{Messerschmitt}(2013)}]{Messerschmitt_SETI_Comms_PwrEff_2013}
{Messerschmitt}, D. 2013, End-to-end interstellar communication system design
  for power efficiency, Tech. rep., University of California at Berkeley.
\newblock \url{https://arxiv.org/abs/1305.4684v2}

\bibitem[{Messerschmitt(2012)}]{Messerschmitt_Spread_Spectrum_2012}
Messerschmitt, D.~G. 2012, Acta Astronautica, 81, 227–238,
  \dodoi{10.1016/j.actaastro.2012.07.024}

\bibitem[{Messerschmitt(2015)}]{Messerschmitt_SS_SETI_Comms_2015}
---. 2015, Acta Astronautica, 107, 20–39,
  \dodoi{10.1016/j.actaastro.2014.11.007}

\bibitem[{{Morrison}(2018)}]{Morrison_Cyclo_2018}
{Morrison}, I. 2018, in {Towards an All-Sky Radio Telescope for SETI}.
\newblock
  \url{http://www.jodrellbank.manchester.ac.uk/media/eps/jodrell-bank-centre-for-astrophysics/news-and-events/2018/wide-field-seti/presentations/MORRISON---PRESENTATION.pdf}

\bibitem[{Nan {et~al.}(2011)Nan, Li, Jin, Wang, Zhu, Zhu, Zhang, Yue, \&
  Qian}]{NAN_FAST_2011}
Nan, R., Li, D., Jin, C., {et~al.} 2011, International Journal of Modern
  Physics D, 20, 989–1024, \dodoi{10.1142/s0218271811019335}

\bibitem[{Newburgh {et~al.}(2016)Newburgh, Bandura, Bucher, Chang, Chiang,
  Cliche, Davé, Dobbs, Clarkson, Ganga, Gogo, Gumba, Gupta, Hilton, Johnstone,
  Karastergiou, Kunz, Lokhorst, Maartens, Macpherson, Mdlalose, Moodley,
  Ngwenya, Parra, Peterson, Recnik, Saliwanchik, Santos, Sievers, Smirnov,
  Stronkhorst, Taylor, Vanderlinde, Vuuren, Weltman, \&
  Witzemann}]{Newburgh_HIRAX_2016}
Newburgh, L.~B., Bandura, K., Bucher, M.~A., {et~al.} 2016, in Ground-based and
  Airborne Telescopes VI, ed. H.~J. Hall, R.~Gilmozzi, \& H.~K. Marshall, Vol.
  9906, International Society for Optics and Photonics (SPIE), 2039 -- 2049,
  \dodoi{10.1117/12.2234286}

\bibitem[{{Ng} {et~al.}(2017){Ng}, {Vanderlinde}, {Paradise}, {Klages},
  {Masui}, {Smith}, {Bandura}, {Boyle}, {Dobbs}, {Kaspi}, {Renard}, {Shaw},
  {Stairs}, \& {Tretyakov}}]{CHIME_Ng_2017}
{Ng}, C., {Vanderlinde}, K., {Paradise}, A., {et~al.} 2017, in 2017 XXXIInd
  General Assembly and Scientific Symposium of the International Union of Radio
  Science (URSI GASS), 1--4, \dodoi{10.23919/URSIGASS.2017.8105318}

\bibitem[{Oosterloo {et~al.}(2010)Oosterloo, Verheijen, \& van
  Cappellen}]{Apertif_2010_Oosterloo}
Oosterloo, T., Verheijen, M., \& van Cappellen, W. 2010, The latest on Apertif.
\newblock \doarXiv{1007.5141}

\bibitem[{{Pingel} \& {Pisano}(2018)}]{Pingel_GBT_FLAG_2018}
{Pingel}, N.~M., \& {Pisano}, D.~J. 2018, in 2018 2nd URSI Atlantic Radio
  Science Meeting (AT-RASC), 1--1, \dodoi{10.23919/URSI-AT-RASC.2018.8471298}

\bibitem[{Price {et~al.}(2018)Price, MacMahon, Lebofsky, Croft, DeBoer,
  Enriquez, Foster, Gajjar, Gizani, Hellbourg, \& et~al.}]{Price_2018}
Price, D.~C., MacMahon, D. H.~E., Lebofsky, M., {et~al.} 2018, Publications of
  the Astronomical Society of Australia, 35, e041, \dodoi{10.1017/pasa.2018.36}

\bibitem[{{Price} {et~al.}(2020){Price}, {Enriquez}, {Brzycki}, {Croft},
  {Czech}, {DeBoer}, {DeMarines}, {Foster}, {Gajjar}, {Gizani}, {Hellbourg},
  {Isaacson}, {Lacki}, {Lebofsky}, {MacMahon}, {Pater}, {Siemion}, {Werthimer},
  {Green}, {Kaczmarek}, {Maddalena}, {Mader}, {Drew}, \& {Worden}}]{Price_2020}
{Price}, D.~C., {Enriquez}, J.~E., {Brzycki}, B., {et~al.} 2020, \aj, 159, 86,
  \dodoi{10.3847/1538-3881/ab65f1}

\bibitem[{Rajwade {et~al.}(2017)Rajwade, Pingel, Black, Ruzindana, Burnett,
  Jeffs, Warnick, Pisano, Lorimer, Prestage, \& et~al.}]{GBT_FLAG_Rajwade_2017}
Rajwade, K.~M., Pingel, N.~M., Black, R.~A., {et~al.} 2017, Proceedings of the
  International Astronomical Union, 13, 398–399,
  \dodoi{10.1017/s1743921317009012}

\bibitem[{Roshi(2021)}]{NGAT_White_Paper_2021}
Roshi, D.~A. 2021, The Future of the Arecibo Observatory: The Next Generation
  Arecibo Telescope White Paper, v2.
\newblock \url{http://www.naic.edu/NGAT/NGAT_WhitePaper_v2_01022021.pdf}

\bibitem[{Selina(2019)}]{ngVLA_Sys_Design_Selina_2019}
Selina, R.~J. 2019, Next Generation VLA System Reference Design: Volume 1
  System Design.
\newblock
  \url{https://ngvla.nrao.edu/system/media_files/binaries/241/original/Volume_1.pdf?1568298579}

\bibitem[{Sheikh {et~al.}(2020)Sheikh, Siemion, Enriquez, Price, Isaacson,
  Lebofsky, Gajjar, \& Kalas}]{Sheikh_2020}
Sheikh, S.~Z., Siemion, A., Enriquez, J.~E., {et~al.} 2020, The Astronomical
  Journal, 160, 29, \dodoi{10.3847/1538-3881/ab9361}

\bibitem[{{Sheikh} {et~al.}(2019){Sheikh}, {Wright}, {Siemion}, \&
  {Enriquez}}]{Sheikh_2019}
{Sheikh}, S.~Z., {Wright}, J.~T., {Siemion}, A., \& {Enriquez}, J.~E. 2019,
  \apj, 884, 14, \dodoi{10.3847/1538-4357/ab3fa8}

\bibitem[{Shostak(2000)}]{Shostak_2000}
Shostak, S. 2000, Acta Astronautica, 46, 649 ,
  \dodoi{https://doi.org/10.1016/S0094-5765(00)00027-8}

\bibitem[{{Siegel}(2018)}]{Siegel_2018}
{Siegel}, E. 2018, {The Drake Equation Is Broken; Here's How To Fix It},
  Forbes Magazine

\bibitem[{{Siemion} {et~al.}(2015){Siemion}, {Benford}, {Cheng-Jin},
  {Chennamangalam}, {Cordes}, {Falcke}, {Garrington}, {Garrett}, {Gurvits},
  {Hoare}, {Korpela}, {Lazio}, {Messerschmitt}, {Morrison}, {O'Brien},
  {Paragi}, {Penny}, {Spitler}, {Tarter}, \& {Werthimer}}]{Siemion_2015}
{Siemion}, A., {Benford}, J., {Cheng-Jin}, J., {et~al.} 2015, in Advancing
  Astrophysics with the Square Kilometre Array (AASKA14), 116.
\newblock \doarXiv{1412.4867}

\bibitem[{{Siemion} {et~al.}(2011){Siemion}, {Werthimer}, {Anderson}, {Chen},
  {Cobb}, {Cordes}, {Filiba}, {Foster}, {Gowda}, {Korpela}, {Lebofsky},
  {Little}, {Mallard}, {Spitler}, \& {Wagner}}]{Siemion_SETI_Developments_2011}
{Siemion}, A. P.~V., {Werthimer}, D., {Anderson}, D., {et~al.} 2011, in 2011
  XXXth URSI General Assembly and Scientific Symposium, 1--4,
  \dodoi{10.1109/URSIGASS.2011.6051263}

\bibitem[{{Siemion} {et~al.}(2012){Siemion}, {Bower}, {Foster}, {McMahon},
  {Wagner}, {Werthimer}, {Backer}, {Cordes}, \& {van
  Leeuwen}}]{Siemion_2012ApJ}
{Siemion}, A. P.~V., {Bower}, G.~C., {Foster}, G., {et~al.} 2012, \apj, 744,
  109, \dodoi{10.1088/0004-637X/744/2/109}

\bibitem[{Siemion {et~al.}(2013)Siemion, DEMOREST, KORPELA, MADDALENA,
  WERTHIMER, COBB, HOWARD, LANGSTON, LEBOFSKY, MARCY, \&
  TARTER}]{Siemion_2013_SETI_Kepler_TurboSETI}
Siemion, A. P.~V., DEMOREST, P., KORPELA, E., {et~al.} 2013, Astrophysical
  Journal, 767, 1

\bibitem[{SKAO(2016)}]{SKA1_Survey_Perf_2016}
SKAO. 2016, {SURVEY PERFORMANCE CALCULATIONS},
  \url{http://skacontinuum.pbworks.com/w/file/fetch/119268513/SKA-TEL-SKO-01-0000645-SurveyPerformanceCalculations-part-1-signed.pdf},
  SKA Organisation

\bibitem[{Staveley-Smith {et~al.}(1996)Staveley-Smith, Wilson, Bird, Disney,
  Ekers, Freeman, Haynes, Sinclair, Vaile, Webster, \&
  et~al.}]{Staveley_Smith_1996}
Staveley-Smith, L., Wilson, W.~E., Bird, T.~S., {et~al.} 1996, Publications of
  the Astronomical Society of Australia, 13, 243–248,
  \dodoi{10.1017/S1323358000020919}

\bibitem[{{Steigenberger} {et~al.}(2019){Steigenberger}, {Thoelert}, \&
  {Montenbruck}}]{Steigenberger_GPS_EIRP_2019}
{Steigenberger}, P., {Thoelert}, S., \& {Montenbruck}, O. 2019, GPS and GLONASS
  Satellite Transmit Power: Update for IGS repro3, Tech. rep., German Space
  Operations Center.
\newblock \url{http://acc.igs.org/repro3/TX_Power_20190711.pdf}

\bibitem[{Tarter(2001{\natexlab{a}})}]{Tarter_SETI_2001}
Tarter, J. 2001{\natexlab{a}}, Annual Review of Astronomy \& Astrophysics, 39,
  511, \dodoi{10.1146/annurev.astro.39.1.511}

\bibitem[{Tarter(2001{\natexlab{b}})}]{Tarter_SETI_SPIE_2001}
Tarter, J.~C. 2001{\natexlab{b}}, in The Search for Extraterrestrial
  Intelligence (SETI) in the Optical Spectrum III, ed. S.~A. Kingsley \&
  R.~Bhathal, Vol. 4273, International Society for Optics and Photonics (SPIE),
  93 -- 103, \dodoi{10.1117/12.435361}

\bibitem[{{Taylor} {et~al.}(2016){Taylor}, {Nolan}, {Rivera-Valentın},
  {Richardson}, {Rodriguez-Ford}, {Zambrano-Marin}, {Howell}, \&
  {Schmelz}}]{Taylor_Arecibo_Radar_2016}
{Taylor}, P.~A., {Nolan}, M.~C., {Rivera-Valentın}, E.~G., {et~al.} 2016, in
  47th Lunar and Planetary Science Conference (2016), Universities Space
  Research Association.
\newblock \url{https://www.hou.usra.edu/meetings/lpsc2016/pdf/2534.pdf}

\bibitem[{Tingay {et~al.}(2016)Tingay, Tremblay, Walsh, \&
  Urquhart}]{Tingay_2016}
Tingay, S.~J., Tremblay, C., Walsh, A., \& Urquhart, R. 2016, The Astrophysical
  Journal, 827, L22, \dodoi{10.3847/2041-8205/827/2/l22}

\bibitem[{Tingay {et~al.}(2018{\natexlab{a}})Tingay, Tremblay, \&
  Croft}]{Tingay_2018_Galactic_Anticenter}
Tingay, S.~J., Tremblay, C.~D., \& Croft, S. 2018{\natexlab{a}}, The
  Astrophysical Journal, 856, 31, \dodoi{10.3847/1538-4357/aab363}

\bibitem[{{Tingay} {et~al.}(2013){Tingay}, {Goeke}, {Bowman}, {Emrich}, {Ord},
  {Mitchell}, {Morales}, {Booler}, {Crosse}, {Wayth}, {Lonsdale}, {Tremblay},
  {Pallot}, {Colegate}, {Wicenec}, {Kudryavtseva}, {Arcus}, {Barnes},
  {Bernardi}, {Briggs}, {Burns}, {Bunton}, {Cappallo}, {Corey}, {Deshpande},
  {Desouza}, {Gaensler}, {Greenhill}, {Hall}, {Hazelton}, {Herne}, {Hewitt},
  {Johnston-Hollitt}, {Kaplan}, {Kasper}, {Kincaid}, {Koenig}, {Kratzenberg},
  {Lynch}, {Mckinley}, {Mcwhirter}, {Morgan}, {Oberoi}, {Pathikulangara},
  {Prabu}, {Remillard}, {Rogers}, {Roshi}, {Salah}, {Sault}, {Udaya-Shankar},
  {Schlagenhaufer}, {Srivani}, {Stevens}, {Subrahmanyan}, {Waterson},
  {Webster}, {Whitney}, {Williams}, {Williams}, \&
  {Wyithe}}]{MWA_Tinguay_2013_PASA}
{Tingay}, S.~J., {Goeke}, R., {Bowman}, J.~D., {et~al.} 2013, \pasa, 30, e007,
  \dodoi{10.1017/pasa.2012.007}

\bibitem[{Tingay {et~al.}(2018{\natexlab{b}})Tingay, Kaplan, Lenc, Croft,
  McKinley, Beardsley, Crosse, Emrich, Franzen, Gaensler, Horsley,
  Johnston-Hollitt, Kenney, Morales, Pallot, Steele, Trott, Walker, Wayth,
  Williams, \& Wu}]{Tingay_2018_Oumuamua}
Tingay, S.~J., Kaplan, D.~L., Lenc, E., {et~al.} 2018{\natexlab{b}}, The
  Astrophysical Journal, 857, 11, \dodoi{10.3847/1538-4357/aab359}

\bibitem[{Torchinsky {et~al.}(2017)Torchinsky, Broderick, Gunst, Faulkner, \&
  van Cappellen}]{Torchinsky_2017_MFAA}
Torchinsky, S.~A., Broderick, J.~W., Gunst, A., Faulkner, A.~J., \& van
  Cappellen, W. 2017, SKA Aperture Array Mid Frequency Science Requirements.
\newblock \doarXiv{1610.00683}

\bibitem[{Tremblay \& Tingay(2020)}]{MWA_Tremblay_Vela_2020}
Tremblay, C.~D., \& Tingay, S.~J. 2020, Publications of the Astronomical
  Society of Australia, 37, e035, \dodoi{10.1017/pasa.2020.27}

\bibitem[{{van Ardenne} {et~al.}(2009){van Ardenne}, {Bregman}, {van
  Cappellen}, {Kant}, \& {Bij de Vaate}}]{VanArdenne_2009}
{van Ardenne}, A., {Bregman}, J.~D., {van Cappellen}, W.~A., {Kant}, G.~W., \&
  {Bij de Vaate}, J.~G. 2009, Proceedings of the IEEE, 97, 1531,
  \dodoi{10.1109/JPROC.2009.2021594}

\bibitem[{Vanderlinde {et~al.}(2019)Vanderlinde, Liu, Gaensler, Bond, Hinshaw,
  Ng, Chiang, Stairs, Brown, Sievers, Mena, Smith, Bandura, Masui, Spekkens,
  Belostotski, Dobbs, Turok, Boyle, Rupen, Landecker, Pen, \&
  Kaspi}]{CHORD_vanderlinde_2019}
Vanderlinde, K., Liu, A., Gaensler, B., {et~al.} 2019, {The Canadian Hydrogen
  Observatory and Radio- transient Detector (CHORD)},  Zenodo,
  \dodoi{10.5281/zenodo.3765414}

\bibitem[{{Wang} {et~al.}(2018){Wang}, {Ruf}, {Block}, {McKague}, \&
  {Gleason}}]{Wang_GPS_EIRP_ION_2018}
{Wang}, T., {Ruf}, C., {Block}, B., {McKague}, D., \& {Gleason}, S. 2018, in
  Proceedings of the 31st International Technical Meeting of the Satellite
  Division of The Institute of Navigation (ION GNSS+ 2018), 2879--2890,
  \dodoi{10.33012/2018.16101}

\bibitem[{{Warnick} {et~al.}(2016){Warnick}, {Maaskant}, {Ivashina},
  {Davidson}, \& {Jeffs}}]{Warnick_2016}
{Warnick}, K.~F., {Maaskant}, R., {Ivashina}, M.~V., {Davidson}, D.~B., \&
  {Jeffs}, B.~D. 2016, Proceedings of the IEEE, 104, 607,
  \dodoi{10.1109/JPROC.2015.2491886}

\bibitem[{{Wayth} {et~al.}(2018){Wayth}, {Tingay}, {Trott}, {Emrich},
  {Johnston-Hollitt}, {McKinley}, {Gaensler}, {Beardsley}, {Booler}, {Crosse},
  {Franzen}, {Horsley}, {Kaplan}, {Kenney}, {Morales}, {Pallot}, {Sleap},
  {Steele}, {Walker}, {Williams}, {Wu}, {Cairns}, {Filipovic}, {Johnston},
  {Murphy}, {Quinn}, {Staveley-Smith}, {Webster}, \&
  {Wyithe}}]{MWA2_Wayth_2018}
{Wayth}, R.~B., {Tingay}, S.~J., {Trott}, C.~M., {et~al.} 2018, \pasa, 35, 33,
  \dodoi{10.1017/pasa.2018.37}

\bibitem[{{Weber}(2000)}]{Weber_MIT_LL_Radar_2000}
{Weber}, M. 2000, in 9th Conference on Aviation, Range, and Aerospace
  Meteorology (ARAM), American Meteorological Society.
\newblock
  \url{https://archive.ll.mit.edu/mission/aviation/publications/publication-files/ms-papers/Weber_2000_ARAM_MS-14191_WW-10147.pdf}

\bibitem[{{Welch} {et~al.}(2009){Welch}, {Backer}, {Blitz}, {Bock}, {Bower},
  {Cheng}, {Croft}, {Dexter}, {Engargiola}, {Fields}, {Forster},
  {Gutierrez-Kraybill}, {Heiles}, {Helfer}, {Jorgensen}, {Keating}, {Lugten},
  {MacMahon}, {Milgrome}, {Thornton}, {Urry}, {van Leeuwen}, {Werthimer},
  {Williams}, {Wright}, {Tarter}, {Ackermann}, {Atkinson}, {Backus}, {Barott},
  {Bradford}, {Davis}, {DeBoer}, {Dreher}, {Harp}, {Jordan}, {Kilsdonk},
  {Pierson}, {Randall}, {Ross}, {Shostak}, {Fleming}, {Cork}, {Vitouchkine},
  {Wadefalk}, \& {Weinreb}}]{ATA_Welch_2009_IEEEProc}
{Welch}, J., {Backer}, D., {Blitz}, L., {et~al.} 2009, Proceedings of the IEEE,
  97, 1438, \dodoi{10.1109/JPROC.2009.2017103}

\bibitem[{Wlodarczyk-Sroka {et~al.}(2020)Wlodarczyk-Sroka, Garrett, \&
  Siemion}]{Wlodarczyk_Sroka_2020}
Wlodarczyk-Sroka, B.~S., Garrett, M.~A., \& Siemion, A. P.~V. 2020, Monthly
  Notices of the Royal Astronomical Society, 498, 5720–5729,
  \dodoi{10.1093/mnras/staa2672}

\bibitem[{{Wright} {et~al.}(2018{\natexlab{a}}){Wright}, {Gelino}, \&
  {Participants}}]{Wright_2018_NASA}
{Wright}, J.~T., {Gelino}, D., \& {Participants}. 2018{\natexlab{a}}, arXiv
  e-prints, arXiv:1812.08681.
\newblock \doarXiv{1812.08681}

\bibitem[{{Wright} {et~al.}(2018{\natexlab{b}}){Wright}, {Kanodia}, \&
  {Lubar}}]{Wright_2018AJ}
{Wright}, J.~T., {Kanodia}, S., \& {Lubar}, E. 2018{\natexlab{b}}, \aj, 156,
  260, \dodoi{10.3847/1538-3881/aae099}

\bibitem[{{Zarb-Adami} {et~al.}(2010){Zarb-Adami}, {Faulkner}, {de Vaate},
  {Kant}, \& {Picard}}]{Zarb-Adami_2010}
{Zarb-Adami}, K., {Faulkner}, A., {de Vaate}, J. G.~B., {Kant}, G.~W., \&
  {Picard}, P. 2010, in 2010 IEEE International Symposium on Phased Array
  Systems and Technology, 883--890, \dodoi{10.1109/ARRAY.2010.5613258}

\end{thebibliography}

\clearpage
\appendix
\vspace{-3mm}

\section{The SETI Haystack Fraction as a Search Metric} \label{sec:HaystackFraction}

\citet{Wright_2018AJ} developed an ET search metric which is a function of nine search dimensions: sensitivity, central frequency, distance and angular position (3 spatial dimensions), TX bandwidth, repetition rate, polarization, and modulation.  By analogy to searching for a ``Needle'' in the ``Cosmic Haystack'', the ``Haystack Fraction'' as defined below was evaluated for various past SETI efforts.  The survey\footnote{The term ``scan'' is used in the paper, but we use ``survey'' here to avoid confusion with raster scans over a set of pointing directions for TX or RX.  A survey may have many scans.  Wright's repetition time $T_{rep}$ will generally correspond to $T_{TXscan}$ in our model.} and total haystack volumes are integrations of sensitivity over all search dimensions, and the haystack fraction $\Phi_{H1}$ is their ratio, as follows: 

\begin{equation}  \label{eq:eqn101}
V_{H1\!-\!survey}(EIRP_{MIN}) = \int_{\Gamma_{HV}} S(d;EIRP_{MIN},\phi_{MIN}) \; d\Gamma_{HV}
\end{equation}
\begin{equation}  \label{eq:eqn102}
V_{H1\!-\!total}(EIRP_{MIN}) = \int_{\Gamma_{HV}} S_{MAX}(EIRP_{MIN}) \; d\Gamma_{HV}
\end{equation}
\begin{equation}  \label{eq:eqn103}
\Phi_{H1} = V_{H1\!-\!survey}/V_{H1\!-\!total}
\end{equation}

\noindent where $\Gamma_{HV}$ is the full haystack integration volume over all search dimensions.  The sensitivity functions over range d ($0 < d < d_{H\!-\!MAX}$) are defined as

\begin{equation}  \label{eq:eqn104}
S(d;EIRP_{MIN}) = min(4\pi d^2/EIRP_{MIN},1/\phi_{MIN})
\end{equation}
\begin{equation}  \label{eq:eqn105}
S_{MAX}(d;EIRP_{MIN}) = 4\pi d^2/EIRP_{MIN}
\end{equation}

\vspace{0mm}

\noindent  The maximum range of interest is $d_{H\!-\!MAX}$\footnote{Note this is different from detection range $d_{MAX}$ described earlier in (\ref{eq:eqn9}), which is designated as critical range $d_{crit}$ by Wright.} (e.g. 10 kpc), $EIRP_{MIN}$ is a reference EIRP (e.g. $10^{13}\,$W; the range of EIRPs included is $EIRP_{MIN}$ to $\infty$), and $\phi_{MIN}$ is the minimum detectable flux level (assumed constant, but could be a function of frequency and TX waveform parameters).  $S(d;EIRP_{MIN})$ starts at 0 and follows a parabola to a detection range $d_{crit}$ and then is constant at a value of $1/\phi_{MIN}$ out to $d_{H\!-\!MAX}$.  $S_{MAX}(d;EIRP_{MIN})$ follows the same parabola all the way to $d_{H\!-\!MAX}$.  If a survey (search effort) covers the full range of all search dimensions, the haystack fraction will go to 1 as $\phi_{MIN} \rightarrow 0$.  The reader is referred to the paper for more details.

On the plus side, Wright's haystack volume provides a unified, quantitative means of comparing searches with widely differing characteristics.  It demonstrates that receivers need to offer high sensitivity and wide FOV, and detection algorithms must accommodate a wide range of TX characteristics (and not just offer point solutions).  The Haystack fraction numbers suggest that we have only just begun to cover the search space.

On the minus side, the haystack volumes as defined may be hard to interpret or assess as a search metric. Some issues:

\begin{itemize}
\item{Bin values\footnote{Let us call an ``integration bin'' an N-D voxel of sufficiently small dimensions. The ``bin value'' equals the integrand at the voxel center. The bin values are summed in N dimensions to obtain the integral.} in the integration have no indication of relative likelihood, so much of the volume contributing to the sum could have zero or very low probability.  Consider:
\begin{itemize}
\item{The survey and total haystack volumes as defined are heavily weighted toward high frequencies. A key question: would the likelihood of an ET transmission between 0-1 GHz be the same as over 101-102 GHz, or should these bands be weighted differently? }
\item{The volumes are heavily weighted toward wide TX bandwidths. Would ET's choice of a transmission bandwidth between 10 to 20 MHz be 10 million times more likely than of a bandwidth of 1 Hz? }
\item{Regarding repetition rate, based on energy arguments, continuous reception at a given EIRP level over long periods should be much less likely than intermittent reception.\footnote{If we assume ET is scanning over a large solid angle, continuous receptions would imply low gain $G_{TX}$ so that so that $P_{TX} = EIRP/G_{TX} \ge EIRP_{MIN}/G_{TX}$ could be extraordinarily large.}  }
\end{itemize}
To balance contributions over the various haystack variable ranges, we need to appropriately weight the integrand, most appropriately with an assumed \textit{a priori} multidimensional joint probability distribution.  If variables are correlated, this can be appropriately modelled.  Clearly \textit{a priori} distributions are unknown and must be hypothesized.  We may introduce biases with our assumptions, but the current haystack functions make the implicit assumption that all bins are equally likely. 
\item{The choice of $d_{H\!-\!MAX}$ has a dramatic effect on the haystack volumes: do we want ranges well beyond those we expect to detect to dominate our metric? }
\item{While the integrand $S(d;EIRP_{MIN},\phi_{MIN})$} elegantly combines detections over a continuum of EIRP and d combinations, and defines an ideal haystack corresponding to $\phi_{MIN}\rightarrow 0$, it's unclear whether the volume integral has the right characteristics.  For example, we might expect $V_{H1\!-\!survey}(EIRP_{MIN},\phi_{MIN})$ to reduce to a variation of the Drake FOM for some conditions.\footnote{For total TX bandwidth less than the instrument bandwidth and the special case $d_{H\!-\!MAX=d_{crit}}$, the dominant 6D haystack term ($V_{12}\Omega$ or $V_{22}\Omega$ in the paper) is proportional to ${EIRP_{MIN}}^{3/2} \; \Omega _{Total} \; {\phi_{MIN}}^{-5/2} \; BW_{Total} \; \Delta\nu = DFM \; {EIRP_{MIN}}^{3/2} \; \Delta\nu/\phi_{MIN}$.  The extra ${EIRP_{MIN}}^{3/2}$ and $\Delta\nu$ factors are reasonable.  The additional $1/\phi_{MIN}$ factor makes the units work out but the justification is not obvious.}}
\item{There is a subtle issue, possibly just of interpretation, regarding the number of observations and solid angle.  For $V_{H1\!-\!survey}$, integration is over the total solid angle surveyed $\Omega_{total}=N_{obs}\Omega_{FOV}$, while for $V_{H1\!-\!total}$, integration is over $4\pi$. Generally past SETI efforts have had $\Omega_{total} < 4\pi$, but has $\Omega_{total} > 4\pi$ been considered?  To avoid confusion, we should keep $N_{obs}$ and $\Omega_{FOV}$ separate, with $\Omega_{FOV}\le4\pi$.}
\end{itemize} 

\subsection{Haystack Function Example}

As an example, consider the haystack functions for an unresolved narrow-band case where we assume the only unknown TX parameter is TX center frequency $\nu$, and $BW_{TX}<\Delta\nu< BW_{total}<BW_{H\!-\!MAX}$, where  $BW_{TX}$, $\Delta\nu$, $BW_{total}=\nu_{MAX}\!-\!\nu_{MIN}$, and $BW_{H\!-\!MAX}=\nu_{H\!-\!MAX}\!-\!\nu_{H\!-\!MIN}$ are transmit bandwidth, resolution bandwidth, instrument bandwidth and haystack bandwidth, respectively. Let $\Omega=N_{obs}\,\Omega_{FOV}$ and ${d_{crit}}^2=EIRP_{MIN}/(4 \pi \phi_{MIN})$. Again $d_{crit}$ is the detection range corresponding to $EIRP_{MIN}$.  After integrating $\Omega$ and instrument bandwidth, and ignoring edge effects, the survey haystack becomes:

\begin{equation}  \label{eq:eqn111}
V_{H1\!-\!survey} = N_{obs}\,\Omega_{FOV}\,BW_{total} \int_{0}^{d_{H\!-\!MAX}} S(d;EIRP_{MIN},\phi_{MIN})\, d^2 \, dd 
\end{equation}
\begin{equation}  \label{eq:eqn112} \nonumber
= N_{obs}\,\Omega_{FOV}\,BW_{total} \left[ \int_{0}^{d_{crit}} \frac{4 \pi d^4\,dd}{EIRP_{MIN}} + \int_{d_{crit}}^{d_{H\!-\!MAX}} \frac{d^2\,dd}{\phi_{MIN}}  \right]
\end{equation}
\begin{equation}  \label{eq:eqn113} \nonumber
= N_{obs}\,\Omega_{FOV}\,BW_{total} \left[\frac{4 \pi {d_{crit}}^5}{5\,EIRP_{MIN}} + \frac{{d_{H\!-\!MAX}}^3 - {d_{crit}}^3} {3\,\phi_{MIN}}  \right]
\end{equation}
\begin{equation}  \label{eq:eqn114} \nonumber
= N_{obs}\,\Omega_{FOV}\,BW_{total} \left[\frac{{d_{H\!-\!MAX}}^3 - \frac{2}{5}{d_{crit}}^3} {3\,\phi_{MIN}}  \right]
\end{equation}

\noindent We note that if $d_{crit}\approx100\,$pc and $d_{H\!-\!MAX}=10\,$kpc as suggested, then $d_{crit}\ll d_{H\!-\!MAX}$ and $V_{H1\!-\!survey}$ is barely a function of $EIRP_{MIN}$.  Unless $d_{H\!-\!MAX}$ is very close to $d_{crit}$, we get:

\begin{equation}  \label{eq:eqn115}
V_{H1\!-\!survey} \approx N_{obs}\,\Omega_{FOV}\,BW_{total} \left[\frac{{d_{H\!-\!MAX}}^3} {3\,\phi_{MIN}}  \right]
\end{equation}

\noindent Meanwhile the total haystack volume with $\Omega=4\pi$ is

\begin{equation}  \label{eq:eqn116}
V_{H1\!-\!total} = 4\pi\,BW_{H\!-\!MAX} \int_{0}^{d_{H\!-\!MAX}} \frac{4 \pi d^4\,dd}{EIRP_{MIN}} 
\end{equation}
\begin{equation}  \label{eq:eqn117} \nonumber
= 4\pi\,BW_{H\!-\!MAX} \frac{4 \pi}{EIRP_{MIN}} \frac{ {d_{H\!-\!MAX}}^5}{5} = 4\pi\,BW_{H\!-\!MAX} \frac{1}{{d_{crit}}^2\,\phi_{MIN}} \frac{ {d_{H\!-\!MAX}}^5}{5} 
\end{equation}

\noindent After some arranging, the haystack fraction becomes

\begin{equation}  \label{eq:eqn118}
\Phi_{H1\!-\!total} = \frac{V_{H1\!-\!survey}}{V_{H1\!-\!total}} = N_{obs}\,\frac{5}{3}\,\frac{{d_{crit}}^2}{{d_{H\!-\!MAX}}^2}
\frac{\Omega_{FOV}}{4\pi} \frac{BW_{total}}{BW_{H\!-\!MAX}}
\end{equation}

\noindent We note that an arbitrary selection of $d_{H\!-\!MAX}$ will produce a rather arbitrary haystack fraction!  This is troubling: we would hope that there would be little sensitivity to $d_{H\!-\!MAX}$ provided it is large enough to cover potential detection ranges.  No matter how perfect a receiver is with realistic lower bounds of $\phi_{MIN}$, the haystack fraction will never be close to unity for arbitrary $d_{H\!-\!MAX}$.  The haystack fraction as defined hardly seems to be a reliable measure of search fraction. 

The ratio between haystack volumes for two systems A and B looks a bit more intuitive as cascade of parameter ratios:

\begin{equation}  \label{eq:eqn119}
\frac{V_{H1\!-\!survey.A}}{V_{H1\!-\!survey.B}} = 
\frac{N_{obs.A}}{N_{obs.B}} \frac{{\phi_{MIN.A}}^{-1}}{{\phi_{MIN.B}}^{-1}} \frac{\Omega_{FOV.A}}{\Omega_{FOV.B}} \frac{BW_{total.A}}{BW_{total.B}}
\end{equation}

\noindent One observation: if the $\phi_{MIN}$ values above had a $-\frac{3}{2}$ exponent, both numerator and denominator would be the respective Drake FOMs. 

A similar analysis can be done for fully resolved signals. Given the issues identified above, we propose alternative Haystack formulations below.

\subsection{Modified Sensitivity Haystack Function}

First, in spite of reservations due to the issues just identified, we can suggest modifications to the original haystack functions in (\ref{eq:eqn101}) to (\ref{eq:eqn105}).  Define a volume integral for observation i and a specific combination the TX parameters $\gamma_{TX}\in\Gamma_{TX}$ as follows:

\begin{equation}  \label{eq:eqn121}
V_{H2\!-\!obs}(i;EIRP_{MIN},\phi_{MIN},\gamma_{TX}) = \int_{V} S_i(d;EIRP_{MIN},\phi_{MIN},\gamma_{TX})  \, dV
\end{equation}

\noindent Integration is over a full sphere of radius $d_{H\!-\!MAX}$ but we assume $S_i$ has support only over a solid angle $\Omega_{FOV}$. Define a combined survey volume which applies a weighting of the \textit{a priori} distribution $f_{\Gamma_{TX}}(\gamma_{TX})$, integrates over all combinations of TX parameters, and sums all observations together, as follows:

\begin{equation}  \label{eq:eqn122}
V_{H2\!-\!survey}(EIRP_{MIN}) = \sum_{i=1}^{N_{obs}} \; \int_{\Gamma_{TX}} \; V_{H2\!-\!obs}(i;EIRP_{MIN},\gamma_{TX}) \, f_{\Gamma_{TX}}(\gamma_{TX}) \, d\gamma_{TX}
\end{equation}
\begin{equation}  \nonumber 
= \sum_{i=1}^{N_{obs}} \; E_{[\Gamma_{TX}]}  \left[ V_{H2\!-\!obs}(i;EIRP_{MIN},\gamma_{TX}) \right]
\end{equation}

\noindent We note that such a weighting produces the expected value\footnote{We denote $E_{[\Gamma_{TX}]}$ for the expected value over all TX parameters $\Gamma_{TX}$.} of $V_{H2\!-\!obs}$ over all possible combinations of TX parameters.  Therefore, $V_{H2\!-\!survey}$ represents the ``parameter ensemble'' average of the volume integral of sensitivity, summed over all observations.

The ideal haystack is just the volume integral of the $S_{MAX}$ function, and the haystack fraction is the ratio as before:

\begin{equation}  \label{eq:eqn123}
V_{H2\!-\!total}(EIRP_{MIN}) = \int_{V} S_{MAX}(d;EIRP_{MIN})  \, dV
\end{equation}

\begin{equation}  \label{eq:eqn124}
\Phi_{H2} = V_{H2\!-\!survey}/V_{H2\!-\!total}
\end{equation}

\vspace{1mm}

\noindent The integration for $V_{H2\!-\!total}$ is again over a sphere of radius $d_{H\!-\!MAX}$.

This formulation is easier to interpret: the haystack is an average volume integral of sensitivity, with units of $(W/m^2)\,m^3=W\!-\!m$.  The probability distribution weighting makes us less sensitive to how the haystack dimensions or limits are defined, and removes the units associated with TX parameters.  For example, if we limit $\Gamma_{TX}$ to be a subset of the total (e.g. narrow-band signals of less than 10 Hz bandwidth over the instrument bandwidth), we can evaluate how well we are doing for a certain more limited class of signals.  We can examine how each observation contributes to the haystack fraction, and we can sum observations from separate SETI efforts to realize a composite haystack fraction.  Results from multiple algorithms (e.g. one in the frequency domain, another in the time domain) can be combined as well\footnote{If two algorithms produce results that are independent (no overlap), they can be simply summed.  Otherwise  correlations in performance need to taken into account.}.

The use of integrated sensitivity is intuitively appealing as the result is a linear sum of average sensitivities from each cubic meter in the sphere.  However, the issues with the choice of $d_{H\!-\!MAX}$ and behavior of the haystack fraction remain.  It is unclear what averaged sensitivity and the haystack ratio tell us about the probability of detecting ET.  

\subsection{Detection Volume and Star Count Haystack Functions}

It should be possible to find a different quantity to integrate over volume with better characteristics.  A first possible measure is the detection volume:

\begin{equation}  \label{eq:eqn131}
V_{Det\!-\!obs}(i;EIRP_0,\gamma_{RXi},\gamma_{TX}) = \int_{V} P_D(d;EIRP_0,\gamma_{RXi},\gamma_{TX})  \, dV \approx \frac{1}{3} \, \Omega_{FOV} \, {d_{MAX}}^3
\end{equation}

\noindent Of course, $P_D$ varies between 0 and 1. $\gamma_{TX}\in\Gamma_{TX}$ is an instance within the set of transmit parameters as before, while $\gamma_{RXi}$ is an instance within the set of receiver parameters for observation i. Note that some RX parameters will be specific to an observation, including the steering direction and $T_{sys}$. 

We can similarly calculate the expected number of stars within the detection range.  The haystack volume associated with a single observation is:

\begin{equation}  \label{eq:eqn132}
N_{STAR\!-\!obs}(i;EIRP_0,\gamma_{RXi},\gamma_{TX}) = \int_{V}  \rho_{STAR} \, P_D(d;EIRP_0,\gamma_{RXi},\gamma_{TX})  \, dV 
\approx \frac{1}{3} \, \rho_{STAR} \, \Omega_{FOV} \, {d_{MAX}}^3
\end{equation}

\noindent We can compute expected values over $\Gamma_{TX}$ and sum all observations as before:

\begin{equation}  \label{eq:eqn133}
N_{STAR\!-\!survey}(EIRP_0) = \sum_{i=1}^{N_{obs}} \; \int_{\Gamma_{TX}} \; N_{STAR\!-\!obs}(i;EIRP_0,\gamma_{RXi},\gamma_{TX}) \, f_{\Gamma_{TX}}(\gamma_{TX}) \, d\gamma_{TX}
\end{equation}
\begin{equation}  \nonumber 
= \sum_{i=1}^{N_{obs}} \; E_{[\Gamma_{TX}]} \left[ N_{STAR\!-\!obs}(i;EIRP_0,\gamma_{RXi},\gamma_{TX}) \right] = E_{[\Gamma_{TX}]} \left[SFOM1(\gamma_{TX}) \right]
\end{equation}

\noindent Note that $\rho_{STAR}$ and $d_{MAX}$ may be functions or range and steering direction. Per section \ref{ssec:NominalEIRP}, we use a reference value for $EIRP_0$ (e.g. $10^{13}W$) to provide a benchmark value of $N_{STAR}$, and we can scale results by $(EIRP/EIRP_0)^{3/2}$ otherwise.  The haystack function $N_{STAR\!-\!survey}$ ends up being equivalent to the TX ensemble average of $SFOM1 = N_{obs} N_{STAR}$ from  (\ref{eq:eqn21}).

\subsection{The Detection Count Haystack Function}

While total detection volume or star-observations could be useful work-unit measures for an ET search, it is not obvious how to define the ideal total haystack volume metrics for these quantities, which would then determine corresponding haystack fractions. We propose that the best measure of ET search performance is the expected number of detections that would be produced during a survey effort.  We can obtain this with proper scaling of star-observations, as follows:

\begin{equation}  \label{eq:eqn141}
N_{Det\!-\!obs}(i;EIRP_0,P_{civTX},\gamma_{RXi},\gamma_{TX}) = \frac{T_{TXdwell}\,P_{civTX}}{T_{TXscan}}\,\int_{V}  \rho_{STAR} \, P_D(d;EIRP_0,\gamma_{RXi},\gamma_{TX})  \, dV 
\end{equation}
\begin{equation}  \nonumber 
= \frac{T_{TXdwell}\,P_{civTX}}{T_{TXscan}}\,N_{STAR\!-\!obs}(i;EIRP_0,\gamma_{RXi},\gamma_{TX})
\approx \frac{T_{TXdwell}\,P_{civTX}}{T_{TXscan}}\,\frac{1}{3} \, \rho_{STAR} \, \Omega_{FOV} \, {d_{MAX}}^3
\end{equation}

\clearpage
\noindent The survey haystack function is:
\begin{equation}  \label{eq:eqn142}
N_{Det\!-\!survey}(EIRP_0,P_{civTX}) = \sum_{i=1}^{N_{obs}} \; \int_{\Gamma_{TX}} \; N_{Det\!-\!obs}(i;EIRP_0,P_{civTX},\gamma_{RXi},\gamma_{TX}) \, f_{\Gamma_{TX}}(\gamma_{TX}) \, d\gamma_{TX}
\end{equation}
\begin{equation}  \nonumber 
= \sum_{i=1}^{N_{obs}} \; E_{[\Gamma_{TX}]} \left[ N_{Det\!-\!obs}(i;EIRP_0,P_{civTX},\gamma_{RXi},\gamma_{TX}) \right] 
\end{equation}
\begin{equation}  \nonumber 
= P_{civTX}\, \sum_{i=1}^{N_{obs}} \; E_{[\Gamma_{TX}]} \left[ \frac{T_{TXdwell}\,N_{STAR\!-\!obs}(i;EIRP_0,\gamma_{RXi},\gamma_{TX})}{T_{TXscan}} \right]
\end{equation}

We also propose that a reasonable haystack fraction for a search effort is the fraction of searching performed compared to that required to achieve a single detection, averaged over the TX ensemble.  The $N_{Det\!-\!survey}$ volume is exactly this.  Therefore, if we let $N_{Det\!-\!total} = 1$ over all combinations of TX parameters, the haystack fraction will simply be 

\begin{equation}  \label{eq:eqn145}
\Phi_{Det}=N_{Det\!-\!survey}
\end{equation}

An update to the Haystack Boundary table from \citep{Wright_2018AJ} appears in Table \ref{tbl:TableA1} and summarizes the necessary search dimensions.  The distance parameter $d_{H\!-\!MAX}$ covers all distances from 0 to $\infty$ and no longer needs to be specified, as $P_D(d)\rightarrow 0$ beyond $d_{MAX}$.  Note that frequency drift rate has been added, per \citet{Siemion_2013_SETI_Kepler_TurboSETI} and  \citet{Sheikh_2019}.  The TX dwell time $T_{TXdwell}$ has also been added.  The limits shown are hypothesized, and might be narrowed considerably with more analysis of algorithms and link budgets. The fact that variables are intertwined (correlated) makes the haystack model somewhat more complex.

\begin{deluxetable*}{|p{40mm}|c|c|c|p{70mm}|}
\tablecaption{\textbf{Summary of Revised Haystack Dimensions}}
\tablewidth{0pt}
\label{tbl:TableA1}
\tablehead{
\colhead{\textbf{Dimension}} & \colhead{\textbf{Symbol}}  & \colhead{\textbf{Lower Bound}} & \colhead{\textbf{Upper Bound}} & \colhead{\textbf{Comments}} 
}
\startdata
\textbf{Volume parameters} & V &  &  & \\ \hline		
Distance & $d_{H\!-\!MAX}$ & 0 & $\infty$ & Detection range $d_{MAX}$ is determined by RX and TX parameters.  Assume uniform \textit{a priori} distribution over all space, so TX density only depends on $\rho_{STAR}.$ \\ \hline	
Solid angle coverage & $\Omega$ & 0 & $4\pi$ & RX scan should be over full $4\pi$ in WFS, but limited to $\Omega_{FOV}$ for any one observation. $\Omega_{FOV}$ will be a function of frequency.  Distinct from $\Omega_{total} = N_{obs} \Omega_{FOV}$  \\ \hline
\textbf{Transmit Parameters} & $\Gamma_{TX}$ &  &  &  \\ \hline
Effective Isotropic Radiated Power & EIRP & $10^{13}\,$W &  & Leave as a reference value $EIRP_0=10^{13}\,$W, so EIRP is left out of the haystack integration. Could alternatively consider a truncated power law distribution.\\ \hline
TX center frequency & $\nu$ & 10 MHz & 115 GHz & FFOM peaks over 300-700 MHz, and declines rapidly above 2 GHz.  Assume \textit{a priori} distribution is strongly frequency-dependent, favoring lower frequencies. \\ \hline
TX Bandwidth	 & $BW_{TX}$ & 1 Hz & 20 MHz & Narrow-band ($<\!10$ KHz) is the focus of current analysis, but wideband pulse train waveforms are possible. Form of the \textit{a priori} distribution is unclear, but certainly is not uniform.\\ \hline
TX Dwell Time & $T_{TXdwell}$ & 1 minute & 1 hour & Amount of time spent in one TX look direction.  Active transmission time is $T_{TX}=\delta_{TX}\,T_{TXdwell}$. where $\delta_{TX}$ is the TX duty cycle.\\ \hline
Repetition Period & $T_{rep}$, $T_{TXscan}$ & Continuous & 1 year & TX scans over $4\pi$ in time $T_{TXscan}$ in model for WFS.  Continuous reception is unlikely due to average TX power limitations. EIRP, $T_{TXscan}$, and $T_{TXdwell}$ will be correlated. Could also use $N_{TXscan}=T_{TXscan}/T_{TXdwell}$ as a measure of repetition period. \\ \hline
Polarization fraction & $\eta_{pol}$ & 0 & 1 & 0=unpolarized, 1=completely polarized. Equals 1 for most modern receivers (Stokes I, full coverage), so this can generally be ignored in the analysis. \\ \hline
Normalized drift rate & $\dot{\nu}_{norm}$ & $-200$ nHz & $200$ nHz & $\dot{\nu}_{norm}=\dot{\nu}/\nu$. Doppler drift rate is due to TX-RX relative accelerations.  The cause is astrophysical.  Possibly compensated by ET. \\ \hline
\enddata
\end{deluxetable*}

\subsection{Conclusions on Haystack Functions}

To sum up:

\begin{itemize}
\item {Issues with the Wright haystack functions (\ref{eq:eqn101}) to (\ref{eq:eqn105}) were identified, suggesting that the haystack fraction is unreliable as currently defined. }
\item{A new paradigm was proposed which weights a desired FOM measure according to the \textit{a priori} joint pdf of the TX parameters.  The prior pdfs may be interpreted as an ``ET behavior model'' and specifies what combinations of TX parameters are likely or unlikely.  This reduces much of the haystack integration to an expected value operation over the TX parameters.  This approach leads to reasonable scaling and removes TX parameter units from the haystack value.  The haystack can be readily interpreted as a volume integral of the FOM measure summed over all observations and averaged over the ensemble of TX parameters.}
\item{A haystack function $N_{Det\!-\!survey}$ was defined for a SETI effort based on the expected number of ET detections that that the survey should generate with certain assumed benchmark parameters like EIRP and $P_{civTX}$.  We define the reference ideal detection count to be unity for any set of TX parameters (though of course we would like more), so the search fraction $\Phi_{Det}$ is simply $N_{Det\!-\!survey}$}.
\end{itemize}

More work is needed to evaluate the revised haystack fraction of past SETI surveys.  The hardest part may be establishing reasonable \textit{a priori} pdfs over the full range of possible TX parameters.  There may be many arguments that produce very different but justifiable pdfs.  As a first step, it may be useful to evaluate haystacks that apply to a limited range of TX parameters, e.g. narrow-band signals with a limited range of TX bandwidths, to gain experience before considering the full TX parameter space.

\clearpage

\section{SETI Search with Limited Computation} \label{sec:LimitedComp}

We have defined the detection rate DPY assuming a set of spatial and waveform parameters.  Spatial parameters include range d and steer direction ($\alpha,\delta$), where $\alpha$=Right Ascension and  $\delta$=Declination. The set of transmit parameters is suggested in Table \ref{tbl:TableA1}, but for simplicity let us assume $\Gamma_{TX}$ is limited to frequency $\nu$ and frequency rate $\dot{\nu}$, as is typically done in narrow-band SETI searches.  Ideally the search process will span the range of these parameters.  Given hardware limitations and finite computation, the search ranges must be limited, and it will take analysis and expert opinion to decide the best search dimensions and limits. 

To guide the choice of search ranges, consider the expected number of detections $\overline{N_{Det}}$ for a survey with $N_{obs}$ observations over all RX sites.  Per (\ref{eq:eqn132}):

\begin{equation}  \label{eq:eqn201}
\overline{N_{Det}} = P_{civTX}\, \sum_{i=1}^{N_{obs}} \; E_{[\Gamma_{TX}]} \left[ \frac{T_{TXdwell}\,N_{STAR\!-\!obs}(i;EIRP_0,\gamma_{RXi},\gamma_{TX})}{T_{TXscan}} \right]
\end{equation}

\noindent The search is over discrete bins in frequency and frequency rate, so the expected value is done discretely over two levels of summation:

\begin{equation}  \label{eq:eqn202}
\overline{N_{Det}} = \frac{T_{TXdwell}\,P_{civTX}}{T_{TXscan}} \sum_{i=1}^{N_{obs}} \; \sum_{j=1}^{N_F} \\
 \sum_{k=1}^{N_R} \; N_{STAR}(\alpha_i,\delta_i,\nu_j,\dot{\nu_k}) \\
 \; p_1(\nu_j) \; p_2(\nu_j,\dot{\nu_k})
\end{equation}

\noindent $p_1$, and $p_2$, are discrete \textit{a priori} probabilities associated with frequency bin j and frequency rate bin k.  Assume these are related to prior distributions as follows:

{\addtolength{\leftskip}{10 mm}
\setlength{\parindent}{0 mm}

$ $

$p_1(\nu_j) = f_1(\nu_j)\;\Delta\nu$, \hspace{1.5cm} $f_1(\nu)$ = \textit{a priori} frequency density

$ $

$p_2(\nu_j,\dot{\nu_k}) = f_2(\dot{\nu_k}/\nu_j)\;\Delta\dot{\nu}$, \hspace{5mm}$f_2(\dot{\nu}_{norm})$ = \textit{a priori} normalized frequency-rate density, $\dot{\nu}_{norm} = \dot{\nu}/\nu$, 

$ $

}

\noindent where $\Delta\nu$ is the frequency bin width and  $\Delta\dot{\nu}\approx\Delta\nu/\tau$ \citep{Siemion_2013_SETI_Kepler_TurboSETI} is the frequency-rate bin width.  We assume that $\nu_j$ and $\dot{\nu}_{norm-k}$ (which is related to astrophysical accelerations \citep{Sheikh_2019}) are mutually independent.  It follows that: 

\begin{equation}  \label{eq:eqn30b}
\overline{N_{Det}} =  \frac{P_{civTX}\,AFOM\,AvgFOM^{3/2}\,EIRP^{3/2}}{N_{TXscan}\,3\,(8 \pi k_B)^{3/2}} \sum_{i=1}^{N_{obs}} \; \rho_{STAR}(\alpha_i,\delta_i) \; \\
\sum_{j=1}^{N_F} \; FFOM(\alpha_i,\delta_i,\nu_j) \; f_1(\nu_j) \\ 
 \sum_{k=1}^{N_R} \;  f_2(\dot{\nu_k}/\nu_j) \; \Delta\nu \; \Delta\dot{\nu}
\end{equation}

We will have $N_{obs} N_F N_R$ space/frequency/frequency-rate bins to compute over $T_{obs}$, and perhaps $N_F N_R$ frequency/frequency-rate bins to compute in real time.  The question: how can one best allocate coverage to maximize $\overline{N_{Det}}$ within a computation budget?  Clearly this is a trade space; we need to examine each bin contribution and choose coverage so as to maximize the above summation. Some notes:

\begin{itemize}
\item{We should choose our observation sequence $(\alpha_i,\delta_i)$ based on the $T_{sky}$-adjusted star density, as reflected by star density-FFOM product $[\rho_{STAR}(\alpha_i,\delta_i)\,FFOM(\alpha_i,\delta_i,\nu_j)]$.  (Note that FFOM is a function of $T_{sys}$ which is in turn a function of $T_{sky}(\alpha_i,\delta_i)$.) The density $\rho_{STAR}(\alpha_i,\delta_i)$ should be a representative value near $d_{MAX}$.}
\item{Assume $FFOM(\alpha_i,\delta_i,\nu_j)$=$FFOM(\nu_j)$  for now.  The frequency bin contribution is $FFOM(\nu_j) \, f_1(\nu_j)$, so FFOM is scaled by the \textit{a priori} frequency distribution $f_1(\nu)$.  We have never observed ET, so we can only hypothesize $f_1(\nu)$.  Several schools of thought:}
\begin{itemize}
\item{FFOM strategy: If one believes $f_1(\nu)$=$constant$, then we should choose $N_F$ bins which best span the FFOM($\nu$) peak.}
\item{Water Hole strategy: If $f_1(\nu)$ is bandpass near the ``Water Hole'' (1420-1662 MHz), then FFOM($\nu$) is largely irrelevant, and we should choose bins near those frequencies.}
\item{Big straddle strategy: We could choose to cover from  $\sim$100 MHz to the water hole as a compromise.  FFOM may vary over an order of magnitude in this case, or less with aperture arrays depending on how station beamforming is done.}
\item{Otherwise, we could evaluate functions like a truncated power law distribution ($f_1(\nu)=k\nu^\beta$ over a certain band, with k a normalizing constant) and examine the $FFOM(\nu_j) \, f_1(\nu_j)$ product.  Since $\beta$ would presumably need to be greater than 2 (implying a very strong bias toward high frequencies) to equalize the downslope of FFOM above 2 GHz, it may be hard to justify high frequencies based on detection rate arguments. We presume that ET has concluded this also.}
\end{itemize}
\item{The frequency-rate bin contribution is $f_2(\dot{\nu_k}/\nu_j)$.  There are two differing strategies for choosing $N_R$:}
\begin{itemize}
\item{Span the expected rates due to astrophysics: A list of astrophysical phenomena which would cause TX-RX relative accelerations and the resulting maximum drift rates was explored by \citet{Sheikh_2019}, and subsequent work may  estimate the resulting $f_2(\dot{\nu}_{norm})$ distribution. One might guess this will be a Gaussian distribution centered on Earth's frequency rate which will be truncated by the choice of $N_R$. We need to choose $N_R$ so the tail areas are minimal. (Implicit in the earlier DPY derivation was that $N_R$ will be chosen large enough so as to cover virtually all possible cases.)}
\item{Assume that ET will adjust frequency rate: If a civilization is sufficiently sophisticated to be conducting Active SETI, it may choose to de-chirp its transmission so that it will be received at a low drift rate in an appropriate galactic frame of reference.  This would mean that $N_R$ could be much lower and search would be simplified. }
\end{itemize}
\item{Since the frequency rate effect is proportional to frequency, we might best choose $N_R$ to be proportional to frequency:  $N_R$=$N_R(\nu)\approx(\nu/\nu_0)\,N_R(\nu_0)$.  This would imply a trapezoidal search region in $\nu$ and $\dot{\nu}$ instead of a rectangular search region. $\overline{N_{Det}}$ might be further enhanced by choice of an arbitrary shape defined by ranking the bin contributions and taking the best set that fits within the computation budget.}
\item{The selection of $\Delta\nu$ affects $N_F N_R$ dramatically.  At a budgeted level of computation, we may be willing to trade SNR (as influenced by $AvgFOM(\Delta\nu)$) for more frequency or frequency-rate coverage so as to maximize $\overline{N_{Det}}$.}
\end{itemize}

We can see from the above discussion that there can be multiple legitimate arguments for the choice ET might make regarding transmit parameters.  Each will significantly affect prior distributions and therefore the optimum search parameters.  

We may also note that if ET has been conducting SETI for a long time (as might be expected before an Active SETI effort were to begin), ET would recognize classes of signals that would be easier to detect and distinguish from astrophysical signals.  This might favor very narrow-band signals that would not normally occur in the natural world, and discourage broad-band signals that might be confused with diffuse emitters, pulsars or fast radio bursts. The prior densities for transmit bandwidth would be biased accordingly.  Likewise, if ET understands that lower frequencies should have a higher detection rate according to FFOM, one might reasonably assume that ET would favor lower frequencies, and the frequency prior density would be biased to reflect this.  Therefore, without better information, own own intuition may be the best guide to ET's likely choices for transmit parameters.

\clearpage
\section{Additional References} \label{sec:References}

References about systems listed in Tables \ref{tbl:Table1} and \ref{tbl:Table2} may be found below.

\begin{deluxetable*}{|l|c|} [ht]
\tablecaption{Radio Astronomy System References}
\tablewidth{0pt}
\tablecolumns{2}
\label{tbl:Table7}
\tablehead{
\colhead{\textbf{Name}} & \colhead{\textbf{Reference}} 
}
\startdata
Parkes Observatory & \citet{Hobbs_2020} \\ \hline 
Green Bank Telescope & \url{https://www.gb.nrao.edu/scienceDocs/GBTpg.pdf} \\ \hline 
Parkes Multibeam & \citet{Staveley_Smith_1996}, \citet{Chippendale_2016} \\ \hline 
GBT FLAG PAF & \citet{GBT_FLAG_Rajwade_2017}, \citet{Pingel_GBT_FLAG_2018} \\ \hline 
Allen Telescope Array & \citet{ATA_Welch_2009_IEEEProc} \\ \hline 
Giant Metrewave RT & \citet{GMRT_Gupta_2017}  \\ \hline 
Arecibo Telescope & \url{https://www.naic.edu/ao/scientist-user-portal/astronomy/astronomer-guide} \\ \hline 
Jansky Very Large Array & \url{https://library.nrao.edu/evla.shtml}, \url{https://library.nrao.edu/public/memos/evla/EVLAM_195.pdf} \\ \hline 
FAST 500m Aperture & \citet{NAN_FAST_2011} \\ \hline 
Westerbork APERTIF & \citet{Apertif_2010_Oosterloo} \\ \hline 
MeerKAT 64 & \citet{MeerKAT_Lehmensiek_2014}, \citet{MeerKAT_Jonas_2016} \\ \hline 
ASKAP & \url{https://www.atnf.csiro.au/projects/askap/specs.html}, \citet{SKA1_Survey_Perf_2016}, \citet{ASKAP_McConnell_2020_Survey} \\ \hline
MeerKAT Extension 84 & \citet{MeerKAT_Ext_2020_SOW} \\ \hline 
CHORD & \citet{CHORD_vanderlinde_2019} \\ \hline 
Next Generation VLA & \citet{ngVLA_Sys_Design_Selina_2019} \\ \hline 
SKA1 Mid (197 Dish) & \citet{SKA1_Baseline_Dewdney_2015}, \citet{McPherson_SKA1_2018}, \citet{Braun_2019}, \citet{SKA1_Survey_Perf_2016}   \\ \hline 
L-Band Array of Small Arrays & \citet{Lynch_2018} \\ \hline 
Next Gen Arecibo Telescope & \citet{NGAT_White_Paper_2021} \\ \hline 
DSA-2000 & \citet{Hallinan_DSA2000_2019} \\ \hline 
SKA2 Mid Dish & \citet{Braun_2019} \\ \hline 
SKA2 Mid MFAA & \citet{Torchinsky_2017_MFAA}, \citet{Gunst_MFAA_2020} \\ \hline 
CHIME & \citet{CHIME_Amiri_2018} \\ \hline 
HIRAX & \citet{Newburgh_HIRAX_2016} \\ \hline 
\textbf{Murchison} Widefield Array 2 & \citet{MWA2_Wayth_2018}, \citet{MWA_Tinguay_2013_PASA}, \citet{MWA_Tremblay_Vela_2020} \\ \hline 
SKA1-Low & \citet{SKA1_Baseline_Dewdney_2015}, \citet{McPherson_SKA1_2018}, \citet{Braun_2019}   \\ \hline 
SKA2-Low &  \citet{Braun_2019}  \\ \hline 
\enddata
\end{deluxetable*}

\vspace{-20mm}

\begin{deluxetable*}{|l|c|} [ht]
\tablecaption{EIRP References}
\tablewidth{0pt}
\tablecolumns{2}
\label{tbl:Table8}
\tablehead{
\colhead{\textbf{System}} & \colhead{\textbf{Reference}} 
}
\startdata
GPS & \citet{Steigenberger_GPS_EIRP_2019,Wang_GPS_EIRP_ION_2018} \\ \hline 
FM Radio & \url{https://www.fcc.gov/media/radio/fm-station-classes} \\ \hline 
DTV  & \url{https://data.fcc.gov/download/incentive-auctions/OET-69/Baseline_Data_and_Maps_2013July.pdf} \\ \hline
ASR-9 ATC Radar & \citet{Weber_MIT_LL_Radar_2000} \\ \hline
Intelsat & \citet{INTELSAT_16QAM_2005}, \url{https://www.intelsat.com/fleetmaps/} \\ \hline
Arecibo Radar & \citet{Taylor_Arecibo_Radar_2016,Arecibo_430MHz_Manual_2001}, \url{http://www.naic.edu/aisr/sas/transmitter/trans-home.html}, \\ 
 & \url{http://www.naic.edu/~nolan/radar/} \\ \hline
\enddata
\end{deluxetable*}

\end{document}